\documentclass[a4paper,11pt]{article}
\usepackage{a4wide}
\usepackage{amsmath,amssymb,amsthm,mathtools,array,booktabs}
\usepackage{graphicx,hyperref}
\usepackage{xcolor}
\usepackage{cancel}
\usepackage{tikz}
\usepackage{amscd}
\usepackage{latexsym,cite}
\usepackage{subcaption}
\usepackage{ytableau}
\usepackage{scalerel}
\usepackage{calc}
\input xy
\xyoption{all}

\hypersetup{
  unicode,
  bookmarksnumbered,
  linktoc = all,
  pdfborderstyle = {/S/U/W 0.5}
}

\def\cO{\mathcal{O}}

\def\cA{\mathcal{A}}

\newcommand{\Af}{{A_\infty}}
\newcommand{\bP}{{\mathbb{P}}}

\numberwithin{equation}{section}

\begin{document}
\thispagestyle{empty}
\begin{flushright}
preprint
\end{flushright}
\vspace{1cm}
\begin{center}
{\LARGE\bf Hybrid models for homological projective duals and noncommutative 
resolutions} 
\end{center}
\vspace{8mm}
\begin{center}
{\large Jirui Guo\footnote{{\tt jrguo@tsinghua.edu.cn}}  and Mauricio 
Romo\footnote{{\tt 
mromoj@tsinghua.edu.cn}}}
\end{center}
\vspace{6mm}
\begin{center}
Yau Mathematical Sciences Center, Tsinghua University, Beijing, 100084, China
\end{center}
\vspace{15mm}

\begin{abstract}
\noindent
We study hybrid models arising as homological projective duals (HPD) of certain
projective embeddings $f:X\rightarrow\mathbb{P}(V)$ of Fano manifolds $X$. More 
precisely, the category of B-branes of such hybrid models corresponds to the 
HPD category of the embedding $f$. B-branes on these hybrid models can 
be seen as global matrix factorizations over some compact space $B$ or, 
equivalently, as the derived category of the sheaf of $\mathcal{A}$-modules on 
$B$, where  $\mathcal{A}$ is a sheaf of $A_{\infty}$-algebras. This latter 
interpretation corresponds to a noncommutative resolution of $B$. We compute 
explicitly the algebra $\mathcal{A}$ by several methods, for some specific 
class of hybrid models. If the target space of the hybrid model is a global 
orbifold, $\mathcal{A}$ takes the form of a smash product of an 
$A_{\infty}$-algebra with a finite group. However, this is not the case in 
general because the orbifold group can only be defined locally. One needs to 
treat the target space as an algebraic stack in such cases. We apply our results 
to the HPD of $f$ corresponding to a Veronese embedding of projective space and 
the projective embedding of Fano complete intersections in $\mathbb{P}^{n}$.
\end{abstract}
\newpage
\setcounter{tocdepth}{3}
\tableofcontents
\setcounter{footnote}{0}

\section{Introduction}

It is known that hybrid models provide realizations of a series of 
two-dimensional superconformal field theories which can be obtained from certain 
phases of gauged linear sigma models (GLSM) \cite{Witten:1993yc}. Roughly 
speaking, a hybrid model is a two-dimensional $\mathcal{N}=(2,2)$ 
supersymmetric field theory whose target space is of the form 
$Y=\mathrm{Tot}(\mathcal{E}\rightarrow B)$ for some holomorphic vector bundle 
(or an orbibundle, in general) $\mathcal{E}$ over the base space $B$ where the 
fields interact via a superpotential $W\in H^{0}(\mathcal{O}_{Y})$ which is a 
holomorphic function on the total space. The hybrid model can be viewed as a 
family of Landau-Ginzburg (LG) models fibred over the base space. Further 
sufficient conditions (but not necessary) in $Y$ and $W$ guarantee that these 
models RG flows to SCFTs and also makes them tractable as quantum field 
theories (QFT), see for instance \cite{Aspinwall:2009qy,Bertolini:2013xga}.

Recently, it is found that homological projective dual (HPD) 
\cite{kuznetsov2007homological} of certain projective embeddings can be 
described by hybrid models. This was found in 
mathematics \cite{ballard2017homological,rennemo2017fundamental} and a physics 
formulation is presented in\footnote{The first appeareance of HPD in the 
context of dynamics of GLSMs can be found in \cite{Caldararu:2007tc}.} 
\cite{Chen:2020iyo}: if a GLSM 
$\mathcal{T}_{X}$ for a projective morphism 
$f:X\rightarrow\mathbb{P}(V)$ is known, one can build up an extended GLSM 
$\mathcal{T}_{\mathcal{X}}$ such that the Higgs branch of one of its phases 
gives rise to the HPD of $f:X\rightarrow\mathbb{P}(V)$. In the abelian 
cases, this construction provides a very explicit characterization of the HPD 
of $f:X\rightarrow\mathbb{P}(V)$. Indeed, the Higgs 
branch of the phase of $\mathcal{T}_{\mathcal{X}}$, relevant to the question, 
is a hybrid model.

These hybrid models, with target space $Y=\mathrm{Tot}(\mathcal{E}\rightarrow 
B)$ can be view as (orbifold) LG models over affine charts of $B$ in the cases 
that $B$ is smooth, or if it has the structure of a global orbifold. This gives 
rise to the global structure of a noncommutative resolution or more generally, 
the B-brane category becomes the derived category of sheaves of 
$\mathcal{A}$-modules for some sheaf of algebras $\mathcal{A}$. Let us 
illustrate 
this in a well known example.

For a LG model with quadratic superpotential $W_{LG}$, it was shown in 
\cite{Kapustin:2002bi} that the category of B-branes (homotopy category 
of matrix factorizations) $MF(W_{LG})$ is equivalent to the derived category of 
finite dimensional Clifford modules, where the Clifford algebra is defined by 
the Hessian of the superpotential $W_{LG}$. In addition, if a $\mathbb{Z}_2$ 
orbifold is present that leaves $W_{LG}$ invariant, then the category 
of B-branes of this LG orbifold $MF(W_{LG},\mathbb{Z}_2)$ is equivalent to the 
derived category of finite dimensional modules of the even subalgebra of the 
corresponding Clifford algebra. Consequently, the category of matrix 
factorizations of a $\mathbb{Z}_2$-orbifold hybrid model with superpotential 
quadratic along the fiber coordinates is equivalent to the derived category of 
the sheaf of modules of the sheaf of even parts of a Clifford algebra over the 
base $B$. This is 
the case of the HPD category of the degree $2$ Veronese embedding
$\mathbb{P}(V)\hookrightarrow\mathbb{P}(\mathrm{Sym}^{2}V)$ 
\cite{kuznetsov2008derived}. Thus, the hybrid model orbifold can be viewed as a 
noncommutative resolution of $B$.

In this work, we generalize the idea and construct an explicit correspondence 
between hybrid models and noncommutative spaces. Consider first the case where 
$B$ can be described locally by affine charts and $Y$ is a global orbifold such 
that the orbifold group $G$ acts trivially on $B$. Then, at a generic 
point in 
the base $p\in B$ of the hybrid model, we can model its dynamics by a LG 
orbifold. Denote the category of B-branes of this LG orbifold as $MF(W,G)$, 
where $G$ is the orbifold group. We first study the $\Af$-algebra 
$\cA_{D0}=\mathrm{End}(\mathcal{B}_{D0})$ associated with the endomorphism 
algebra of a $D0$-brane $\mathcal{B}_{D0}\in MF(W)$ of the LG model. The 
algebra $\cA_{D0}$ takes the form of an $A_{\infty}$ algebra with a finite 
number of generators (as an algebra) $\psi_{i}$ that satisfy the higher 
products relations given in (\ref{algrels1}) and (\ref{algrels2}) (for a 
homogeneous $W$) and for general elements \eqref{relsgeneral1} and 
\eqref{relsgeneral2}\footnote{We also consider the case of inhomogenous 
$W$. The $A_{\infty}$ multiplication rules are given in \eqref{eqsinhomog1} and 
\eqref{eqsinhomog2}.}. It then 
sets up the equivalence between matrix factorizations and $\Af$-modules of 
$\cA_{D0} \sharp G$, where $\sharp$ denotes the 
smash product (the mathematical approach toward this equivalence can be found 
in \cite{tu2014matrix}), more precisely
\begin{eqnarray}
MF(W,G)\cong D(Mod-\cA_{D0} \sharp G),
\end{eqnarray}
where the appeareance of the derived category is a consequence that $MF(W,G)$ 
is taken to be the homotopy category. We then use this equivalence to relate a 
hybrid model to a noncommutative resolution 
\begin{eqnarray}
D(Y,W)\cong D(B, \cA_{D0} \sharp G),
\end{eqnarray}
i.e. the 
derived category of sheaf of $\cA_{D0} \sharp G$-modules over $B$. 

The more general case, for example when $G$ acts on $B$, or more precisely, 
when $B$ has orbifold singularities and/or it cannot be written as a global 
orbifold still has a similar structure. In such a case we have that $D(Y,W)$ is 
equivalent to the derived category of $\mathcal{A}$-modules for some 
sheaf of $A_{\infty}$-algebras $\mathcal{A}$, defined over the algebraic stack 
$Y$. We 
studied this case in detail in section \ref{sec:hybrid}.

These results can be used to study HPD of several spaces. As 
mentioned above, the GLSM construction realizes the HPDs as hybrid 
orbifold models, which can be identified with noncommutative 
resolutions, or derived categories of sheaves of $A_\infty$-modules, as the 
equivalence suggests. Therefore, given a projective embedding engineered by an 
abelian GLSM, the hybrid model describing the HPD can be read off following 
\cite{Chen:2020iyo}. One can then use the correspondence discussed in this paper 
to give a noncommutative geometric decription of the HPD. More precisle we 
apply these results to the follwoing families of examples
\begin{enumerate}
 \item \textbf{HPD of degree $d$ Veronese embedding of $\mathbb{P}^{n}$}: The 
HPD of the degree $d$ Veronese embedding 
$\mathbb{P}(V)\hookrightarrow\mathbb{P}(\mathrm{Sym}^{2}V)$, $V\cong 
\mathbb{C}^{n+1}$ was studied in \cite{Chen:2020iyo,ballard2014derived} and is 
found to be given by a hybrid LG orbifold i.e. its target space $Y$ can be 
written as a global orbifold, specifically
\[
Y=\mathrm{Tot}\left( \mathcal{O}\left( -\frac{1}{d} \right)^{\oplus(n+1)} 
\rightarrow \mathbb{P}^{{n+d \choose d}-1} \right)/\mathbb{Z}_d
\]
where $\mathcal{O}\left( -\frac{1}{d} \right)$ denotes an orbibundle (see 
appendix \ref{app:orbibundle}) and a superpotential of degree $d$ in the fiber 
coordinates. In section \ref{sec:HPDVeronese} we show that the B-brane 
category of this hybrid model can be written as
\begin{equation}\label{eqver}
D(\bP^{{n+d \choose d}-1},\cA_{0}\sharp \mathbb{Z}_d) = \langle 
\cA^{(1-C_{d,n})},\cdots,\cA^{(-1)},\cA^{(0)}\cong\mathcal{A}_{0} \rangle,
\end{equation}
where $\cA_{0}\cong \mathcal{A}_{D0}$ is the $A_{\infty}$-algebra described 
above. We also describe explicitly the components of the semiorthogonal 
decomposition (predicted in \cite{Chen:2020iyo}), in (\ref{eqver}), in terms of 
$\cA_{0}\sharp \mathbb{Z}_d$-modules.
 \item \textbf{HPD of Fano hypersurfaces}: The 
HPD of the degree $d$ Fano hypersurface inside $\mathbb{P}^{n}$, embedded 
naturally, is analyzed in section \ref{sec:Fanohyper}. This case, analyzed in 
\cite{Chen:2020iyo,ballard2014derived}, is similar to the HPD of the degree $d$ 
Veronese embedding, since the target space of the HPD is also a hybrid LG 
orbifold with target space given by a global orbifold, namely
\[
Y=\mathrm{Tot}\left( \cO_{\check{\mathbb{P}}^{n}}^{\oplus(n+1)}\oplus 
\cO_{\check{\mathbb{P}}^{n}}(-1) \rightarrow \check{\bP}^n \right)/\mathbb{Z}_d 
\]
and superpotential
\begin{equation}
W = F_{d}(x) + p \sum_{i=0}^n x_i y_i,
\end{equation}
(see section \ref{sec:Fanohyper} for a detailed description of the 
variables). The B-brane category of this hybrid model has a (dual) Lefschetz 
decomposition proposed in \cite{Chen:2020iyo}:
\begin{equation}
D(Y,W) = \langle \mathcal{B}_{n-1}(1-n), 
\mathcal{B}_{n-2}(2-n),\cdots, \mathcal{B}_2(-2), \mathcal{B}_1(-1), 
\mathcal{B}_0 \rangle,
\end{equation}
in section \ref{sec:Fanohyper} we describe this decomposition explicitly using 
the Lefschetz decomposition of the Fano hypersurface, induced by the natural 
embedding. In addition we describe $D(Y,W)$ as a noncommutative resolution
\begin{equation}
D(Y,W) \cong D(\check{\mathbb{P}}^{n},\cA_{0}\sharp \mathbb{Z}_d),
\end{equation}
giving very explicit constructions for the cases of degrees $d=2,3$.

\item \textbf{HPD of Fano complete intersections}: The 
HPD of Fano complete intersections was studied in 
\cite{Chen:2020iyo}. For an intersection of hypersurfaces of degree $d_{i}$, 
$i=1,\ldots, k$ on $\mathbb{P}^{n}$, the HPD was found to be given by a 
hybrid model with target space
\[
Y=\mathrm{Tot}\left( \mathcal{O}(-1,0)^{\oplus(n+1)} \oplus \mathcal{O}(1,-1) 
\rightarrow \mathrm{W}\mathbb{P}(d_1,\cdots,d_k) \times \check{\mathbb{P}}^n 
\right)
\]
and superpotential
\[
W = \sum _{\alpha=1}^k p_{\alpha} F_{d_{\alpha}}(x) +  p\sum_{i=0}^n x_i y_i,
\]
details of the notation can be found in section \ref{sec:HPDCIII}. In this 
case, the main difficulty lays on the fact that, in general, $Y$ cannot be 
written as a global orbifold, but as a local one. In special cases such as 
$d_{i}=d$ for all $i$, we can write it as a global orbifold. As far as we are 
aware this is the less studied case in the literature. In section 
\ref{sec:hybrid} we describe the sheaf of algebra $\mathcal{A}$ in this case, 
as a sheaf of algebras over the algebraic stack $Y$ and we apply it, in section 
\ref{sec:HPDCIII}, to the HPD of Fano complete intersections.

\end{enumerate}

This paper is organized as follows. We review the basic facts about GLSMs for 
HPD and $\Af$-algebras in sections \ref{sec:section2} and \ref{sec:LGAinf} 
respectively. In section \ref{sec:AinfLG}, we set up the relationship between 
matrix factorizations and $\Af$-modules. We first find the structure of the 
$\Af$-algebra $\cA_{D0}$ by various means (deformation theory of 
$\mathcal{B}_{D0}$ and effective superpotential, $\Af$-homomorphism), then 
we propose a functor realizing the equivalence between the category of matrix 
factorizations and the derived category of $\Af$-modules of $\cA_{D0}$ and 
sketch its generalization to the orbifold case $\cA_{D0} \sharp G$. We provide 
checks of this proposal in appendix \ref{app:checks}. We then apply this 
correspondence in section \ref{sec:examples} to describe the HPD of degree $d$ 
Veronese embedding of projective space, Fano hypersurfaces and complete 
intersections in projective spaces as noncommutative spaces with the structure 
sheaf given by the corresponding sheaf of $\Af$-algebras. The same result was 
obtained for Veronese embeddings by summing over the ribbon trees in 
\cite{ballard2014derived}, we review this method in appendix \ref{app:ribbon}. 
Finally, in section \ref{sec:functor} we give some details of the functor 
$D(B,\mathcal{A}-Mod)\rightarrow D(\mathcal{X})$, i.e. of the 
embedding of the noncommutative resolution category into the derived category 
of 
the universal hyperplane section of $X$.

\section{\label{sec:section2}Lightning review of GLSMs for HPD}

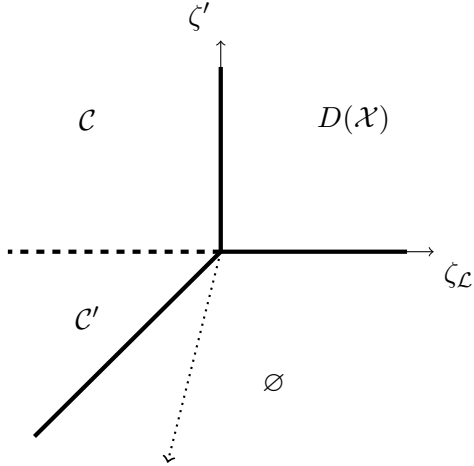
\begin{figure}
\centering
 \begin{tikzpicture}[scale=0.70]
  	\draw[thin,->] (0,0) -- (4,0) 
node[anchor=north west]{$\zeta_{\mathcal{L}}$};
	\draw[thin,->] (0,0) -- (0,4) node[anchor=south east]{$\zeta'$};
	\draw [ultra thick] (0,0) -- (0,3.5);
	\draw [ultra thick] (0,0) -- (3.5,0);
	\draw [ultra thick] (0,0) -- (-3.5,-3.5);
	\draw [ultra thick, dashed] (0,0) -- (-4,0);
	\draw [thick, dotted, ->] (0,0) -- (-1,-4);
    \node at (2.5,2.5) {$D(\mathcal{X})$};
    \node at (-2.5,2.5) {${\cal C}$};
    \node at (-2.5,-1.25) {${\cal C}'$};
    \node at (1,-2.5) {$\varnothing$};
 \end{tikzpicture}
 \caption{Higgs branches of the GLSM 
$\mathcal{T}_{\mathcal{X}}$}\label{fig:phase}
{\footnotesize \begin{flushleft} The theory $\mathcal{T}_{\mathcal{X}}$ has a 
geometric phase realizing the universal hyperplane section $\mathcal{X}$ and a 
LG phase realizing the HPD category $\mathcal{C}$. When $D(Y_{\zeta_{l}\ll 
-1},W_{\zeta_{l}\ll -1})$ is empty, $\mathcal{C}' \cong \mathcal{C}$, otherwise 
$\mathcal{C}'$ is a subcategory of $\mathcal{C}$. The dashed arrow shows the 
direction of RG flow. \end{flushleft} }
 \end{figure}

In this section, we review the construction of homological projective duals 
(HPD) of projective morphisms $f:X\rightarrow \mathbb{P}(S)$ (where $S\cong 
\mathbb{C}^{n+1}$) proposed in \cite{Chen:2020iyo}. We refer the reader to 
\cite{Chen:2020iyo} for the details of definitions and notations. The 
construction of \cite{Chen:2020iyo} assumes that we have a gauged linear sigma 
model (GLSM) construction for $f:X\rightarrow \mathbb{P}(S)$, i.e. a GLSM having 
a geometric phase corresponding to a Higgs branch\footnote{We do not need to 
assume the GLSM is nonanomalous.} that RG flows to a nonlinear sigma model 
(NLSM) whose target space is the image of $f$ in $\mathbb{P}(S)$. Denote that 
GLSM by
\begin{eqnarray}
\mathcal{T}_{X}=(G,\rho_{m}: G\rightarrow GL(V),W,t_{\mathrm{ren}},R).
\end{eqnarray}
We make some remarks about the tuple $\mathcal{T}_{X}$:
\begin{itemize}
\item The gauge group $G$ is taken to be a compact Lie group and 
$W\in\mathrm{Sym}(V^{\vee})^{G}$ denotes the superpotential.

 \item The FI-theta parameter $t_{\mathrm{ren}}$ is renormalized and depends on 
the energy scale for the case of anomalous GLSMs. In the following, to avoid 
cluttering we will simply denote $t:=t_{\mathrm{ren}}$ having in mind that $t$ 
may depend on the energy scale.
\item The (vector) R-charge assignment $R$ will not play an important role in 
our discussions below. Moreover, it is only well defined in the IR and depends 
on 
the phase. Hence we will leave it unspecified.  
\item For simplicity, in the discussion below, we will assume that the gauge 
group $G$ gets classically broken to a finite subgroup in every phase. This is 
always true if $G$ is abelian and $\rho_{m}$ is faithful. In the case that $G$ 
is nonabelian this is usually not true (see for example \cite{Hori:2006dk}) 
however, the GLSM phases are still well defined theories and we expect that our 
analysis can be carried out. All the examples we will cover in this work 
corresponds to $G$ abelian. Some comments and conjectures for nonabelian $G$ 
can be found in \cite{Chen:2020iyo}.
\end{itemize}
The morphism $f$ must be base point free, hence $f$ defines 
a line bundle $\mathcal{L}$ over $X$ given by
\begin{eqnarray}
\mathcal{L}=f^{*}\mathcal{O}_{\mathbb{P}(S)}(1).
\end{eqnarray}
Then, since there is a corresponding character $\chi\in 
\mathrm{Hom}(G,\mathbb{C}^{*})$ to 
$\mathcal{L}$ be there exist a distinguished 
$U(1)_{\mathcal{L}}\subset G$ (with associated FI-theta parameter 
$t_{\mathcal{L}}=\zeta_{\mathcal{L}}-i\theta_{\mathcal{L}}$) associated to 
$\mathcal{L}$ (see \cite{Chen:2020iyo} for more details). The  
components of $f$ transforms under $g\in G$ by multiplication by $\chi(g)$ 
i.e. they have homogeneous weight under $U(1)_{\mathcal{L}}$ and 
corresponds to 
sections of $\mathcal{L}$. In the following we assume that the aforementioned 
geometric phase is 
a 
pure Higgs phase\footnote{It is always possible in the cases $X$ is Fano or 
Calabi-Yau (CY) to set this phase being located at $\zeta_{\mathcal{L}}\gg 1$. 
These  are the cases we will cover in this work. The 
generalization is straightforward.} and its category of B-branes will be denoted 
by $D(X_{\zeta_{\mathcal{L}}\gg 1}):=D^{b}Coh(X_{\zeta_{\mathcal{L}}\gg 1})$, 
if this phase is located at 
$\zeta_{\mathcal{L}}\gg 1$. As we vary the parameter $\zeta_{\mathcal{L}}$ we 
find, in general that the phase at $\zeta_{\mathcal{L}}\ll -1$ has a Higgs 
branch whose category of B-branes we denote\footnote{The space 
$D(Y_{\zeta_{\mathcal{L}}\ll -1},W_{\zeta_{\mathcal{L}}\ll -1})$ denotes the 
category of B-branes of a hybrid Landau-Ginzburg (LG) model with target space 
$Y_{\zeta_{\mathcal{L}}\ll -1}$ and superpotential $W_{\zeta_{\mathcal{L}}\ll 
-1}:=W|_{Y_{\zeta_{\mathcal{L}}\ll -1}}$. The details on how such category 
arises in the current context can be 
found in \cite{Chen:2020iyo}.} $D(Y_{\zeta_{\mathcal{L}}\ll 
-1},W_{\zeta_{\mathcal{L}}\ll -1})$ and a mixed Coulomb-higgs branch that splits 
into several isolated vacua, whose categories of B-branes we denote as 
$E_{1},\ldots,E_{k}$. Both categories of B-branes at the different values of 
$\zeta_{\mathcal{L}}$ are expected to be related by 
\cite{ballard2019variation,Hori:2013ika,hori2019notes,Clingempeel:2018iub}
\begin{eqnarray}\label{equivTX}
\langle D(Y_{\zeta_{\mathcal{L}}\ll -1},W_{\zeta_{\mathcal{L}}\ll 
-1}),E_{1},\ldots,E_{k} \rangle\cong D(X_{\zeta_{\mathcal{L}}\gg 1})
\end{eqnarray}
The equivalence (\ref{equivTX}) is realized at the level of the GLSM via the so 
called window categories. They are defined entirely via the UV datum (or GLSM 
datum) i.e. the tuple $\mathcal{T}_{X}$. Defining a B-brane $\mathcal{B}$ in 
the GLSM requires to specify a 
representation $\rho_{M}: G\rightarrow GL(M)$. If we denote $q^{\mathcal{L}}$ 
the weight of $\rho_{M}$ restricted to $U(1)_{\mathcal{L}}$, then we 
define two conditions on the weights $q^{\mathcal{L}}$:
\begin{eqnarray}\label{constrwin}
&&\textbf{Small window}:\qquad|\theta_{\mathcal{L}}+2\pi q^{\mathcal{L}}|<\pi 
\mathrm{min}(N_{\mathcal{L},\pm}),\nonumber\\
&&\textbf{Big window}:\qquad|\theta_{\mathcal{L}}+2\pi q^{\mathcal{L}}|<\pi 
\mathrm{max}(N_{\mathcal{L},\pm}),
\end{eqnarray}
where $N_{\mathcal{L},\pm}:=\sum_{a}(Q_{a}^{\mathcal{L}})^{\pm}$, 
$(x)^{\pm}:=(|x|\pm x)/2$ and $Q_{a}^{\mathcal{L}}$ are the weights of 
$\rho_{m}$ restricted to $U(1)_{\mathcal{L}}$. 
Therefore we have the definition of the window subcategories by the constraints 
(\ref{constrwin}): $\mathcal{W}^{\mathcal{L}}_{+,b}$ (resp. 
$\mathcal{W}^{\mathcal{L}}_{-,b}$) corresponds to the objects $\mathcal{B}$ 
such that the weights $q^{\mathcal{L}}$ of $\rho_{M}$ satisfy the big (resp. 
small) window constraint for 
$b=\lfloor\frac{\theta_{\mathcal{L}}}{2\pi}\rfloor$. The we have 
\begin{eqnarray}
D(Y_{\zeta_{\mathcal{L}}\ll -1},W_{\zeta_{\mathcal{L}}\ll 
-1})\cong\mathcal{W}^{\mathcal{L}}_{-,b}\hookrightarrow 
D(X_{\zeta_{\mathcal{L}}\gg 1})\cong\mathcal{W}^{\mathcal{L}}_{+,b} 
\end{eqnarray}
for any $b\in\mathbb{Z}$. It is straightforward to see that in the nonanomalous 
case, $N_{\mathcal{L},+}=N_{\mathcal{L},-}$ and small and big window categories 
become equivalent giving an equivalence of categories
\begin{eqnarray}
D(Y_{\zeta_{\mathcal{L}}\ll -1},W_{\zeta_{\mathcal{L}}\ll 
-1})\cong\mathcal{W}^{\mathcal{L}}_{b}\cong 
D(X_{\zeta_{\mathcal{L}}\gg 1})
\end{eqnarray}
where we dropped the $\pm$ index. This equivalence, via window categories is 
known as the grade restriction rule and was originally proposed for $G$ abelian 
and nonanomalous GLSMs in \cite{Herbst:2008jq} and later rigurously generalized 
to anomalous GLSMs and nonabelian $G$ in 
\cite{halpern2015derived,ballard2019variation}, the physical aspects of these 
generalizations where first studied in 
\cite{Hori:2013ika,hori2019notes} for nonanomalous and nonabelian GLSMs (plus 
some aspects of the anomalous case only for $G=U(1)$) and for anomalous and 
abelian GLSMs in \cite{Clingempeel:2018iub}. Our presentation of the window 
categories (\ref{constrwin}) is based on \cite{Clingempeel:2018iub}. Before 
moving on to the construction of the GLSM containing the HPD of $f:X\rightarrow 
\mathbb{P}(S)$ we illutrate this construction with a few examples:
\begin{itemize}
 \item Consider the case $\mathcal{T}_{X}=(U(1),\rho_{m}:U(1)\rightarrow 
GL(\mathbb{C}^{n+2}),W=p_{0}F_{d}(x),t,R)$ where 
$(p_{0},x_{0},\ldots,x_{n})\in \mathbb{C}^{n+2}$, $\rho_{m}$ is defined as
\begin{eqnarray}
\rho_{m}(\lambda)\cdot 
(p_{0},x_{0},\ldots,x_{n})=(\lambda^{-d}p_{0},\lambda x_{0},\ldots,\lambda 
x_{n}),\qquad \lambda\in U(1)
\end{eqnarray}
and $F_{d}(x)\in\mathbb{C}[x_{0},\ldots,x_{n}]$ is homogeneous of degree 
$d\leq n+1$ satisfying $dF_{d}^{-1}(0)=\{ 0\}$. Then
\begin{eqnarray}
X_{\zeta\gg 1}=\{F_{d}=0,p_{0}=0\}\cap Y_{\zeta\gg 1},\qquad  Y_{\zeta\gg 
1}=\mathcal{O}(-d)\rightarrow \mathbb{P}^{n}
\end{eqnarray}
then, the function $f$ corresponds to the natural embedding of $X_{\zeta\gg 1}$ 
in $\mathbb{P}^{n}$ and $\mathcal{L}=\mathcal{O}_{X}(1)$. The analysis of the 
window categories gives 
\cite{ballard2019variation,Hori:2013ika,hori2019notes,Clingempeel:2018iub}
\begin{eqnarray}
D(\mathbb{C}^{n+1}/\mathbb{Z}_{d},F_{d})\cong\mathcal{W}^{\mathcal{L}}_{-,b}
\hookrightarrow 
D(X_{\zeta\gg 1})\cong\mathcal{W}^{\mathcal{L}}_{+,b} 
\end{eqnarray}
and the equivalence (\ref{equivTX}) becomes
\begin{eqnarray}
\langle 
D(\mathbb{C}^{n+1}/\mathbb{Z}_{d},F_{d}),E_{1},\ldots, 
E_{n+1-d}\rangle\cong
D(X_{\zeta\gg 1}).
\end{eqnarray}

\item Consider the case $\mathcal{T}_{X}=(U(1),\rho_{m}:U(1)\rightarrow 
GL(V),W,t,R)$ where
\begin{eqnarray}
V=\mathrm{Sym}^{d}\mathbb{C}^{n+1}\oplus 
\mathrm{Sym}^{d}\mathbb{C}^{n+1}\oplus \mathbb{C}^{n+1},
\end{eqnarray}
denote $(p_{1},\ldots,p_{{n+d\choose d}},y_{1},\ldots,y_{{n+d\choose 
d}},x_{0},\ldots,x_{n})\in V$, the representation $\rho_{m}$ is defined by
\begin{eqnarray}
\rho_{m}(\lambda)\cdot 
(p,y,x)=(\lambda^{-d}p,\lambda^{d}y,\lambda x),\qquad \lambda\in U(1)
\end{eqnarray}
\begin{eqnarray}
W=\sum_{j=1}^{{n+d\choose d}}p_{j}(y_{j}-f_{j}(x)),
\end{eqnarray}
where $f_{j}(x)\in\mathbb{C}[x_{0},\ldots,x_{n}]$ are the monomials of degree 
$d$ in $x$. In this case $X_{\zeta\gg 1}\cong \mathbb{P}^{n}$, but the function 
$f:\mathbb{P}^{n}\rightarrow\mathbb{P}(\mathrm{Sym}^{d}\mathbb{C}^{n+1})$ 
becomes the degree $d$ Veronese embedding. However, an analysis of this GLSM 
shows that the small window is empty and the equivalence (\ref{equivTX}) becomes
\begin{eqnarray}
\langle 
E_{1},\ldots, 
E_{n+1}\rangle\cong
D(\mathbb{P}^{n})\cong\mathcal{W}^{\mathcal{L}}_{+,b}. 
\end{eqnarray}
Ondeed, this GLSM is equivalent to the GLSM 
$\mathcal{T}_{\mathbb{P}^{n}}=(U(1),\rho_{m}:U(1)\rightarrow 
GL(\mathbb{C}^{n+1}),W\equiv 0,t,R)$, where $\rho_{m}(\lambda)$ 
acts with weight $1$ on every variable \cite{Caldararu:2017usq}, i.e., to the 
usual GLSM for $\mathbb{P}^{n}$. However, the HPD is expected to depend on $f$ 
\cite{kuznetsov2007homological}. This is reflected in the following 
construction of the extended GLSM $\mathcal{T}_{\mathcal{X}}$, which is the 
GLSM containing the HPD. The extended GLSM $\mathcal{T}_\mathcal{X}$ depends on 
whether it is induced from $\mathcal{T}_X$ or $\mathcal{T}_{\mathbb{P}^n}$.
\end{itemize}
Starting 
from $\mathcal{T}_{X}$, we define an extension $\mathcal{T}_{\mathcal{X}}$ of 
$\mathcal{T}_{X}$ given by
\begin{eqnarray}\label{exGLSM}
\mathcal{T}_{\mathcal{X}}=(\widehat{G}=G\times U'(1),\hat{\rho}_{m}: 
\widehat{G}\rightarrow GL(V\oplus V'),\widehat{W},\widehat{R}),
\end{eqnarray}
where $V'$ is a representation of $U'(1)\times U(1)_{\mathcal{L}}\subseteq 
\widehat{G}$ with weights $(-1,-Q)\oplus(1,0)^{\oplus(n+1)}$, where 
$Q\in\mathbb{Z}$ is the weight of the character $\chi$ defined above. Denoting 
the 
coordinates of $V'$ as $(p,s_{0},\ldots,s_{n})$, the superpotential 
$\widehat{W}$ is given by
\begin{eqnarray}
\widehat{W}=W+p\sum_{j=0}^{n}s_{j}f_{j}(x),
\end{eqnarray}
where $f_{j}(x)$ are the components of the image of the map $f$. The GLSM 
$\mathcal{T}_{\mathcal{X}}$ is identified with the GLSM of the universal 
hyperplane section\footnote{We recall the reader that  $\mathcal{X}$ is defined 
as the fiber product $X\times_{\mathbb{P}(S)}\mathcal{H}\subset X\times 
\mathbb{P}(S^{\vee})$ where $\mathcal{H}=\{ (u,v)\in \mathbb{P}(S)\times 
\mathbb{P}(S^{\vee})|v(u)=0\}\subset \mathbb{P}(S)\times 
\mathbb{P}(S^{\vee})$ is the incidence divisor.} $\mathcal{X}$ of $X$: Its 
Higgs branch deep in the first 
quadrant of the FI parameter of $U'(1)\times U(1)_{\mathcal{L}}$ 
corresponds to a NLSM with target space $\mathcal{X}$. The phase space of 
$(\zeta',\zeta_{\mathcal{L}})$ takes generically the form specified in Figure 
\ref{fig:phase}, as analyzed in \cite{Chen:2020iyo}. In Figure 
\ref{fig:phase} we have specified the B-brane categories on the Higgs branches 
in every phase, which are the relevant branches for determining the HPD of 
$f:X\rightarrow\mathbb{P}(S)$.

Keeping 
$\zeta'\gg 1$ and varying $\zeta_{\mathcal{L}}$ leads to the following 
equivalence of categories
\begin{eqnarray}\label{HPDC}
\mathcal{C}=D(\widehat{Y}_{\zeta_{\mathcal{L}}\ll 
-1},\widehat{W}_{\zeta_{\mathcal{L}}\ll 
-1})\cong\widehat{\mathcal{W}}^{\mathcal{L}}_{-,b}\hookrightarrow 
D(\mathcal{X}_{\zeta_{\mathcal{L}}\gg 
1})\cong\widehat{\mathcal{W}}^{\mathcal{L}}_{+,b},
\end{eqnarray}
where the categories $\widehat{\mathcal{W}}^{\mathcal{L}}_{\pm,b}$ are defined 
analogously 
to $\mathcal{W}^{\mathcal{L}}_{\pm,b}$, but in the GLSM 
$\mathcal{T}_{\mathcal{X}}$. The category $\mathcal{C}$ is 
identified with the HPD category of $f:X\rightarrow\mathbb{P}(S)$ i.e. the 
proposal of \cite{Chen:2020iyo,ballard2017homological,rennemo2017fundamental} 
is that the subcategory of $D(\mathcal{X})$ corresponding to the small 
window category is equivalent to $\mathcal{C}$. We proceed 
to illustrate $\mathcal{T}_{\mathcal{X}}$ and $\mathcal{C}$ in the example of a 
degree $d$ Veronese embedding (the example of a Fano hypersurface is reviewed 
in detail in section \ref{sec:Fanohyper}.
\begin{itemize}
 \item Consider the GLSM $\mathcal{T}_{\mathcal{X}}$ corresponding the the 
degree $d$ Veronese embedding. Then, $\widehat{G}=U'(1)\times 
U(1)_{\mathcal{L}}$. The weights of the representation $\hat{\rho}_{m}: 
\widehat{G}\rightarrow GL(V\oplus V')$ are given in the following table 
 \[
\begin{array}{ccccccccccc}
& x_0 & x_1 & \cdots & x_n & p & s_1 & \cdots & s_{{n+d\choose 
d}} \\
U(1)_{\mathcal{L}} & 1 & 1 & \cdots & 1 & -d & 0 & \cdots & 0 \\
U'(1) & 0 & 0 & \cdots & 0 & -1 &   1 & \cdots & 1
\end{array}
\]
We remark that here, the fields $(p,s_{j})$ span the representation $V'$ and 
the 
fields $x_{i}$ span the representation $V$, where we simplified the model by 
integrating 
out the massive fields $(y,p)$ of the original representation. The 
superpotential $\widehat{W}$ then becomes
\begin{eqnarray}
\widehat{W}=p\sum_{j=1}^{{n+d\choose 
d}}s_{j}f_{j}(x).
\end{eqnarray}
Then, the Higgs branch in the region $\zeta_{\mathcal{L}}\ll-1$ and $\zeta'\gg 
1$ becomes the hybrid model with target space
\begin{equation}
Y_{LG}=\mathrm{Tot}\left( \mathcal{O}\left( -\frac{1}{d} \right)^{\oplus(n+1)} 
\rightarrow \mathbb{P}^{{n+d \choose d}-1} \right)/\mathbb{Z}_d
\end{equation}
and superpotential
\begin{equation}
W_{LG}=\sum_{j=1}^{{n+d\choose 
d}}s_{j}f_{j}(x)
\end{equation}
then, the category $\mathcal{C}$, corresponding to the HPD of the degree $d$ 
Veronese embedding of $\mathbb{P}^{n}$, is given by 
$\mathcal{C}=D(Y_{LG},W_{LG})$. In this particular family of examples one can 
show $\mathcal{C}\cong \mathcal{C}'$ \cite{Chen:2020iyo}. We will revisit the 
category $\mathcal{C}=D(Y_{LG},W_{LG})$ in section \ref{sec:HPDVeronese}
\end{itemize}

Everything 
can be carried over when 
taking linear sections of $X$, but in this work we will be mainly interested in 
the HPD of $X$.

Using this proposal, we can express $\mathcal{C}$ as the category of B-branes 
on a Higgs branch that can be described as a fibered LG model i.e. a hybrid 
model, which usually will have the 
characteristics of a \emph{good 
hybrid} in the sense of \cite{Aspinwall:2009qy,Bertolini:2013xga}, making it 
very tractable.

\section{\label{sec:LGAinf}Lightning review of $A_{\infty}$ algebras and their 
relation to open topological strings}

In this section we present the useful definitions and results that relate 
$A_{\infty}$ to the relevant physical systems we are going to need in the 
subsequent sections. Let us start with the definition of $A_{\infty}$ algebra 
(our main reference is \cite{keller2001introduction} but other useful 
sources are \cite{stasheff86,penkava1994a,lefevre2003infini}). 
\\\\
\textbf{Definition}. An $A_{\infty}$ algebra over a field $\mathbb{K}$ Consist 
of a $\mathbb{Z}$-graded $\mathbb{K}$-vector space $A$
\begin{eqnarray}
A=\bigoplus_{p\in\mathbb{Z}}A^{p}
\end{eqnarray}
endowed with homogeneous $\mathbb{K}$-linear maps\footnote{The grading of 
$A^{\otimes n}$ is given by $(A^{\otimes 
n})^{p}=\bigoplus_{i_{1}+\ldots+i_{n}=p}A^{i_{1}}\otimes\cdots\otimes 
A^{i_{n}}$.}
\begin{eqnarray}
m_{n}:A^{\otimes n}\rightarrow A,\qquad n\geq 1
\end{eqnarray}
of degree $2-n$ satisfying the relations
\begin{eqnarray}\label{ainftyrels}
\sum_{r+s+t=n}(-1)^{r+st}m_{u}(\mathbf{1}^{\otimes r}\otimes m_{s}\otimes 
\mathbf{1}^{\otimes t})=0,\qquad n\geq 1
\end{eqnarray}
where $u=r+t+1$ and $s\geq 1$, $r,t\geq 0$.
\\\\
Let us make a few important remarks. First, note that (\ref{ainftyrels}) 
implies $m_{1}\circ m_{1}=0$ hence, $m_{1}$ is a differential. Second, the maps 
in the tensor products, such as in (\ref{ainftyrels}) are subject to the 
Koszul sign rule:
\begin{eqnarray}
(f\otimes g)(a\otimes b)=(-1)^{|g||a|}f(a)\otimes g(b),
\end{eqnarray}
where we assume $a$ is a homogeneous element of degree $|a|$, and 
$|g|$ denotes the degree of the map $g$. We define next a morphism of 
$A_{\infty}$ algebras.
\\\\
\textbf{Definition}. A morphism $f:A\rightarrow B$ between $A_{\infty}$ 
algebras consists of a family of maps 
\begin{eqnarray}
f_{n}:A^{\otimes n}\rightarrow B
\end{eqnarray}
of degree $1-n$ satisfying
\begin{eqnarray}\label{ainftyrelsmor}
\sum_{r+s+t=n}(-1)^{r+st}f_{u}(\mathbf{1}^{\otimes r}\otimes m^{A}_{s}\otimes 
\mathbf{1}^{\otimes 
t})=\sum_{l=1}^{n}\sum_{I=n}(-1)^{\epsilon(l)}m^{B}_{l}(f_{i_{1}}
\otimes\cdots\otimes f_{i_{l}}),\qquad n\geq 1
\end{eqnarray}
where $u=r+t+1$, $s\geq 1$, $r,t\geq 0$ and the second sum over $I=n$ means sum 
over all decompositions $i_{1}+\ldots +i_{l}=n$ (with $i_{k}\geq 1$). The sign 
$\epsilon(l)$ is given by
\begin{eqnarray}
\epsilon(l)=(l-1)(i_{1}-1)+(l-2)(i_{2}-1)+\ldots+2(i_{l-2}-1)+(i_{l-1}-1).
\end{eqnarray}
Finally, we denote $m^{A,B}_{n}$ the maps of $A$ and $B$, respectively.
\\\\
Note that the map $f_{1}$ induces a map $f_{1,*}$ between the cohomologies
\begin{eqnarray}
f_{1,*}:H(A)\rightarrow H(B),
\end{eqnarray}
where $H(A)$ ($H(B)$) denotes the cohomology of the differential $m^{A}_{1}$ 
($m^{B}_{1}$). Then, a morphism is called quasi-isomorphism if $f_{1,*}$ is an 
isomorphism and is called strict if $f_{i}=0$ for all $i\neq 1$.
\\\\
\textbf{Definition}. An $A_{\infty}$-module over $A$ is given by a 
$\mathbb{Z}$-graded vector space $M$ endowed with maps 
\begin{eqnarray}
m^{M}_{n}:M\otimes A^{\otimes n-1}\rightarrow M,\qquad n\geq 1
\end{eqnarray}
of degree $2-n$ satisfying
\begin{eqnarray}
\sum_{r+s+t=n}(-1)^{r+st}m^{M}_{u}(\mathbf{1}^{\otimes r}\otimes 
\tilde{m}_{s}\otimes 
\mathbf{1}^{\otimes t})=0\qquad, n\geq 1
\end{eqnarray}
where $u=r+t+1$, $s\geq 1$, $r,t\geq 0$ and 
\begin{eqnarray}
m^{M}_{u}(\mathbf{1}^{\otimes r}\otimes 
\tilde{m}_{s}\otimes 
\mathbf{1}^{\otimes t})=\begin{cases}
                   m^{M}_{u}(\mathbf{1}^{\otimes r}\otimes m_{s}\otimes 
\mathbf{1}^{\otimes t}), & \text{if } 
r>0\\
                   m^{M}_{u}( 
m^{M}_{s}\otimes 
\mathbf{1}^{\otimes t}). &  
\text{if } r=0
                  \end{cases}
\end{eqnarray}
\\\\
There is an alternative construction of the $A_{\infty}$-algebra known as the 
bar construction. Consider a $\mathbb{Z}$-graded $\mathbb{K}$-vector space $V$ 
and the tensor algebra
\begin{eqnarray}
T^{\bullet}V:=\bigoplus_{n\geq 1}V^{\otimes n}.
\end{eqnarray}
Then, any coderivation $b:T^{\bullet}V\rightarrow T^{\bullet}V$ can be written 
in terms of degree $1$ maps $b_{n}:V^{\otimes}\rightarrow V$. Explicitly, by 
denoting $b_{n,u}$ the component of $b$ mapping $V^{\otimes n}$ to $
V^{\otimes u}$, we can write
\begin{eqnarray}
b_{n,u}=\sum_{r+s+t=n,r+t+1=u}\mathbf{1}^{\otimes r}\otimes 
b_{s}\otimes 
\mathbf{1}^{\otimes t},\qquad r,t\geq 1,s\geq 1.
\end{eqnarray}
Imposing $b^{2}=0$ is equivalent to the conditions
\begin{eqnarray}\label{binftyrels}
\sum_{r+s+t=n}b_{u}(\mathbf{1}^{\otimes r}\otimes b_{s}\otimes 
\mathbf{1}^{\otimes t})=0,\qquad n\geq 1
\end{eqnarray}
where $u=r+t+1$ and $s\geq 1$, $r,t\geq 0$. Then, if we identify $V=A[1]$, 
where $A[1]$ is the grading shift $(A[1])^{p}=A^{p+1}$ and we denote the 
natural degree $-1$ map $h:A\rightarrow A[1]$, then if we write 
\begin{eqnarray}
m_{n}=h^{-1}\circ b_{n}\circ h^{\otimes n}
\end{eqnarray}
or equivalently\footnote{Here we used that the inverse of $h^{\otimes n}$ is 
$(-1)^{\frac{n(n-1)}{2}}(h^{-1})^{\otimes n}$. We remak that there are 
different ways to define $b_{n}$ in terms of $m_{n}$, leading to different sign 
conventions.}
\begin{eqnarray}
b_{n}=(-1)^{\frac{n(n-1)}{2}}h\circ m_{n}\circ (h^{-1})^{\otimes n}.
\end{eqnarray}
The relations $(\ref{binftyrels})$ are equivalent to $(\ref{ainftyrels})$. An 
$A_{\infty}$-algebra $A$ is called minimal if $m_{1}\equiv 0$ and is called 
strictly unital if there exists a degree $0$ element $1_{A}\in A^{0}$ satisfying
\begin{eqnarray}\label{unital cdts}
&& m_{1}(1_{A})=0,\nonumber\\
&& m_{2}(1_{A}\otimes a)=m_{2}(a\otimes 1_{A})=a,\nonumber\\
&& m_{i}(a_{1}\otimes\cdots\otimes a_{i})=0, \text{ \ if any\ 
}a_{k}=1_{A}\qquad 
i>2
\end{eqnarray}
for all $a,a_{1},\ldots, a_{i}\in A$. Moreover, if $A$ is equipped with a 
bilinear form $\langle 
\cdot,\cdot\rangle:A\otimes A\rightarrow \mathbb{C}$, then $A$ is called cyclic 
(w.r.t. $\langle 
\cdot,\cdot\rangle$) if it satisfies
\begin{eqnarray}
\langle a_{0},b_{n}(a_{1}\otimes\cdots \otimes 
a_{n}))\rangle=(-1)^{(|a_{0}|+1)(|a_{1}|+\ldots+|a_{n}|+n)}\langle 
a_{1},b_{n}(a_{2}\otimes\cdots \otimes 
a_{0}))\rangle,
\end{eqnarray}
where $a_{i}\in A$ are homogeneous elements.\\\\
We have the following important theorem 
\cite{kadeishvili1980homology,kontsevich2001homological}:
\\\\
\textbf{Theorem}. Any $A_{\infty}$-algebra $(A,m_{n})$ is 
$A_{\infty}$-quasi-isomorphic to a minimal $A_{\infty}$-algebra called a 
minimal model for $A$. Moreover this minimal model can be taken to be 
$(H(A),m^{H}_{n})$ which is unique up to $A_{\infty}$-isomorphism and satisfies 
\begin{enumerate}
 \item The map $f_{1}: H(A)\rightarrow A$ is given by the inclusion map.
 \item The map $m_{2}^{H}$ is given by the map induced by $m_{2}$.
\end{enumerate}
Then, this theorem plus the conditions \eqref{ainftyrelsmor} for $A_{\infty}$ 
morphisms applied to the inclusion map $\iota:H(A)\rightarrow A$ give us a way 
to recursively determine the products $m_{n}^{H}$ from the knowledge of 
$(A,m_{n})$. Let us write some of these relations to ilustrate this point 
(recall that $m^{H}_{1}\equiv 0$):
\begin{eqnarray}\label{relsinclusion}
\iota\circ m_{2}^{H} &=& m_{2}(\iota\otimes \iota)+m_{1}\circ f_{2},\nonumber\\
\iota\circ m_{3}^{H} &=& f_{2}(m^{H}_{2}\otimes \mathbf{1})-f_{2} 
(\mathbf{1}\otimes m^{H}_{2})+m_{2}(\iota\otimes f_{2})-m_{2}(f_{2}\otimes 
 \iota)+m_{1}\circ f_{3},\nonumber\\
\vdots  
\end{eqnarray}
so, the maps $f_{n}:H(A)^{\otimes n}\rightarrow A$ and the higher products 
$m^{H}_{n}$ can be determined 
recursively (see for example \cite{Aspinwall:2004bs}).

In the case of topological strings we will be interested in 
$A_{\infty}$-categories, which are defined as follows
\\\\
\textbf{Definition}. A $A_{\infty}$-category $\mathcal{A}$ with objects 
$\mathrm{Ob}(\mathcal{A})$ consists of the data
\begin{enumerate}
 \item For all $A,B\in\mathrm{Ob}(\mathcal{A})$ the space 
$\mathrm{Hom}_{\mathcal{A}}(A,B)$ is a $\mathbb{Z}$-graded vector space.
\item For all $n\geq 1$ and any set of objects $A_{0},\ldots, 
A_{n}\in\mathrm{Ob}(\mathcal{A})$ there exists a degree $2-n$ map
\begin{eqnarray}\label{mnAcat}
m_{n}:\mathrm{Hom}_{\mathcal{A}}(A_{n-1},A_{n})\otimes 
\mathrm{Hom}_{\mathcal{A}}(A_{n-2},A_{n-1})\otimes\cdots 
\mathrm{Hom}_{\mathcal{A}}(A_{0},A_{1})\rightarrow\mathrm{Hom}_{\mathcal{A}}(A_{
0},A_{n})\nonumber
\end{eqnarray}
satisfying
\begin{eqnarray}
\sum_{r+s+t=n}(-1)^{r+st}m_{u}(\mathbf{1}^{\otimes r}\otimes m_{s}\otimes 
\mathbf{1}^{\otimes 
t})=0.
\end{eqnarray}
\end{enumerate}

We also define\\\\
\textbf{Definition}. An $A_{\infty}$-functor between $A_{\infty}$-categories 
$\mathcal{A}_{1}$ and $\mathcal{A}_{2}$ consists of the data
\begin{enumerate}
 \item A map $\mathcal{F}:\mathrm{Ob}(\mathcal{A}_{1})\rightarrow 
\mathrm{Ob}(\mathcal{A}_{2})$.
\item For all $n\geq 1$ and any set of objects $A_{0},\ldots, 
A_{n}\in\mathrm{Ob}(\mathcal{A}_{1})$ there exists a degree $1-n$ map
\begin{eqnarray}
\mathcal{F}_{n}:\mathrm{Hom}_{\mathcal{A}_{1}}(A_{n-1},A_{n})\otimes 
\mathrm{Hom}_{\mathcal{A}_{1}}(A_{n-2},A_{n-1})\otimes\cdots 
\mathrm{Hom}_{\mathcal{A}_{1}}(A_{0},A_{1})\rightarrow\mathrm{Hom}_{\mathcal{A}_
{2}} (\mathcal{F}(A_ {
0}),\mathcal{F}(A_{n}))\nonumber
\end{eqnarray}
satisfying conditions analogous to (\ref{ainftyrelsmor}).
\end{enumerate}

More precisely, in topological string theory we encounter cyclic, unital and 
minimal 
$A_{\infty}$-categories\footnote{We say an $A_{\infty}$-category has strict 
identities if, for every $A\in \mathrm{Ob}(\mathcal{A})$ there is a degree $0$ 
element $1_{A}\in \mathrm{Hom}_{\mathcal{A}}(A,A)$ satisfying the conditions  
(\ref{unital cdts}),  whenever a map $m_{n}$ can be consistently inserted as 
defined in (\ref{mnAcat}).} and we take the field 
$\mathbb{K}=\mathbb{C}$ from now on. It is easy to see that the 
$A_{\infty}$-category of a single object is equivalent to an 
$A_{\infty}$-algebra. Next we move on to explain how these structures arise in 
topological strings. For simplicty we consider a worldsheet with disk topology 
and 
boundary conditions characterized by a single D-brane $\mathcal{D}$. Upon 
topological twist, this configuration has a single scalar nilpotent supercharge 
$\mathbf{Q}$. The 'off-shell' space of open strings stretching from 
$\mathcal{D}$ to itself is given by a graded vector space, which we denote 
$V_{\mathcal{D}}$ and there is an action of $\mathbf{Q}$ in this vector space, 
hence we can take the cohomology 
\begin{eqnarray}
\mathrm{End}(\mathcal{D}):=H_{\mathbf{Q}}(V_{\mathcal{D}}),
\end{eqnarray}
which is the 
space of physical states of the topological strings stretching between 
$\mathcal{D}$ and itself. If we denote 
$\psi_{a}$ the elements of $\mathrm{End}(\mathcal{D})$, their disc correlators 
encode the Stasheff conditions (\ref{ainftyrels}). More precisely, the disk 
correlator of two elements (the boundary topological metric), denoted $\langle 
\psi_{a},\psi_{b}\rangle$, equips $\mathrm{End}(\mathcal{D})$ with an 
(nondegenerate) inner product. In \cite{Hofman:2000ce,Herbst:2004jp} it is 
found that the relation 
between the 
disk correlators with more than two insertions and the maps $b_{k}$
\begin{equation}\label{opencorrB}
B_{i_0 i_1\cdots i_k} := (-1)^{|a_{1}|+\ldots+|a_{k-1}|+k-1} \left\langle 
\psi_{i_0} 
\psi_{i_1} P \int \psi^{(1)}_{i_2} \cdots \int \psi^{(1)}_{i_{k-1}} \psi_{i_k} 
\right\rangle = \langle \psi_{i_{0}},b_k(\psi_{i_1},\cdots,\psi_{i_{k}})\rangle,
\end{equation}
where $\psi^{(1)}_{a}$ denotes the 1-form descendants of $\psi_{a}$. The 
correlators (\ref{opencorrB}) are defined using an appropriate regulator 
\cite{Herbst:2004jp} and they satisfy a cyclicity condition:
\begin{equation}
B_{i_0 i_1\cdots i_k} =(-1)^{(|a_{m}|+1)(|a_{0}|+\ldots+|a_{k-1}|+k)}B_{i_k 
i_0\cdots i_{k-1}}.
\end{equation}
It is important to remark that on the right hand side of (\ref{opencorrB}), the 
operators $\psi_{a}$ should be considered in the space 
$\mathrm{End}(\mathcal{D})[1]$. In other words, the graded space $A$ is 
identified with $\mathrm{End}(\mathcal{D})$. Hence, up to a sign that, in 
general depends on the degree of the insertions, we can identify
\begin{equation}
B_{i_0 i_1\cdots i_k} \sim  \langle 
m_k(\psi_{i_0},\cdots,\psi_{i_{k-1}}), \psi_{i_k} \rangle. 
\end{equation}
In general, for a SCFT we 
can define a trace function\footnote{This is just the (twisted) correlator on 
the sphere. See for example the review \cite{Warner:1993zh}.}
\begin{equation}
\gamma:A\rightarrow \mathbb{C}
\end{equation}
of degree $-\hat{c}=-\frac{c}{3}$, where $c$ is the central charge of the 
SCFT. Then the inner product can be written as
\begin{equation}
\langle\cdot,\cdot\rangle:A\otimes A\rightarrow \mathbb{C},\qquad 
(\psi_{a},\psi_{b})\mapsto 
\langle\psi_{a},\psi_{b}\rangle=\gamma(m_{2}(\psi_{a},\psi_{b})).
\end{equation}
Then, we simply write the relation:
\begin{equation}
B_{i_0 i_1\cdots i_k} = \gamma \left( 
m_2(m_k(\psi_{i_0},\cdots,\psi_{i_{k-1}}), \psi_{i_k}) \right),
\end{equation}
where the sign is hidden in $\gamma$.
When considering multiple branes, 
this structure becomes an $A_{\infty}$-category. For instance, in the case of a 
SCFT defined by the NLSM with a CY target space $X$, the category of B-branes 
(topological open strings in the B-model) is equivalent to $DCoh(X)$, the 
derived category of coherent sheaves on $X$ 
\cite{Sharpe:1999qz,Douglas:2000gi,Aspinwall:2001pu,Aspinwall:2009isa} 
and an $A_{\infty}$ structure on this category has been derived from physics 
and 
mathematical point of view 
\cite{Tomasiello:2001yq,Lazaroiu:2001nm,lunts2010uniqueness,bondal1990enhanced}. 
Analogous results also exist in the case of $G$-equivariant 
categories of matrix factorizations $MF_{G}(W)$, when $G$ is a finite abelian 
group and $W$ is a quasi-homogeneous polynomial 
\cite{Herbst:2004jp,Carqueville:2009ay,murfet2019constructing}.

 
\section{$\Af$-algebras associated with Landau-Ginzburg 
models}\label{sec:AinfLG}

In this section we will apply the results reviewed in section \ref{sec:LGAinf} 
to the specific case of LG orbifolds. We begin by reviewing the physics 
approach of categories of matrix factorizations, arising as B-branes on LG 
orbifolds. Fix a vector 
space $\mathbb{V}$ of rank $N$ with coordinates denoted by $x_{i}$, 
$i=1,\ldots,N$. We specify a left R-symmetry given by a $\mathbb{C}^{*}_{L}$ 
action on $\mathbb{V}$ with weights $q_{i}\in\mathbb{Q}\cap (0,1)$. The 
\emph{orbifold group} will be specified by a finite abelian group $G$ and a 
representation $\rho_{orb}:G\rightarrow GL(\mathbb{V})$. We specify a 
\emph{superpotential}, that is a holomorphic, $G$-invariant function 
$W:\mathbb{C}^{N}\rightarrow \mathbb{C}$, $W\in 
\mathbb{C}[x_{1},\ldots,x_{N}]$. As an $\mathcal{N}=(2,2)$ theory, the 
LG orbifold is specified by the data
\begin{eqnarray}
  \label{lgdata}
(W,G,\rho_{orb},\mathbb{C}^{*}_{L}),
\end{eqnarray}
but we impose some extra requirement on (\ref{lgdata}). In order for 
the vector R-symmetry to be nonanomalous, we require $W$ to 
be quasi-homogeneous, of weight $1$ under the $\mathbb{C}^{*}_{L}$ action
i.e.~$W(\lambda^{q_{i}}\phi_{i})=\lambda W(\phi_{i})$ \cite{Vafa:1988uu} (this 
implies that $W$ has charge $2$ under the vector R-symmetry). Moreover, $W$ 
being quasi-homogeneous implies $dW^{-1}(0)=\{ 0\}$, i.e. $W$ is compact, in 
the 
sense that it defines a compact SCFT in the IR. Quasi-homogeneity of $W$ 
guarantees that we always have the symmetry $x_{j}\rightarrow e^{ 2 
i\pi q_{i}}x_{j}$. If $d$ denotes the lowest nonzero integer such that 
$dq_{i}\in \mathbb{Z}$ for all $i$, then this specifies a $\mathbb{Z}_{d}$ 
action generated by $J=\mathrm{diag}(e^{ 2i\pi 
q_{1}},\ldots,e^{ 2i\pi q_{N}})$. Denote by $\mathrm{Aut}(W)$ the group of 
diagonal automorphisms of $W$, i.e.
\begin{eqnarray}
\mathrm{Aut}(W)=\left\{ \mathrm{diag}(e^{ 2\pi i \lambda_{1}},\ldots,e^{ 2\pi i 
\lambda_{N}})\in U(1)^{N}:W(e^{2\pi i\lambda_{i}}x_{i})=W(x_{i})  
\right\}.
\end{eqnarray}
we then say an orbifold group $G$ 
is \emph{admissible} if it satisfies
\begin{eqnarray}\label{Gadm}
\langle J\rangle\subseteq G\subseteq \mathrm{Aut}(W),
\end{eqnarray}
and we will require this condition. B-type D-branes $\mathcal{B}$ in LG 
orbifolds are characterized in terms of matrix factorizations of $W$ 
\cite{Kapustin:2002bi,Brunner:2003dc}. More precisely, $\mathcal{B}$ consists 
of the data 
\begin{equation}\label{mfbbb}
\mathcal{B}=(M,\sigma,Q,R_{M},\rho_{M}),
\end{equation}
where $M$ (the Chan-Paton space) is a free 
$\mathbb{C}[x_{1},\ldots,x_{N}]$-module, $\sigma$ is an involution on 
$M$, inducing a $\mathbb{Z}_2$-grading (so we can write 
$M=M_{0}\oplus M_{1}$, with $\sigma 
M_{i}=(-1)^{i}M_{i}$) and $Q(x)$ is a 
$\mathbb{Z}_2$-odd endomorphism on $M$ satisfying
\begin{equation}
  \label{lg-mf}
  Q^2=W\cdot\mathrm{id}_{M}.
\end{equation}
Under the vector R-charge, $W$ has 
charge $2$: $W(\lambda^{2q_i}x_i)=\lambda^2W(x_i)$ with the charges $q_i$ 
of the left R-symmetry. Therefore, by 
(\ref{lg-mf}), $Q$ must have vector R-charge $1$. This defines a 
compatible representation $R_{M}:U(1)_{V}\rightarrow GL(M)$ of 
the vector R-symmetry, satisfying:
\begin{equation}
 R_{M}(\lambda)Q(\lambda^{2q_i}x_i)R_{M}^{-1}
(\lambda)=\lambda Q(x_i),
\end{equation}
as well as another compatible representation of $G$, 
$\rho_{M}:G\rightarrow GL(M)$ satisfying
\begin{equation}
  \label{def-orbmat}
  \rho_{M}(g)^{-1}Q(\rho_{orb}(g)\cdot x_j)\rho_{M}(g)=Q
(x_ {j} ).
\end{equation}
Given a pair of B-branes 
$\mathcal{B}_{i}=(M^{(i)},\sigma_{i},Q_{i},R^{(i)}_{M},\rho^{(i)}_{M})$, 
$i=1,2$, we can define the space of morphisms betwen them, 
$Hom(\mathcal{B}_{1},\mathcal{B}_{2})$ as graded morphisms
\begin{equation}
 \Psi:M^{(1)}\rightarrow M^{(2)},
\end{equation}
i.e. $\Psi\in 
V_{r_{1},r_{2}}:=Mat_{r_{1},r_{2}}(\mathbb{C}[x_{1},\ldots,x_{N}])$, the space 
of $r_{1}\times r_{2}$ matrices with coefficients in 
$\mathbb{C}[x_{1},\ldots,x_{N}]$, where $r_{i}=\mathrm{rk}(M^{(i)})$ satisfying
\begin{equation}
D_{12}\circ\Psi:= Q_{2}\Psi-\sigma_{2}\Psi\sigma_{1} Q_{1}=0
\end{equation}
modulo $D_{12}$-exact morphisms. The differential $D_{12}$ can be identified 
with the conserved supercharge $\mathbf{Q}$ of the worldsheet theory on the 
open string stretching between $\mathcal{B}_{1}$ and $\mathcal{B}_{2}$. 
Therefore we can denote
\begin{equation}
Hom(\mathcal{B}_{1},\mathcal{B}_{2})=H_{D_{12}}(V_{r_{1},r_{2}}).
\end{equation}
The space $Hom(\mathcal{B}_{1},\mathcal{B}_{2})$ 
is $\mathbb{Z}_{2}$-graded and we denote its homogeneous components, and 
elements, as
\begin{equation}\label{morphismsMF}
Hom(\mathcal{B}_{1},\mathcal{B}_{2})=H^{0}(\mathcal{B}_{1},\mathcal{B}_{2}
)\oplus H^{1}(\mathcal{B}_{1},\mathcal{B}_{2}),\qquad\phi_{i}\in 
H^{0}(\mathcal{B}_{1},\mathcal{B}_{2}),\qquad 
\psi_{i}\in H^{1}(\mathcal{B}_{1},\mathcal{B}_{2}).
\end{equation}
The category $MF(W,G)$ of objects $\mathcal{B}$ with morphisms defined as
\begin{equation}
Hom_{MF(W,G)}(\mathcal{B}_{1},\mathcal{B}_{2}):=Hom(\mathcal{B}_{1},\mathcal{B}_
{2})^{G},
\end{equation}
i.e., $\Psi\in Hom_{MF(W,G)}(\mathcal{B}_{1},\mathcal{B}_{2})$ satisfies
\begin{equation}
\rho_{M^{(2)}}(g)^{-1}\Psi(\rho_{orb}(g)\cdot 
x_{i})\rho_{M^{(1)}}(g)=\Psi( x_{i}),
\end{equation}
will be referred as the category of B-branes on the LG orbifold. This category 
also has a grading that we will review next.

\subsection*{Gradings}

The category $MF(W,G)$ defined above has a natural $\mathbb{Q}$-grading given 
by the R-charge. More precisely, it is the fact that the superpotential $W$ is 
quasi-homogeneous that guarantees the existence of this $\mathbb{Q}$-grading 
(because then the vector R-charge is conserved) \cite{Walcher:2004tx}. The 
orbifold by $G$ satisfying 
(\ref{Gadm}) guarantees that the physical states will have integer 
R-charges \cite{Vafa:1989xc} and hence, we can put an integer grading on open 
string states. For a reduced and irreducible matrix factorization 
$\mathcal{B}\in MF(W,G)$, the map 
$\rho_{M}$ satisfies \cite{Walcher:2004tx}
\begin{equation}
\rho_{M}(J)=\sigma \circ R_{M}(e^{i\pi})e^{-i\pi\varphi}
\end{equation}
for some $\varphi \in \frac{2}{d}\mathbb{Z}$. The morphism $\Psi\in 
Hom_{MF(W,G)}(\mathcal{B}_{1},\mathcal{B}_{2})$ has R-charge 
$q_{\Psi}\in\mathbb{Q}$ defined by
\begin{equation}
R_{M^{(2)}}(\lambda)\Psi(\lambda^{2q_{i}} 
x_{i})R_{M^{(1)}}(\lambda)^{-1}=\lambda^{q_{\Psi}}\Psi( x_{i}).
\end{equation}
Then, a $\mathbb{Z}$-grading on $\Psi$ is defined by
\begin{equation}
\mathrm{deg}(\Psi)=\varphi_{2}-\varphi_{1}+q_{\Psi}.
\end{equation}
The category $MF(W,G)$ with this additional grading is known in the mathematics 
literature as the category of graded, $G$-equivariant matrix factorizations 
\cite{orlov2009derived}.

\subsection{Effective superpotential, deformations and $A_{\infty}$ 
structures}\label{sec:effpot}

The category $MF(W,G)$ can be given an $A_{\infty}$ structure 
\cite{caldararu2010curved}, and the higher order products can be read off from 
the 
computation of the unobstructed deformations of the objects  
$\mathcal{B}\in MF(W,G)$, as we will explain in this section, and will become 
useful later. However, it is very convenient to use a description of $MF(W,G)$ 
that follows very closely \cite{Kapustin:2002bi}. Consider first the case of 
a trivial orbifold 
\begin{equation}
G=\mathbf{1},
\end{equation}
then we denote the category just as $MF(W)$. 
Then, in the case $dW^{-1}(0)=\{ 0\}$, this category has a single generator 
\cite{dyckerhoff2011compact} given by the matrix factorization 
$\mathcal{B}_{D0}=(M,\sigma,Q_{D0},R_{M})$ with
\begin{equation}\label{Qd0explicit}
Q_{D0} = \sum_{i=1}^N\left( x_i \overline{\eta}_i +q_{i} 
\frac{\partial W}{\partial x_i} \eta_i \right),
\end{equation}
where the subscript $D0$ is because this matrix factorization is reminiscent to 
the $D0$-brane in \cite{Kapustin:2002bi}. The objects $\overline{\eta}_i$, 
$\eta_i$, $i=1,\ldots, N$ are generators of a Clifford algebra of rank $2N$, 
namely they satisfy the relations
\begin{equation}
\{\overline{\eta}_{i},\eta_j \}=\delta_{i,j}\mathbf{1}\qquad 
\{\overline{\eta}_{i},\overline{\eta}_{j}\}=\{\eta_{i},\eta
_{j}\}=0.
\end{equation}
Then, we can consider 
\begin{equation}
\mathcal{A}_{D0}:=Hom_{MF(W)}(\mathcal{B}_{D0},\mathcal{B}_{D0}),
\end{equation}
which has 
the structure of an $A_{\infty}$-algebra\footnote{In the particular case that 
$W$ 
is homogeneous of degree $2$, then $\mathcal{A}_{D0}$ becomes simply the 
(complex) Clifford algebra $Cl(q)$ associated with the quadratic form 
$q_{ij}:=\partial_{i}\partial_{j}W$ \cite{Kapustin:2002bi,Hori:2000ic}.} 
\cite{dyckerhoff2011compact} and moreover we have the equivalence
\begin{equation}
MF(W)\cong D(\mathrm{Mod}-\mathcal{A}_{D0})
\end{equation}
where $D(\mathrm{Mod}-\mathcal{A}_{D0})$ stands for the derived category of 
$A_{\infty}$-modules over $\mathcal{A}_{D0}$. Given an object $\mathcal{B}\in 
MF(W)$, the module associated to $\mathcal{B}$ is given by 
$M_{\mathcal{B}}:=Hom_{MF(W)}(\mathcal{B}_{D0},\mathcal{B})$ where the maps 
\begin{equation}\label{productsfunct}
m_{n}^{\mathcal{B}}:M_{\mathcal{B}}\otimes \mathcal{A}_{D0}^{\otimes 
n-1}\rightarrow M_{\mathcal{B}}
\end{equation}
come from the $A_{\infty}$ structure of the category $MF(W)$, in particular
\begin{equation}
m_{2}^{\mathcal{B}}:M_{\mathcal{B}}\otimes \mathcal{A}_{D0}\rightarrow 
M_{\mathcal{B}},\qquad 
m_{2}^{\mathcal{B}}(\Psi^{\mathcal{B}},\Psi)=\Psi^{\mathcal{B}}\circ\Psi.
\end{equation}
When we add an orbifold, we expect the following equivalence
\begin{equation}
MF(W,G)\cong D(\mathrm{Mod}-\mathcal{A}_{D0}\sharp G),
\end{equation}
where $\mathcal{A}_{D0}\sharp G$ is the smash product between 
$\mathcal{A}_{D0}$ and the group algebra $\mathbb{C}[G]$, the product in 
$\mathcal{A}_{D0}\sharp G$ is given by \cite{tu2014matrix}
\begin{equation}
(a\sharp g_{1})\cdot(b\sharp g_{2})=(a\cdot g_{1}bg^{-1}_{1}) \sharp g_{1}g_{2},
\end{equation}
hence, studying the algebra $\mathcal{A}_{D0}$ is crucial. The higher order 
products 
$m_{n}:\mathcal{A}_{D0}^{\otimes n}\rightarrow \mathcal{A}_{D0}$ can be 
read off from the effective superpotential $\mathcal{W}_{\mathrm{eff}}$ defined 
by 
\begin{equation}
\mathcal{W}_{\mathrm{eff}} = \mathrm{Tr}\left( \sum_{k=2}^\infty 
\sum_{i_0,i_1,\cdots,i_k} \frac{B_{i_0\cdots i_k}}{k+1} Z_{i_0}Z_{i_1}\cdots 
Z_{i_k} \right),
\end{equation}
the function $\mathcal{W}_{\mathrm{eff}}$ encodes obstructions to the 
boundary deformations of the SCFT, and can be computed as follows. Consider the 
matrix factorization $Q_{D0}$, then our objective is to find a deformed matrix 
factorization
\begin{equation}
Q_{D0}^{\mathrm{def}}=Q_{D0}+\sum_{\vec{m}\in 
B}\alpha_{\vec{m}}u^{\vec{m}},\qquad u^{\vec{m}}:=\prod_{i=1}^{n}u_{i}^{m_{i}}
\end{equation}
where $B\subset \mathbb{N}^{n}$, 
$n=\mathrm{dim}H^{1}(\mathcal{B}_{D0},\mathcal{B}_{D0})$, $\alpha_{\vec{m}}$ 
are fermionic operators and $u_{i}$, $i=1,\ldots,n$ are commutative parameters. 
The matrix factorization satisfies 
\begin{equation}
(Q_{D0}^{\mathrm{def}})^{2}=W\cdot 
\mathrm{id}_{M}+\sum_{i=1}^{n}f_{i}(u)\phi_{i},
\end{equation}
where, using the same notation as (\ref{morphismsMF}), $\phi_{i}\in 
H^{0}(\mathcal{B}_{D0},\mathcal{B}_{D0})$. Then, the critical locus of 
$\mathcal{W}_{\mathrm{eff}}$ coincides, as a set, with 
$f_{1}=f_{2}=\cdots=f_{n}=0$. More precisely, if we identify the variables 
$Z_{i}$ with the parameters $u_{i}$, $Z_{i}\equiv u_{i}$ in 
$\mathcal{W}_{\mathrm{eff}}$, then $d\mathcal{W}_{\mathrm{eff}}^{-1}(0)$ 
coincides with the solutions to the equations $f_{1}=f_{2}=\cdots=f_{n}=0$ 
i.e. we can integrate the equations 
\begin{equation}
\frac{\partial \widetilde{\mathcal{W}}_{\mathrm{eff}}}{\partial u_{i}}=f_{i}(u)
\end{equation}
and $\widetilde{\mathcal{W}}_{\mathrm{eff}}$ coincides with 
$\mathcal{W}_{\mathrm{eff}}$ up to a nonlinear redefinition of the parameters 
$u_{i}$. The operators $\alpha_{\vec{m}}$ are computed iteratively. We can 
summarize this process as follows. Define $|\vec{m}|:=\sum_{i}m_{i}$. We start 
by defining
\begin{equation}
\alpha_{e_{i}}:=\psi_{i}\qquad i=1,\ldots,n
\end{equation}
with $e_{i}$, $i=1,\ldots,n$ the cannonical basis of $\mathbb{N}^{n}$. Then, in 
the first step we write
\begin{equation}
Q_{D0}^{\mathrm{def},(1)}=Q_{D0}+\sum_{i=1}^{n}u_{i}\alpha_{e_{i}}
\end{equation}
and we look at the terms of order $|\vec{m}|=2$ in 
$(Q_{D0}^{\mathrm{def},(1)})^{2}$, denote them 
$\sum_{|\vec{m}|=2}y_{\vec{m}}u^{\vec{m}}$. Then, if $y_{\vec{m}}$ is 
$Q_{D0}$-exact, then we can define an operator $\alpha_{\vec{m}}$ (with 
$|\vec{m}|=2$) such that
\begin{equation}
\beta_{\vec{m}}:=-y_{\vec{m}}=[Q_{D0},\alpha_{\vec{m}}].
\end{equation}
Then, if we denote $B_{2}$ the set of all vectors $\vec{m}$ with $|\vec{m}|=2$ 
such that $y_{\vec{m}}$ is exact, we can write 
\begin{equation}
Q_{D0}^{\mathrm{def},(2)}=Q_{D0}+\sum_{i=1}^{n}u_{i}\alpha_{e_{i}}+\sum_{\vec{m}
\in B_{2} }\alpha_{\vec{m}}u^{\vec{m}}
\end{equation}
and repeat the process to find the operators $\alpha_{\vec{m}}$ with $|m|>2$. 
The process ends when none of the $y_{\vec{m}}$ are $Q_{D0}$-exact.

\subsection{LG model with homogeneous superpotential}\label{sec:Af_homo}

In this section we consider the case of a LG model with chiral superfields 
$x_i, i=1,\cdots,n$. The superpotential $W$ is a homogeneous polynomial in 
$x_i$ 
with degree $d \geq 2$. We will set the orbifold $G$ to be trivial in this 
section, hence the relevant category of B-branes will be $MF(W)$. The 
$D0$-brane of this model is the matrix factorization $\mathcal{B}_{D0}$ 
described in the previous subsection (\ref{Qd0explicit}), therefore
\begin{equation}\label{D0}
Q_{D0} = \sum_{i=1}^n \left( x_i \overline{\eta}_i + \frac{1}{d} \frac{\partial 
W}{\partial x_i} \eta_i \right).
\end{equation}
We want to deduce the multiplication rule of the $\Af$-algebra 
\begin{equation}
\cA_{D0}=\mathrm{End}(\mathcal{B}_{D0}).
\end{equation}
The generators of the ring $H^{1}(\mathcal{B}_{D0},\mathcal{B}_{D0})$ are 
straightforward to compute and given by\footnote{Equation (\ref{generator}) 
denotes a particular representative of the $Q_{D0}$-class of $\psi_{i}$. All 
the computations where the explicit matrix form of $\psi_{i}$ is used do not 
depend on the choice of representative.}
\begin{equation}\label{generator}
\psi_i = i\sqrt{\frac{d(d-1)}{2}}\overline{\eta}_i - \frac{i}{\sqrt{2d (d-1)}} 
\sum_{j=1}^n \frac{\partial^2 W}{\partial x_i \partial x_j} \eta_j,~1 \leq i 
\leq n,
\end{equation}
which satisfy $\{ Q_{D0}, \psi_i \} = 0$. Note that
\begin{equation}\label{genrel}
\{ \psi_i,\psi_j \} = \frac{\partial^2 W}{\partial x_i \partial 
x_j}\mathbf{1},
\end{equation}
where $\mathbf{1}\in H^{0}(\mathcal{B}_{D0},\mathcal{B}_{D0})$ is the 
identity operator and $\{ \psi_i,\psi_j \}\simeq 0$, for $d>2$, i.e.
\eqref{genrel} says that $\{ 
\psi_i,\psi_j \}$ is $Q_{D0}$-exact, because $\{Q_{D0},\eta_i\} = x_i$. 
Hence, any monomial in $x_i$ with positive degree is $Q_{D0}$-exact. We remark 
that the operators (\ref{generator}) generate the whole
$H^{1}(\mathcal{B}_{D0},\mathcal{B}_{D0})$ as a ring (not necessarily as a 
vector space). Indeed they generate the whole space 
$\mathrm{End}(\mathcal{B}_{D0})$. This can be shown, for instance, 
using the explicit form of $Q_{D0}$ and the fact that the Dirac matrices 
$\eta_{i},\bar{\eta}_{i}$, $i=1\ldots,n$ plus the identify $\mathbf{1}$ generate 
the off-shell dg algebra.


Next we propose an explicit expression for the functor from $MF(W)$ to the 
category of modules of $\cA_{D0}$, for the case at 
hand. For any matrix factorization $\mathcal{B}=(M,\sigma_M, Q_M, 
R_M)$, the corresponding $\Af$-module is given by 
$Hom(\mathcal{B}_{D0},\mathcal{B})$.
The $\Af$-module structure is given by the $\Af$-multiplications of the 
$\Af$-category $MF(W)$ as described in (\ref{productsfunct}). Conversely, 
given any $\mathbb{Z}_2$-graded $\Af$-module $\mathbf{N}$ of $\cA_{D0}$, we 
propose that the corresponding matrix factorization is given by $M = \mathbf{N} 
\otimes \mathbb{C}[x_1,\cdots,x_n]$ and
\begin{equation}\label{QM}
Q_M (\phi) = \sum_{k = 1}^{d-1} \sum_{i_1,i_2,\cdots,i_k} m^\mathbf{N}_{k+1} 
(\phi,\psi_{i_1},\cdots,\psi_{i_k}) x_{i_1} x_{i_2} \cdots x_{i_k}
\end{equation}
for $\phi \in M$.
We provide several consistency checks for \eqref{QM} in Appendix A.

The $\Af$ structure of $\mathcal{A}_{D0}$ was constructed explicitly in 
\cite{ballard2014derived}, in the case $W$ homogeneous of degree $d$. The  
$\Af$-algebra relations were found to be given by 
\eqref{2m2} when $d=2$ or \eqref{dm2} and \eqref{dmd} when $d>2$. In 
\cite{ballard2014derived}, this $\Af$ 
structure was proved by summing over the ribbon trees. In the remainder 
of this subsection, we give some alternative derivation of \eqref{2m2}, 
\eqref{dm2} and \eqref{dmd} and we give explicit expressions for the higher 
order
products $m_{d}$, when acting on arbitrary elements of $\mathcal{A}_{D0}$. For 
this purpose we analyze separately the case $d=2$ and $d>2$. In the following 
we write the map 
\begin{equation}
\iota: \mathcal{A}_{D0}=\mathrm{End}(\mathcal{B}_{D0})\rightarrow V_{M_{D0}},
\end{equation}
where $V_{M_{D0}}$ is the space of $\mathrm{rk}(M)$-square matrices with 
values in $\mathbb{C}[x_{1},\ldots,x_{n}]$, i.e. the space of endomorphisms of 
$\mathcal{B}_{D0}$ without taking the homology. We can always give to the 
algebra $V_{M_{D0}}$ a dg algebra structure and will not spoil the 
$A_{\infty}$-relations of $\mathcal{A}_{D0}$. In other words, $V_{M_{D0}}$ is 
the off-shell algebra of open strings and we can always find a dg algebra that 
is $A_{\infty}$-quasi-isomprphic to it (this fact is true for any 
$A_{\infty}$-algebra \cite{lefevre2003infini}). We will denote the image under 
$\iota$ of $\psi_{i}$ in $V_{M_{D0}}$ by $v_{i}$:
\begin{equation}
\iota(\psi_i) = v_i.
\end{equation}
In the following we will also make use of the open disk one-point function 
correlators on the disk, for LG models. This was computed in 
\cite{Kapustin:2003ga} (see \cite{dyckerhoff2012kapustin,Carqueville:2012st} 
for a mathematical treatment) and is given by
\begin{equation}\label{res}
\langle \Phi \rangle = \frac{1}{(2 \pi i)^n} 
\oint_{x_i=0} \frac{\mathrm{Str}\left( \frac{\partial Q_{D0}}{\partial x_1} 
\wedge \cdots \wedge \frac{\partial Q_{D0}}{\partial 
x_n}\Phi\right)}{\frac{\partial W}{\partial x_1} \cdots \frac{\partial 
W}{\partial x_n}} dx_1\cdots dx_n.
\end{equation}
This corresponds to the (B-model) correlator in $D^{2}$ with a single boundary 
insertion $\Phi\in \mathcal{A}_{D0}$ and boundary conditions defined by the 
brane $\mathcal{B}_{D0}$.

\subsubsection{$\mathcal{A}_{D0}$ for $d=2$}

From the relations (\ref{ainftyrelsmor}) and \eqref{relsinclusion} (i.e. 
$f_{2}:\mathcal{A}_{D0}^{\otimes 2}\rightarrow V_{M_{D0}}$):
\begin{equation}
\iota(m_2(\psi_i,\psi_j)) = v_i v_j + \{ Q_{D0}, f_2(\psi_i,\psi_j) \}.
\end{equation}
When $d=2$, $v_i v_j$ has no $Q_{D0}$-exact terms, therefore we can choose $f_2 
(\psi_i,\psi_j) = 0$. Consequently,
\begin{equation}\label{lowdm2}
m_2 (\psi_i,\psi_j) + m_2(\psi_j,\psi_i) = \frac{\partial^2 W}{\partial x_i 
\partial x_j},
\end{equation}
i.e. under the multiplication $m_2$, $\cA_{D0}$ is the same as the Clifford algebra $Cl(n,\mathbb{C})$ with the quadratic form given by the Hessian of $W$. 
\eqref{lowdm2} can also be obtained by computing the correlation functions, 
using (\ref{res}). We illustrate this case with a simple example:

\subsubsection*{Example: $W = x_1 x_2$}

The D0-brane is
\begin{equation}
Q = x_1 \overline{\eta}_1 + x_2 \overline{\eta}_2 + \frac{x_2}{2} \eta_1 + 
\frac{x_1}{2} \eta_2.
\end{equation}
The fermionic open string states are
\begin{equation}
\psi_1 = \overline{\eta}_1 - \frac{1}{2} \eta_2, \quad \psi_2 = \overline{\eta}_2 - \frac{1}{2} \eta_1.
\end{equation}
The bosonic open string states can be taken to be\footnote{A representative 
for $\phi$ can be taken to be, for instance, 
$\bar{\eta}_{1}\bar{\eta}_{2}-\frac{1}{2}\bar{\eta}_{1}\eta_{1}-\frac{1}{2}\bar{
\eta}_{2}\eta_{2}+\frac{1}{4}\eta_{1}\eta_{2}$.} $e =1$ and $\phi$ such 
that $\langle e,\phi\rangle=1$ and $\langle e,e\rangle=\langle 
\phi,\phi\rangle=0$. One can compute
\begin{equation}
\gamma(m_2(\psi_1,\psi_2)) = \langle \psi_1 \psi_2 \rangle = \frac{1}{(2 \pi 
i)^2} \oint_{x_1 = x_2 =0} \frac{\mathrm{Str}\left( \frac{\partial Q}{\partial 
x_1} \frac{\partial Q}{\partial x_2} \psi_1 \psi_2 \right)}{\frac{\partial 
W}{\partial x_1} \frac{\partial W}{\partial x_2}} = 1,
\end{equation}
from which we also deduce
\[
\gamma(m_2(m_2(\psi_1,\psi_2),e)) = 1.
\]
Furthermore,
\[
\gamma(m_2(m_2(\psi_1,\psi_2),\psi_1\psi_2)) = \langle \psi_1 \psi_2 \psi_1 
\psi_2 \rangle = \frac{1}{(2 \pi i)^2} \oint_{x_1 = x_2 =0} 
\frac{\mathrm{Str}\left( \frac{\partial Q}{\partial x_1} \frac{\partial 
Q}{\partial x_2} \psi_1 \psi_2 \psi_1 \psi_2 \right)}{\frac{\partial 
W}{\partial x_1} \frac{\partial W}{\partial x_2}} = 1,
\]
thus
\[
m_2(\psi_1,\psi_2) = e + \phi.
\]
The same computation shows
\[
m_2(\psi_1,\psi_1) = m_2(\psi_2,\psi_2) = 0.
\]

\subsubsection{$d>2$}

Now assume the degree of $W$ is greater than 2. Because $\{ v_i, v_j \} = 
 \frac{\partial^2 W}{\partial x_i \partial x_j}$ is $Q_{D0}$-exact, we can 
take $f_2$ such that
\[
\iota (m_2(\psi_i,\psi_j) + m_2(\psi_j,\psi_i)) = \{ v_i,v_j \} + \{ Q_{D0}, 
f_2(\psi_i,\psi_j)+f_2(\psi_j,\psi_i) \} = 0,
\]
and therefore 
\begin{equation}\label{exterioralg}
m_2(\psi_i,\psi_j) + m_2(\psi_j,\psi_i) = 0,
\end{equation} 
which means that $\cA_{D0}$ is the exterior algebra $\wedge^\bullet \mathbb{C}^n$ under the multiplication $m_2$.

Alternatively, $m_2(\psi_i,\psi_j)$ can be determined by the correlation 
functions (\ref{res}).
First note that the one-point correlator
\begin{equation}\label{corr1}
\langle \psi_{i_1}\cdots \psi_{i_m} \rangle = 0 \quad
\mathrm{if}~ i_s = i_t 
\end{equation}
for any pair of indices $i_s$ and $i_t$. This is simply because $\{ 
\psi_i,\psi_j \}$ and $\psi_i^2$ are $Q_{D0}$-exact and as a consequence the 
correlation function can be rewritten as a sum of correlation functions, each 
involving a $Q_{D0}$-exact operator. A corollary of this 
observation is that
\begin{equation}\label{corr2}
\langle \psi_{i_1}\cdots \psi_{i_m} \rangle = 0 \quad \mathrm{if} ~m>n.
\end{equation}
Now assume that $m \leq n$. The formula (\ref{res}) implies
\begin{equation}
\langle \psi_{i_1}\cdots \psi_{i_m} \rangle = \frac{1}{(2 \pi i)^n} 
\oint_{x_i=0} \frac{\mathrm{Str}\left( \frac{\partial Q_{D0}}{\partial x_1} 
\wedge \cdots \wedge \frac{\partial Q_{D0}}{\partial x_n} \psi_{i_1}\cdots 
\psi_{i_m} \right)}{\frac{\partial W}{\partial x_1} \cdots \frac{\partial 
W}{\partial x_n}} dx_1\cdots dx_n.
\end{equation}
The degree of the denominator of the integrand is $nd-n$. In order to have a 
nonzero result, the numerator of the integrand must have degree $nd-2n$, which 
can only result from the term in $\frac{\partial Q_{D0}}{\partial x_1} \wedge 
\cdots \wedge \frac{\partial Q_{D0}}{\partial x_n} \psi_{i_1}\cdots \psi_{i_m}$ 
proportional to $\eta_1\cdots\eta_n$. But in order to make nonzero contribution 
to the supertrace, there should also be $n$ $\overline{\eta}$'s to contract with 
the $\eta's$, this is possible only when $m=n$. In conclusion,
\begin{equation}\label{corr3}
\langle \psi_{i_1}\cdots \psi_{i_m} \rangle = 0 \quad \mathrm{if} ~m<n,
\end{equation}
and it is straightforward to compute from \eqref{res} that
\begin{equation}\label{corr0}
\langle \psi_1 \psi_2 \cdots \psi_n \rangle = 1
\end{equation}
up to a normalization factor, which implies $\psi_{i_1}\cdots \psi_{i_m}$ is 
dual to $\pm \prod_{j \neq i_s, 1\leq s \leq m}\psi_j$. Combining 
\eqref{corr1}, \eqref{corr2} and \eqref{corr3}, we see that for a monomial $f$ 
in $\psi_i$,
\begin{equation}
\gamma(m_2(m_2(\psi_i,\psi_j),f(\psi_1,\cdots,\psi_n))) = \pm \langle \psi_i 
\psi_j f(\psi_1,\cdots,\psi_n) \rangle 
\end{equation}
vanishes unless $f(\psi_1,\cdots,\psi_n) = \pm \prod_{k \neq i,j}\psi_k$. This 
allows us to conclude that $\iota(m_2(\psi_i,\psi_j))$ only contains the term 
dual to 
$\prod_{k \neq i,j}v_k$, i.e.
\begin{equation}
m_2(\psi_i,\psi_j) = \frac{1}{2}\pi([v_i, v_j]).
\end{equation}
where $\pi:V_{M_{D0}}\rightarrow \mathcal{A}_{D0}$ is the projection onto 
$Q_{D0}$-classes.

Now we compute the higher order multiplications $m_k, k>2$. In principle, 
this can be done by performing the algorithm described in section 
\ref{sec:LGAinf} using \eqref{ainftyrelsmor}. Here we take the physical 
perspective and determine the multiplications by studying the deformations of 
$Q_{D0}$ as reviewed in section \ref{sec:effpot}.

Assume that we deform $Q_{D0}$ using the fermionic generators $\psi_{i}$:
\begin{equation}\label{deformQ}
Q_{D0}^{\mathrm{def}} = Q_{D0} + \sum_{\vec{m}:|\vec{m}|>0} \alpha_{\vec{m}} 
u^{\vec{m}},
\end{equation}
where $\alpha_{e_i} = \psi_i/(\sqrt{d(d-1)})$. In principle, we can consider 
further deformations by other elements of 
$H^{1}(\mathcal{B}_{D0},\mathcal{B}_{D0})$ (and even elements of 
$H^{0}(\mathcal{B}_{D0},\mathcal{B}_{D0})$), but if we are interested in 
extracting the higher order products involving only $\psi_{i}$ operators, this
suffices. Indeed, if the most general first order deformation (i.e. $|m|=1$)
has the form
\begin{equation} 
\sum_{i}u_{i}\psi_{i}+\sum_{\mu}u_{\mu}\Lambda_{\mu}
\end{equation}
where $\Lambda_{\mu}$ denote the operators in 
$H^{1}(\mathcal{B}_{D0},\mathcal{B}_{D0})$ that are not $\psi_{i}$'s, then if 
we set $u_{\mu}=0$, after running the algorithm outlined in 
section \ref{sec:effpot} we will get that $(Q_{D0}^{\mathrm{def}})^{2}$ has the 
form
\begin{equation}\label{squareQdef}
(Q_{D0}^{\mathrm{def}})^{2}=W\cdot\mathbf{1}+\sum_{a}f_{a}(u_i)\phi_{a}
\end{equation}
where $f_{a}$'s will have the following interpretation
\begin{equation} 
f_{a}(u_i)=\frac{\partial \widetilde{W}_{\mathrm{eff}}}{\partial 
u_{a}}\Big|_{u_{\mu}=0}=\sum_{k\geq 2}\sum_{i_{1},\ldots,i_{k}}\langle 
m_{k}(\psi_{i_{1}},\ldots,\psi_{i_{k}}),\Lambda^{D}_{a}\rangle 
u_{i_{1}}\cdots u_{i_{k}}
\end{equation}
where $a$ runs over all operators in $\mathcal{A}_{D0}$ and $\Lambda^{D}_{a}$ 
denotes the operator dual to $\Lambda_{a}\in\{\psi_{i},\Lambda_{\mu}\}$. Hence, 
(\ref{squareQdef}) will contain all the information we need about the higher 
products $m_{k}(\psi_{i_{1}},\ldots,\psi_{i_{k}})$, when we set $u_{\mu}=0$.

Define $\alpha_{i_1\cdots i_s} := \alpha_{e_{i_1} + \cdots + 
e_{i_s}}$, $\beta_{i_1\cdots i_s} :=\beta_{e_{i_1} + \cdots 
+ e_{i_s}}$ and $W_{i_{1}\ldots 
i_{s}}:=\partial_{i_{1}}\cdots\partial_{i_{s}}W$. Then we have
\[
\beta_{ij} = \{ \alpha_i, \alpha_j \} = 
\frac{1}{d(d-1)}W_{ij}.
\]
As $d>2$, $\beta_{ij}$ is $Q_{D0}$-exact: $d (d-1) \beta_{ij} = W_{ij} = \{ 
Q_{D0}, \sum_k W_{ijk} \eta_k \} /(d-2)$. Then we can take $\alpha_{ij} = 
-1\sum_k W_{ijk} \eta_k /(d(d-1)(d-2)) $ to cancel $\beta_{ij}$ in 
$(Q_{D0}^{\mathrm{def}})^{2}$. Then at degree 3, one computes $\beta_{ijk} = - 
W_{ijk}/(d(d-1)(d-2))$ and $\alpha_{ijk} = \sum_l W_{ijkl} 
\eta_l/(d(d-1)(d-2))$. This process continues and at degree $m$ we have
\begin{equation}
\beta_{i_1\cdots i_m} = (-1)^m \frac{W_{i_1\cdots i_m}}{d(d-1)\cdots (d-m+1)}
\end{equation}
and
\begin{equation}
\alpha_{i_1\cdots i_m} = (-1)^{m-1} \frac{\sum_j W_{i_1\cdots i_m j} \eta_j}{d(d-1)\cdots (d-m+1)}
\end{equation}
for $2\leq m \leq d$. In particular, $\beta_{i_1\cdots i_d}$ is not 
$Q_{D0}$-exact and cannot be cancelled by a choice of $\alpha_{i_1\cdots i_d}$. 
As a result,
\begin{equation}\label{Qdef2}
(Q_{D0}^{\mathrm{def}})^{2} = Q_{D0}^2 + \frac{(-1)^d}{d!} \sum_{r_1 + \cdots + 
r_n = d} \frac{\partial^d W}{\partial x_1^{r_1} \cdots \partial x_n^{r_n}} 
u_1^{r_1} u_2^{r_2} \cdots u_n^{r_n}\mathbf{1}.
\end{equation}
This means the obstruction to the deformation is given by the identity operator. 
Therefore $\widetilde{\mathcal{W}}_{\mathrm{eff}}$ takes the form
\begin{equation}\label{Qdef3}
\widetilde{\mathcal{W}}_{\mathrm{eff}}=\sum_{i_{1},\ldots,i_{d}}\langle 
m_{d}(\psi_{i_{1}},\ldots,\psi_{i_{d}}),\Lambda\rangle 
u_{i_{1}}\cdots u_{i_{d}}u_{0}+\mathcal{O}(u_{0}^{2}),\qquad 
\Lambda=\psi_{1}\cdots\psi_{n} 
\end{equation}
where $\Lambda$ is the dual to the identity operator. Due to the 
correlation function \eqref{corr0} and \eqref{Qdef2} together with 
\eqref{Qdef3} we conclude that
\begin{equation}
m_d (\psi_{i_1},\psi_{i_2},\cdots,\psi_{i_d}) + \mathrm{cyclic~permutations} = 
\frac{(-1)^d}{d!} W_{i_1\cdots i_d},
\end{equation}
and
\begin{equation}
m_k (\psi_{i_1},\psi_{i_2},\cdots,\psi_{i_k}) + \mathrm{cyclic~permutations} = 
0\qquad k\neq d
\end{equation}
where the $i_l$'s are not necessarily distinct. Using directly the algorithm 
outlined in \ref{sec:LGAinf}, 
and \eqref{ainftyrelsmor}\footnote{\cite{ballard2014derived} used a different 
approach, namely summing the ribbon trees. We review this idea in the 
appendix \ref{app:ribbon}.} we can further determine exactly all the 
higher order products $m_{k}$. This computation ends up determining the 
$A_{\infty}$ relations as 
\begin{itemize}
\item If $d=2$, $\cA_{D0}$ is a family of Clifford algebras:
\begin{equation}\label{algrels1}
m_2(\psi_i,\psi_j)+m_2(\psi_j,\psi_i) = \frac{\partial^2 W}{\partial x_i 
\partial x_j}.
\end{equation}
\item If $d>2$, $\cA_{D0}$ is an $\Af$-algebra, where $m_2$ is the wedge 
product, $m_k=0$ for $k=1,3,4,\cdots,d-1$ and
\begin{equation}\label{algrels2}
m_d(\psi_{i_1},\psi_{i_2},\cdots,\psi_{i_d}) = \frac{1}{d!}\frac{\partial^d 
W}{\partial x_{i_1} \cdots \partial x_{i_d}}.
\end{equation}
\end{itemize}
We illustrate with an example the computation of $m_{k}$, exactly, for 
$d=3$ (using the relations (\ref{ainftyrelsmor}) and \eqref{relsinclusion}):
\[
\{ v_i, v_j \} = W_{ij} = \{ Q, \sum_k W_{ijk} \eta_k \}.
\]
Because
\[
v_i v_j = \frac{1}{2} \{ v_i,v_j \} + \frac{1}{2}[v_i,v_j] = 
\frac{1}{2} W_{ij} + \frac{1}{2} [v_i,v_j],
\]
we have
\[
m_2(\psi_i,\psi_j) = \frac{1}{2} \pi \left( [v_i,v_j] \right),\quad 
f_2(\psi_i,\psi_j) = -\frac{1}{2} \sum_k W_{ijk} \eta_k,
\]
Thus $m_2(\psi_i,\psi_j)+m_2(\psi_j,\psi_i)=0$.
\[
v_j v_i v_k = \{ v_i, v_j \} v_k - v_i v_j v_k = W_{ij} 
v_k - W_{jk} v_i + v_i v_k v_j
\]
yields
\[
v_i[v_j,v_k] - [v_i,v_j]v_k = \{ Q, -\sum_l W_{jkl} \eta_l 
v_i - \sum_l W_{ijl} v_k \eta_l \}.
\]
Therefore
\[
f_2(\psi_i,m_2(\psi_j,\psi_k)) - f_2(m_2(\psi_i,\psi_j),\psi_k) = \frac{1}{2} 
(\sum_l W_{ijl} v_k \eta_l + \sum_{l} W_{jkl} \eta_l v_i).
\]
\[
\begin{split}
&\iota(m_3(\psi_i,\psi_j,\psi_k)) = f_2(\psi_i,m_2(\psi_j,\psi_k)) - 
f_2(m_2(\psi_i,\psi_j),\psi_k) - v_i f_2(\psi_j,\psi_k) - f_2(\psi_i,\psi_j) 
v_k \\
&= \frac{1}{2} \sum_l (W_{ijl}v_k\eta_l + W_{jkl}\eta_l v_i + 
W_{jkl}v_i\eta_l + W_{ijl} \eta_l v_k) = \frac{1}{2} (W_{ijl} \delta_{lk} 
+ W_{jkl} \delta_{li}) = W_{ijk}.
\end{split}
\]
In conclusion, $m_3(\psi_i,\psi_j,\psi_k) = W_{ijk}$.

Therefore, in general, we can say that all the elements of $\cA_{D0}$ can be 
written as linear combinations of the form ($d>2$)
\begin{equation}\label{lambdaproduct}
\Lambda_{i}=\psi_{1}\wedge\cdots\wedge\psi_{i_{r}},\qquad 
\mathrm{deg}(\Lambda_{i}):=r
\end{equation}
where we defined $\mathrm{deg}(\Lambda_{i})$ for later convenience and 
$\wedge$ denotes the usual skew-symmetric wedge product. The fact that all the 
elements can be write as in the formula (\ref{lambdaproduct}) is just a 
consequence of (\ref{exterioralg}). Now, we can determine 
$m_d(\Lambda_1,\Lambda_2,\cdots,\Lambda_d)$ for $\Lambda_i = \psi_{i_1} \wedge 
\cdots \wedge \psi_{i_{\deg(\Lambda_i)}}$. Since $m_k = 0$ for $k \neq 
2,d$, the relation \eqref{ainftyrels} can be solved by the 
rule\footnote{Notice that $\Lambda_1 m_k(\Lambda_2,\cdots,\Lambda_{k+1}) = 
(-1)^{\deg(\Lambda_1) (\deg(\Lambda_2)+\cdots+\deg(\Lambda_{k+1})+k)} 
m_k(\Lambda_2,\cdots,\Lambda_{k+1}) \Lambda_1$.}
\begin{equation}\label{mkRule}
\begin{split}
&m_k(\Lambda_1,\Lambda_2,\cdots,m_2(\Lambda_i,\Lambda_{i+1}),\cdots,\Lambda_{k+1
}) \\
= & (-1)^{\deg(\Lambda_i) (\deg(\Lambda_{i+1})+\cdots+\deg(\Lambda_{k+1}) )} 
m_2(m_k(\Lambda_1,\cdots,\Lambda_{i-1},\Lambda_{i+1},\cdots,\Lambda_{k+1}), 
\Lambda_i) \\
&+ (-1)^{\deg(\Lambda_{i+1}) (\deg(\Lambda_{i+2})+\cdots+\deg(\Lambda_{k+1}) )} 
m_2(m_k(\Lambda_1,\cdots,\Lambda_i,\Lambda_{i+2},\cdots,\Lambda_{k+1}), 
\Lambda_{i+1}).
\end{split}
\end{equation}
By repeated use of \eqref{mkRule}, we are lead to the conclusion that
\begin{equation}\label{relsgeneral1}
m_d(\psi_{i_0}\wedge\cdots \wedge \psi_{i_{t_1}},\psi_{j_0}\wedge\cdots \wedge 
\psi_{j_{t_2}},\cdots,\psi_{k_0}\wedge\cdots \wedge \psi_{k_{t_d}})
\end{equation}
is equal to the sum
\begin{equation}\label{relsgeneral2}
\begin{split}
&\sum_{a_1=0}^{t_1} \sum_{a_2=0}^{t_2} \cdots \sum_{a_d=0}^{t_d} 
m_d(\psi_{i_{a_1}},\psi_{j_{a_2}},\cdots,\psi_{k_{a_d}}) (-1)^{a_1+\cdots+a_d} 
\cdot \\
&\psi_{i_0}\wedge \cdots \hat{\psi}_{i_{a_1}} \cdots \wedge \psi_{i_{t_1}} 
\wedge \psi_{j_0}\wedge \cdots  \hat{\psi}_{j_{a_2}} \cdots \wedge 
\psi_{j_{t_2}} \wedge \cdots \wedge \psi_{k_0} \wedge \cdots  
\hat{\psi}_{k_{a_d}} \cdots \wedge \psi_{k_{t_d}} 
\end{split}
\end{equation}
up to an overall sign.



We finish this section with a very simple example to illustrate the consistency 
between the $\widetilde{\mathcal{W}}_{eff}$ computation and the relations 
(\ref{algrels1}) and (\ref{algrels2}):

\subsubsection*{Example:~$W=x^d$}

Let's consider the LG model with a single chiral superfield $x$ and a 
superpotential $W = x^d$, $d \geq 2$. The D0-brane is given by
\begin{equation}\label{miniD0}
Q = x\overline{\eta} + x^{d-1}\eta.
\end{equation}
The bosonic open string state is $e = 1$ and the fermionic open string state is 
$\psi = \overline{\eta} - \eta$. One can use the Kapustin-Li formula 
\cite{Kapustin:2003ga} to compute the three-point correlation function
\[
\langle \psi \psi \psi \rangle = \frac{1}{2 \pi i} \oint_{x=0} 
\frac{\mathrm{Str}\left( \frac{dQ}{dx} \psi \psi \psi \right)}{\frac{dW}{dx}} = 
\left\{ \begin{array}{ll}
1, & d=2, \\
0, & d>2.
\end{array} \right.
\]
From the relation
\[
\langle \psi \psi \psi \rangle = \gamma\left( m_2(m_2(\psi,\psi),\psi) \right)
\]
and
\[
\gamma(e)=0,\quad \gamma(\psi) = 1,
\]
we see
\[
m_2(\psi,\psi) = \left\{ \begin{array}{ll}
e, & d=2, \\
0, & d>2.
\end{array} \right.
\]
It was shown in \cite{Govindarajan:2006uy} that the effective superpotential, or disk partition function, of the LG model with Dirichlet boundary condition, which is equivalent to \eqref{miniD0}, is
\[
\mathcal{W}_{\mathrm{eff}} = \mathrm{Tr}\left( \frac{Z^{d+1}}{d+1} \right)
\]
up to a rescaling, where $Z$ is the world volume field dual to $\psi$. From 
this effective superpotential we conclude
\[
\gamma\left( m_2(m_s(\psi^{\otimes s}),\psi) \right) = \left\{ \begin{array}{ll}
1, & s=d, \\
0, & s \neq d.
\end{array} \right.
\]
or equivalently
\[
m_s(\psi^{\otimes s}) = \left\{ \begin{array}{ll}
e, & s=d, \\
0, & s \neq d.
\end{array} \right.
\]
Finally, we remark that in principle we can study the correspondence between 
$MF(W)$ and $D(\mathrm{Mod}-\mathcal{A}_{D0})$ from the point of view of tensor 
products of minimal models. A homogeneous superpotential $W\in\mathbb{C}[x_{1},
\ldots,x_{n}]$ of degree $d$ at a special point in complex structure 
moduli can be seen as the tensor product of $n$ $A_{d-1}$ minimal models. This 
relates to the well known structure of tensor products in matrix factorization 
categories. It will be interesting to study further how this tensor product 
structure translates to the category $D(\mathrm{Mod}-\mathcal{A}_{D0})$ as it 
is 
well known that tensor products of $A_{\infty}$-algebras is rather nontrivial 
\cite{Gaberdiel:1997ia,saneblidze2000diagonal,loday2011diagonal}.

\subsection{Inhomogeneous superpotential}

Consider now a quasi-homogeneous superpotential 
$W\in\mathbb{C}[x_{1},\ldots,x_{N}]$. Then, we write the 
superpotential as a sum of homogeneous polynomial-degree terms:
\begin{equation}
W(x) = \sum_{l=2}^d W^{(l)}(x),
\end{equation}
where each term $W^{(l)}(x)$ has polynomial-degree $l$, i.e. where we assign 
degree $1$ to each variable $x_{i}$. The brane $\mathcal{B}_{D0}$ and the 
fermionic generators of the open string states are still given by \eqref{D0} 
and \eqref{generator} respectively. If we turn on deformations as in 
\eqref{deformQ}, we can use the same argument\footnote{Also one can apply an 
argument based on ribbon trees as reviewed in Appendix \ref{app:ribbon}.} to 
deduce that the obstruction is 
given by
\begin{equation}
\frac{(-1)^l}{l!} \sum_{i_1 + \cdots + i_n = l} \frac{\partial^l 
W^{(l)}}{\partial x_1^{i_1} \cdots \partial x_n^{i_n}} u_1^{i_1} u_2^{i_2} 
\cdots u_n^{i_n}
\end{equation}
at degree $l$. Therefore we have the following multiplications
\begin{equation}\label{Af_inhomo1}
m_2 (\psi_i,\psi_j) + m_2(\psi_j,\psi_i) = \frac{\partial^2 W^{(2)}}{\partial 
x_i \partial x_j}
\end{equation}
and
\begin{equation}\label{Af_inhomo2}
m_l (\psi_{i_1},\psi_{i_2},\cdots,\psi_{i_l}) = \frac{1}{l!} \frac{\partial^l 
W^{(l)}}{\partial x_{i_1} \partial x_{i_2} \cdots \partial x_{i_l}}
\end{equation}
for $3 \leq l \leq d$ and $m_k = 0$ for $k>d$.

\subsection{Landau-Ginzburg Orbifold}\label{sec:LGorb}

So far we have considered LG models with trivial orbifold group. For a LG 
orbifold, there is a finite abelian group $G$ acting on the field space, and 
$W\in\mathbb{C}[x_{1},\ldots,x_{n}]^{G}$. As such, the open string states are 
those invariant under the action of $G$. Let $\psi_1,\cdots,\psi_n$ be the 
degree-one fermionic generators \eqref{generator} in 
$\mathrm{End}_{MF(W)}(\mathcal{B}_{D0})$ of the LG model without orbifolding. 
Note that each $\psi_1$ transforms in a definite representation, when we take 
the $G$ action into account. When we incorporate the orbifold the brane defined 
by the matrix factorization $Q_{D0}$ in \eqref{D0} requires the 
specification of a representation $\rho_{M}$, compatible with $\rho_{orb}$. It 
is easy to see that $\rho_{M}:G\rightarrow GL(M)$ is almost completely fixed by 
its action, via conjugation, over $\eta_{j},\bar{\eta}_{j}$, in the definition 
of $Q_{D0}$. Then $\rho_{M}$ is fixed up to its action on the Clifford 
vacuum $|0\rangle$ (defined by $\eta_{j}|0\rangle=0$ for all $j$). Hence, 
we can label $\mathcal{B}_{D0}^{(a)}$ in the category $MF(W,G)$ by a single 
(one-dimensional) irreducible representation\footnote{These are sometimes 
called the orbit branes related to $\mathcal{B}_{D0}$ 
\cite{Ashok:2004zb,Jockers:2006sm}.}: the representation of $|0\rangle$. Then, 
if $G=\mathbb{Z}_{d}$, we have $a=0,\ldots,d-1$ and we denote 
\begin{equation}\label{tiltingorb}
\mathcal{B}_{orb}:=\bigoplus_{a=0}^{d-1}\mathcal{B}_{D0}^{(a)}\in MF(W,G),
\end{equation}
where the set of branes $\mathcal{B}_{D0}^{(a)}$, $a=0,\ldots,d-1$ form a set 
of 
generators of $MF(G,W)$ \cite{tu2014matrix} and the algebra $\mathcal{A}_{D0}$ 
must be replaced by
\begin{equation}
\mathcal{A}_{orb}=\mathrm{End}_{MF(W,G)}(\mathcal{B}_{orb}).
\end{equation}

The correspondence $MF(W,G)\cong D(Mod-\mathcal{A}_{orb})$ was studied in 
\cite{Ashok:2004zb,Ashok:2004xq}, for homogeneous potentials. Moreover, the 
results in 
\cite{tu2014matrix} implies the following isomorphism of $A_{\infty}$ algebras:
\begin{equation}
\mathcal{A}_{orb}\cong \cA_{D0} \sharp G,
\end{equation}
where $\cA_{D0} \sharp G$, the smash product of $\cA_{D0}$ and $\mathbb{C}[G]$, 
is regarded as an $\Af$-algebra with $m_2(g_1,g_2) = g_1 \cdot g_2$ 
and $m_k\equiv 0$ for $k \neq 2$ ($g_{1},g_{2}\in G$). Then $\mathcal{A}_{orb}$ 
can be regarded as the 
product of two $A_{\infty}$-algebras. We can still use the construction 
introduced in 
section \ref{sec:Af_homo} to set up the correspondence between objects of 
$MF(W,G)$ and the $\Af$-modules over $\cA_{D0} \sharp G$. The difference is 
that the module is not only an $\Af$-module of $\cA_{D0}$, but also a 
$\mathbb{C}[G]$-module, this corresponds to the fact that the Chan-Paton spaces 
of 
the matrix factorizations of LG orbifold all carries a $G$-representation.

Specifically for the case $G = \mathbb{Z}_d$. Because the multiplication 
$m^{\cA_{D0} \sharp \mathbb{Z}_d}_2$ of $\cA_{D0} \sharp \mathbb{Z}_d$ satisfies
\begin{equation}
m^{\cA_{D0} \sharp \mathbb{Z}_d}_2 (a \sharp g_1, b \sharp g_2) = m_2^{\cA_{D0}}(a, g_1 b g_1^{-1}) \sharp (g_1 \cdot g_2),
\end{equation}
an $\cA_{D0} \sharp \mathbb{Z}_d$-module is of the form $\oplus_{i=0}^{d-1} M_i \otimes \rho_{\lambda+i}$, where $\oplus_{i=0}^{d-1} M_i$ is a $\mathbb{Z}_d$-graded $\cA_{D0}$-module and $\rho_l$ denotes the one-dimensional representation of $\mathbb{Z}_d$ with weight $\exp(2 \pi i l/d)$.
 
For example, when $d=2$, $\cA_{D0}$ is a Clifford algebra $Cl(n,\mathbb{C})$, 
hence if $G=\mathbb{Z}_{2}$ we have
\begin{equation}\label{corresp1}
MF(W,\mathbb{Z}_{2})\cong D(Mod-\cA_{D0} \sharp \mathbb{Z}_2).
\end{equation}
The category $MF(W,\mathbb{Z}_{2})$ is very similar to the graded category 
$MF(W)$ but its morphisms are diferent. Because of the $\mathbb{Z}_{2}$ 
orbifold all the morphisms between irreducible objects are either even or odd, 
but not both. This is exactly the category studied in \cite{Bertin:2008osh} and 
so, we can use the results in \cite{Bertin:2008osh} to conclude
\begin{equation}
MF(W,\mathbb{Z}_{2})\cong D(Mod_{\mathbb{Z}_{2}}-Cl(n,\mathbb{C})),
\end{equation}
where $ D(Mod_{\mathbb{Z}_{2}}-Cl(n,\mathbb{C}))$ denotes the derived category 
of graded modules over $Cl(n,\mathbb{C})$. A classical result of 
Atiyah-Bott-Shapiro \cite{atiyah1964clifford} (see also 
\cite{Bertin:2008osh}) establish
\begin{equation}
MF(W,\mathbb{Z}_{2})\cong D(Mod_{\mathbb{Z}_{2}}-Cl(n,\mathbb{C}))\cong 
D(Mod-Cl_{0}(n,\mathbb{C})),
\end{equation}
where $Cl_{0}(n,\mathbb{C})$ denotes the even part of the Clifford 
algebra $Cl(n,\mathbb{C})$. We remark that the category considered in 
\cite{Kapustin:2002bi} is the category $MF(W)$ where the morphisms are odd and 
even, and hence is equivalent to $D(Mod-Cl(n,\mathbb{C}))$, where the modules 
are not graded. Finally, we illustrate the correspondence (\ref{corresp1}) with 
an example. Set $W = \sum_{i=1}^{2m} x_i^2$ and each $x_i$ is 
$\mathbb{Z}_2$-odd. Let $S_+$ and $S_-$ be the spinor representation with left 
and right chirality respectively. The $\cA_{D0} \sharp \mathbb{Z}_2$ module 
$M:=S_+ \otimes \rho_0 \oplus S_- \otimes \rho_1$ corresponds to the matrix 
factorization with $Q_M = \sum_{i=1}^m(x_{2i-1} + i x_{2i}) \overline{\eta}_i + 
\sum_{i=1}^m(x_{2i-1} - i x_{2i}) \eta_i$ and the vacuum being 
$\mathbb{Z}_2$-even. The $\cA_{D0} \sharp \mathbb{Z}_2$ module $S_+ \otimes 
\rho_1 \oplus S_- \otimes \rho_0$ corresponds to the matrix factorization with 
the same $Q_M$ but the vacuum being $\mathbb{Z}_2$-odd ($\overline{\eta}_i$ and 
$\eta_i$ are $\mathbb{Z}_2$-odd in both cases).

\subsection{Hybrid Model}\label{sec:hybrid}

Finally, let us consider the $A_\infty$ structure of the matrix factorizations 
of hybrid models.

Start with the trivial fibration. In this case, the theory under 
consideration is defined on the target space
\begin{equation}\label{VfxVb}
(V_f \times V_b) / G,
\end{equation}
where $V_b = \mathbb{C}^m$ and $V_f = \mathbb{C}^n$ are regarded as the base space and the fiber respectively, and $G$ is the orbifold group acting on $V_b$ and $V_f$. 
Suppose that the base coordinates are $z_i, i=1,\cdots,m$ and the fiber coordinates are $x_j, j=1,\cdots,n$. The superpotential $W(x,z)$ is a $G$-invariant holomorphic function in $x_j$ and $z_i$.

The analogue of the $D0$-brane for the LG models we discussed in previous sections is point-like along each fiber, i.e. it localizes to the base space. We call this brane the reference brane and denote it by $\mathcal{B}_0$. The superpotential can be written as
\[
W = \sum_{l} W^{(l)}(x,z),
\]
where $W{(l)}$ is homogeneous in $x_i$ with degree $l$, therefore the 
endomorphism $Q_{0}\in \mathcal{B}_0$ (see (\ref{mfbbb})) is
\[
Q_0 = \sum_{i=1}^n x_i \bar{\eta}_i + \sum_{i=1}^n \sum_l \frac{1}{l} \frac{\partial W^{(l)}(x,z)}{\partial x_i} \eta_i.
\]
Then, as discussed in section \ref{sec:LGorb}, the $A_\infty$-algebra of the hybrid model on \eqref{VfxVb} is
\[
\mathcal{A}_0 \sharp G,
\]
where $\mathcal{A}_0$ is generated by $\psi_i(z) \in \mathrm{End}_{MF(W)}(\mathcal{B}_0)$ satisfying
\[
m_2 (\psi_i(z),\psi_j(z)) + m_2(\psi_j(z),\psi_i(z)) = \frac{\partial^2 W^{(2)}(x,z)}{\partial 
x_i \partial x_j}, \]
\[
m_l (\psi_{i_1}(z),\psi_{i_2}(z),\cdots,\psi_{i_l}(z)) = \frac{1}{l!} \frac{\partial^l 
W^{(l)}(x,z)}{\partial x_{i_1} \partial x_{i_2} \cdots \partial x_{i_l}}, l>2.
\]

For a hybrid model (for 
details on the precise definition of hybrid models see 
\cite{Bertolini:2013xga}), defined on a space of the form 
\begin{equation}\label{Tot}
Y:=\mathrm{Tot}\left( \mathcal{V} \stackrel{\pi}{\rightarrow} B \right)
\end{equation}
with superpotential $W\in H^{0}(\mathcal{O}_{Y})$, where $\mathcal{V}$ is 
a vector bundle over the base space $B$, we can decompose $Y$ into a set of coordinate patches
\[
Y = \bigcup_i U_i,
\]
such that every patch has the form
\[
U_i = \left( V_f \times V_b \right)/ G_i=\widetilde{U}_{i}/G_{i},
\]
and $\dim V_f = \mathrm{rank}\mathcal{V}, \dim V_b = \dim B$. We use the 
notation $\widetilde{U}_{i}$ to denote the affine space $ V_f \times V_b$. The 
reference brane $\mathcal{B}_0$ is of the form
\[
\xymatrix{
	\mathcal{O} \ar@<0.5ex>[r]^-{X} & \pi^*\mathcal{V} \ar@<0.5ex>[r]^-{X} \ar@<0.5ex>[l]^-{f(X,P)} & \pi^*\wedge^2 \mathcal{V} \ar@<0.5ex>[r]^-{X} \ar@<0.5ex>[l]^-{f(X,P)} & \cdots \ar@<0.5ex>[r]^-{X} \ar@<0.5ex>[l]^-{f(X,P)} & \pi^*\wedge^{n-1}\mathcal{V} \ar@<0.5ex>[r]^-{X} \ar@<0.5ex>[l]^-{f(X,P)} & \pi^*\wedge^n \mathcal{V} \ar@<0.5ex>[l]^-{f(X,P)}
},
\]
where $X$ denotes collectively the coordinates along the fiber of $\mathcal{V}$ and $f(X,P)$ is a map that depends also on the base coordinate $P$ such that $f(X,P) \cdot X = X \cdot f(X,P) = W(X,P)$. 
Then from the discussion above, the reference brane $\mathcal{B}_0$ that is point-like along each fiber gives rise to an $A_\infty$-algebra $\mathcal{A}_{0i} \sharp G_i$ within each coordinate patch, we can define a sheaf of $A_\infty$-algebra $\mathcal{A}$ by
\[
\mathcal{A}(U_i) = \mathcal{A}_0(U_i) \sharp G_i.
\]
If we denote by $x^{(i)}$ and $z^{(i)}$ the local fiber and base coordinates in $U_i$, then $\mathcal{A}_0(U_i)$ is generated by $\psi^{(i)}_s, s=1,\cdots,n$, which satisfy
\[
m_2 (\psi^{(i)}_s,\psi^{(i)}_t) + m_2(\psi^{(i)}_t,\psi^{(i)}_s) = 
\frac{\partial^2 W^{(2)}(x^{(i)},z^{(i)})}{\partial 
x^{(i)}_s \partial x^{(i)}_t}, \]
\[
m_l (\psi^{(i)}_{s_1},\psi^{(i)}_{s_2},\cdots,\psi^{(i)}_{s_l}) = \frac{1}{l!} \frac{\partial^l 
W^{(l)}(x^{(i)},z^{(i)})}{\partial x^{(i)}_{s_1} \partial x^{(i)}_{s_{2}} 
\cdots \partial x^{(i)}_{s_l}}, l>2.
\]
In other words, the algebra 
$\mathcal{A}_0(U_i)=\mathcal{A}_0(\widetilde{U}_i)$ i.e. it is constructed by
ignoring the orbifold structure. In order to understand how these algebras are
glued together as we change charts, we have to be careful with the treatment of 
the orbifold singularities of $B$. It has been proposed that the correct 
mathematical framework to study this problem is to view the GLSM as an 
algebraic stack\footnote{For an introduction to algebraic stacks see for 
example \cite{gomez2001algebraic} and for the specific case of smooth toric 
Deligne-Mumnford stacks, relevant for the examples in this work, see 
\cite{fantechi2010smooth}. A relation between GLSMs and stacks can be found in 
\cite{Pantev:2005zs}.}. In this context the intersection of two patches $U_i 
\cap U_j$ is given by the fibered product over $Y$
\begin{equation}
U_i 
\cap U_j=U_i 
\times_{Y} U_j\cong (V_{f}\times\mathcal{V}_{ij})/G_{ij},
\end{equation}
where $G_{ij}$ is a subgroup of $G_{i}\times G_{j}$ and $\mathcal{V}_{ij}$ is a 
quasiprojective variety.  Therefore, in order to define $\mathcal{A}(U_i 
\cap U_j)$ we have to be more careful, because $V_{f}\times\mathcal{V}_{ij}$ is 
not necessarily an affine space. The main difference with the usual LG orbifold 
case, is that in this case the branes $\mathcal{B}_{0}^{(g)}\in MF(W,G_{ij})$, 
$g\in G_{ij}$ are not necessarily all inequivalent objects. There can be a 
subgroup $H_{ij}\subseteq G_{ij}$ such that
\begin{equation}
\mathcal{B}_{0}^{(0)}\cong \mathcal{B}_{0}^{(h)}\qquad \text{for all \ }h\in 
H_{ij}, 
\end{equation}
then the algebra $\mathcal{A}(U_i \cap U_j)$ is defined by
\begin{equation}
\mathcal{A}(U_i \cap U_j)=\mathrm{End}_{MF(W,G_{ij})}\left(\bigoplus_{g\in 
G_{ij}/H_{ij}}\mathcal{B}_{0}^{(g)}\right).
\end{equation}
The subgroup $H_{ij}$ depends on the specific model we study. We illustrate it 
in an example below. However, we expect that $\mathcal{A}$ has the structure 
of a sheaf of algebras over $Y$ hence we must have the isomorphism
\begin{equation}\label{interisom}
\mathcal{A}(U_i)|_{U_i \cap U_j} \cong \mathcal{A}(U_j)|_{U_i \cap U_j}
\end{equation}
and the inclusions
\begin{equation}\label{interisom}
\mathcal{A}(U_i)|_{U_i \cap U_j} \hookrightarrow \mathcal{A}(U_i \cap 
U_j)\hookleftarrow \mathcal{A}(U_j)|_{U_i \cap U_j}.
\end{equation}
The category of matrix factorizations of the hybrid model defined on $Y$ is thus equivalent to the derived category of sheaves of $\mathcal{A}$-modules. In the next section, we apply these results to homological projective duality.

\subsubsection*{Example: Hybrid model on $\mathrm{Tot}(\mathcal{O}(-1) 
\rightarrow \mathrm{W}\mathbb{P}(2,3))$}

As an illustrative example, we derive the sheaf of $A_\infty$-algebra associated 
with a hybrid model on $Y=\mathrm{Tot}(\mathcal{O}(-1) \rightarrow 
\mathrm{W}\mathbb{P}(2,3))$. This is a simple example where we cannot write $Y$ 
as a global orbifold. Let $p_1$ and $p_2$ be the homogeneous coordinates of the 
weighted projective space, i.e. $(p_1,p_2) \sim (\lambda^2 p_1, \lambda^3 p_2)$. 
Assume that the hybrid model is defined by the superpotential $W = P_1 X^2 + P_2 
X^3$, where $x$ is the fiber coordinate. The reference brane $\mathcal{B}_0$ is 
given by the following matrix factorization
\begin{equation}\label{W23B0}
\xymatrix{
	\mathcal{O} \ar@<0.5ex>[rr]^-{X} & &
\mathcal{O}(-1) \ar@<0.5ex>[ll]^-{P_1 X + P_2 X^2}
}.
\end{equation}
Let $U_i$ be the open set defined by $p_i \neq 0$, then $U_1 \cong 
\mathbb{C}^{2}/\mathbb{Z}_2$ and $U_2 \cong \mathbb{C}^{2}/\mathbb{Z}_3$. The 
local coordinates are $(x_{1},z_1)$ in $U_1$ and $(x_{2},z_2)$ in $U_2$, where  
$x_1$ and $x_2$ denote the 
fiber coordinates. Then $z_{1}$ and $z_{2}$ satisfy $z_1^2 z_2^3 = 1$ on $U_1 
\cap U_2$. In $U_1$, the superpotential 
reads $W = x_1^2 + z_1 x_1^3$ and the generator of $\mathbb{Z}_2$ acts as 
$(x_1,z_1) \mapsto (-x_1,-z_1)$. In $U_2$, the superpotential reads $W = z_2 
x_2^2 + x_2^3$ and the generator of $\mathbb{Z}_3$ acts as $(x_2,z_2) \mapsto 
(\exp(-2\pi i/3) x_2, \exp(-2\pi i/3) z_2)$. In $U_1$, the reference brane 
\eqref{W23B0} can be written as $\mathcal{B}_{01}$:
\[
\xymatrix{
	\mathbb{C}_+ \ar@<0.5ex>[rr]^-{x_1} & &
\mathbb{C}_- \ar@<0.5ex>[ll]^-{x_1 + z_1 x_1^2}
},
\]
where the subscript $\pm$ indicates whether the complex plane is $\mathbb{Z}_2$-even or $\mathbb{Z}_2$-odd. As discussed before, the $A_\infty$-algebra $\mathcal{A}(U_1)$ is the $\mathbb{Z}_2$-invariant subspace of $\mathrm{End}(\mathcal{B}_{01}^{(0)} \oplus \mathcal{B}_{01}^{(1)})$, where $\mathcal{B}_{01}^{(0)} = \mathcal{B}_{01}$ and $\mathcal{B}_{01}^{(1)}$ is obtained from $\mathcal{B}_{01}$ by a $\mathbb{Z}_2$-twist, i.e. $\mathcal{B}_{01}^{(1)}$ is the matrix factorization
\[
\xymatrix{
	\mathbb{C}_- \ar@<0.5ex>[rr]^-{x_1} & &
\mathbb{C}_+ \ar@<0.5ex>[ll]^-{x_1 + z_1 x_1^2}
}.
\]
Therefore we have four independent $\mathbb{Z}_2$-invariant endomorphisms in $\mathrm{End}(\mathcal{B}_{01}^{(0)} \oplus \mathcal{B}_{01}^{(1)})$, namely $\mathrm{id} \sharp \pm 1$ and $\psi_1 \sharp \pm 1$, where $\mathrm{id} \sharp +1 (\mathrm{id} \sharp -1)$ is the identity map on $\mathcal{B}_{01}^{(0)} (\mathcal{B}_{01}^{(1)})$ and $\psi_1 \sharp +1 (\psi_1 \sharp -1)$ is the homomorphism in $\mathrm{Hom}(\mathcal{B}_{01}^{(0)}, \mathcal{B}_{01}^{(1)}) (\mathrm{Hom}(\mathcal{B}_{01}^{(1)}, \mathcal{B}_{01}^{(0)}))$ given by the matrix
\[ \left(
\begin{array}{cc}
0 & -i (1+z_1 x_1) \\ i & 0
\end{array} \right).
\]
The structure of the $A_\infty$-algebra is given by
\[
m_2(\psi_1,\psi_1) = 2,\quad m_3(\psi_1,\psi_1,\psi_1) = z_1.
\]

Similarly, in $U_2$, the reference brane \eqref{W23B0} can be written as $\mathcal{B}_{02}$:
\[
\xymatrix{
	\mathbb{C}_0 \ar@<0.5ex>[rr]^-{x_2} & &
\mathbb{C}_2 \ar@<0.5ex>[ll]^-{z_2 x_2 + x_2^2}
},
\]
where the subscript $a$ indicates that the complex plane has 
$\mathbb{Z}_3$-weight $\exp(2 \pi i a /3)$ for $a=0,1,2 \mathrm{ \ mod \ } 3$. 
The $A_\infty$-algebra $\mathcal{A}(U_2)$ is the $\mathbb{Z}_3$-invariant 
subspace of $\mathrm{End}(\oplus_{a=0,1,2} \mathcal{B}_{02}^{(a)})$, where 
$\mathcal{B}_{02}^{(0)} = \mathcal{B}_{02}$ and $\mathcal{B}_{02}^{(a)}$ is 
obtained from $\mathcal{B}_{02}$ by a $\mathbb{Z}_3$-twist, i.e. 
$\mathcal{B}_{01}^{(a)}$ is the matrix factorization
\[
\xymatrix{
	\mathbb{C}_{a} \ar@<0.5ex>[rr]^-{x_2} & &
\mathbb{C}_{a+2} \ar@<0.5ex>[ll]^-{z_2 x_2 + x_2^2}
}.
\]
We have six independent $\mathbb{Z}_3$-invariant endomorphisms in 
$\mathrm{End}(\oplus_{a=0,1,2} \mathcal{B}_{02}^{(a)})$, namely $\mathrm{id} 
\sharp \exp(2 \pi i a /3)$ and $\psi_2 \sharp \exp(2 \pi i a /3)$ for $a = 0,1,2 
\mathrm{ \ mod \ } 3$, where $\mathrm{id} \sharp \exp(2 \pi i a /3)$ is the 
identity map on $\mathcal{B}_{01}^{(a)}$ and $\psi_2 \sharp \exp(2 \pi i a /3)$ 
is  the homomorphism in $\mathrm{Hom}(\mathcal{B}_{02}^{(a)}, 
\mathcal{B}_{02}^{(a+2)})$ given by the matrix
\[ \left(
\begin{array}{cc}
0 & -i (z_2 + x_2) \\ i & 0
\end{array} \right).
\]
The structure of the $A_\infty$-algebra is given by
\[
m_2(\psi_2,\psi_2) = 2 z_2,\quad m_3(\psi_2,\psi_2,\psi_2) = 1.
\]
It is easy to check that $\psi_1$ and $\psi_2$ constitute a section of 
$\mathcal{O} \oplus \mathcal{O}(2)$. We see that $\mathcal{A}(U_1) = 
\mathcal{A}_0(U_1) \sharp \mathbb{Z}_2$ and $\mathcal{A}(U_2) = 
\mathcal{A}_0(U_2) \sharp \mathbb{Z}_3$, where $\mathcal{A}_0(U_i)$ is generated 
by $\psi_i$.

The intersection $U_1\cap U_{2}$ is given by the fibered product, as indicated 
above, then we find
\[
U_1\cap U_{2}=U_1\times_{Y} U_{2}\cong (\mathbb{C}\times 
\mathbb{C}^{*})/\mathbb{Z}_2 \times \mathbb{Z}_3.
\]
We denote the coordinates of $U_1\cap U_{2}$ as $(x_{12},z_{12})\in 
\mathbb{C}\times 
\mathbb{C}^{*}$. The generator of $\mathbb{Z}_2$ acts as
$(x_{12},z_{12})\rightarrow(-x_{12},-z_{12})$ and the generator of 
$\mathbb{Z}_3$ acts as
$(x_{12},z_{12})\rightarrow(x_{12},\exp(-2\pi i /3)z_{12})$. The relation 
between $(-x_{12},-z_{12})$ and the coordinates in the charts $U_{i}$ is given 
by
\[
z_{1}=z_{12}^{3},\qquad z_{2}=z_{12}^{-2},\qquad x_{12}=x_{1}=x_{2}z_{12}^{-1}.
\]
The superpotential in the intersection can be written as 
$W=x_{12}^{2}+z_{12}^{3}x_{12}^{3}$. Given any 
matrix factorization of $W$ in the intersection $U_1\cap U_{2}$, i.e. an object 
of $MF(W,\mathbb{Z}_2 \times \mathbb{Z}_3)$, defined over $\mathbb{C}\times 
\mathbb{C}^{*}$,  we can define a 
similarity transformation \cite{Walcher:2004tx} $z_{12}^{s}\mathbf{1}$, where 
$\mathbf{1}$ is the identity on the Chan-Paton space and $s\in \mathbb{Z}$. In 
$U_1\cap U_{2}$ clearly $z_{12}^{s}\mathbf{1}$ is invertible for any $s$ and it 
leaves invariant the endomorphism $Q$ but it changes $\rho_{M}$, shifting all 
its 
weights simultaneously by $(-1)^{s}\exp(-2\pi i s/3)$. So, it is clear from the 
argument above, the matrix factorization 
$\mathcal{B}^{(0)}_{01}|_{U_1\cap U_2}$, given by:
\[
\xymatrix{
	\mathbb{C}_{(+,0)} \ar@<0.5ex>[rr]^-{x_{12}} & &
\mathbb{C}_{(-,0)} \ar@<0.5ex>[ll]^-{x_{12} + z^{3}_{12} x_{12}^2}
}.
\]
where the subscript $\mathbb{C}_{(a,b)}$ labels the weights of the $\mathbb{Z}_2 
\times \mathbb{Z}_3$ representation, is equivalent to 
$\mathcal{B}^{(1)}_{01}|_{U_1\cap U_2}$, given by:
\[
\xymatrix{
	\mathbb{C}_{(-,0)} \ar@<0.5ex>[rr]^-{x_{12}} & &
\mathbb{C}_{(+,0)} \ar@<0.5ex>[ll]^-{x_{12} + z^{3}_{12} x_{12}^2}
}.
\]
Moreover, we can also show 
$\mathcal{B}_{01}^{(0)}|_{U_1\cap U_{2}}\cong 
\mathcal{B}_{02}^{(a)}|_{U_1\cap U_{2}}$, for any $a$ using the 
similarity transformation
\[
\left(
\begin{array}{cc}
0 & x_{12}+z^{3}_{12} x_{12}^2 \\ x_{12} & 0
\end{array} \right) = 
\left(
\begin{array}{cc}
z_{12} & 0 \\ 0 & 1
\end{array} \right)~
\left(
\begin{array}{cc}
0 & z_{12}^{-1} x_{12} + z_{12}^{2}x_{12}^2 \\ z_{12}x_{12} & 0
\end{array} \right)~
\left(
\begin{array}{cc}
z_{12}^{-1} & 0 \\ 0 & 1
\end{array} \right)
\]
composed with the transformation $z_{12}^{s}\mathbf{1}$ for appropriately 
chosen $s$. The induced transformation on the open string morphism, namely
\[
\left(
\begin{array}{cc}
0 & -i(1+z_1 x_1) \\ i & 0
\end{array} \right) = 
\left(
\begin{array}{cc}
z_2^{-\frac{1}{2}} & 0 \\ 0 & 1
\end{array} \right)~
\left(
\begin{array}{cc}
0 & -i(z_2 + x_2) \\ i & 0
\end{array} \right)~
\left(
\begin{array}{cc}
1 & 0 \\ 0 & z_2^{-\frac{1}{2}}
\end{array} \right)
\]
let us conclude that 
\[
\mathcal{A}(U_1)|_{U_{1}\cap U_{2}}\cong \mathcal{A}(U_1)|_{U_{1}\cap U_{2}} 
\]
as expected. We also conclude that $\mathcal{A}(U_1\cap U_2)$ is thus 
generated by a single element $\psi_{12}$ and the restriction map of the sheaf 
$\mathcal{A}$ is given by
\[
\begin{array}{cc}
\mathcal{A}(U_1) \rightarrow \mathcal{A}(U_1\cap U_2) & \mathcal{A}(U_2) 
\rightarrow \mathcal{A}(U_1\cap U_2) \\
\mathrm{id} \sharp \pm \mapsto \mathrm{id} & \mathrm{id} \sharp 
e^{2 \pi i a/3} \mapsto \mathrm{id}  \\
\psi_1 \sharp \pm \mapsto \psi_{12}  & \psi_2 \sharp e^{2 \pi i 
a/3} \mapsto \psi_{12}
\end{array}
\]

\section{\label{sec:examples}Examples of categories of B-branes on HPD 
phases}

In this section we apply the results from the previous section to HPD 
constructed from GLSMs introduced in \cite{Chen:2020iyo}. The Higgs branch 
category 
$\mathcal{C}$ defined in (\ref{HPDC}) takes the form
\begin{eqnarray}
\mathcal{C}=D(\widehat{Y}_{\zeta_{\mathcal{L}}\ll 
-1},\widehat{W}_{\zeta_{\mathcal{L}}\ll 
-1}),
\end{eqnarray} 
which generically corresponds to a hybrid model as the ones reviewed in 
section \ref{sec:hybrid}. We 
 analyze the following examples in detail:
\begin{itemize}
 \item  Degree $d$ Veronese embeddings.
\item Fano complete intersections in $\mathbb{P}^{n}$.
\end{itemize}

\subsection{HPD of Veronese embedding}\label{sec:HPDVeronese}

As reviewed in section \ref{sec:section2}, the HPD category (\ref{HPDC})  of 
degree-$d$ Veronese embedding of $\mathbb{P}^n= \bP(V)$ ($\dim 
V=n+1$), can be described by the category of B-branes 
on hybrid model with target space\footnote{The notation $\cO(m)$ with $m\in 
\mathbb{Q}$ denotes an orbibundle over $\mathbb{P}^{{n+d \choose d}-1}$. See 
\cite{Chen:2020iyo} or appendix \ref{app:orbibundle} a short review or 
\cite{adem_leida_ruan_2007} for 
details.} \cite{Chen:2020iyo}
\begin{eqnarray}\label{targethyb}
\mathrm{Tot}\left( \mathcal{O}\left( -\frac{1}{d} \right)^{\oplus(n+1)} \rightarrow \mathbb{P}^{{n+d \choose d}-1} \right)/\mathbb{Z}_d
\end{eqnarray} 
with superpotential
\begin{eqnarray}\label{superhyb}
W = \sum_{a=1}^{n+d \choose d} s_a f_a(x),
\end{eqnarray}
where $f_a$'s are the degree $d$ monomials in $x_i$, $i=0,\cdots,n$ and 
$s_{a}$ are homogeneous coordinates in $\mathbb{P}^{{n+d \choose d}-1}$. One 
can interpret the category of matrix factorizations of this LG model as the 
derived category of a noncommutative space $D(\mathbb{P}^{{n+d \choose d}-1}, 
\cA_0\sharp \mathbb{Z}_d)$, i.e. the category of sheaves of modules over the 
sheaf of $\Af$-algebras $\cA_0\sharp \mathbb{Z}_d$. Here, when restricted to a single fiber, the sheaf of $\Af$-algebra $\cA_0$ is the algebra of endomorphisms of the $D0$-brane we defined previously, so $\cA_0$ localizes to the base projective space. Let us denote by $Q_0$ the matrix factorization 
corresponding to the reference brane $\mathcal{B}^{(0)}_0$ given by
\begin{equation}
Q_0 = \sum_i \left(x_i \overline{\eta}_i + \frac{1}{d} \frac{\partial 
W}{\partial x_i} \eta_i \right),
\end{equation}
where we have chosen the trivial 
$\mathbb{Z}_{d}$ representation 
for the Clifford vaccum. Then, at a generic point $p\in \mathbb{P}^{{n+d 
\choose d}-1}$, $\mathcal{A}_{0,p}$ is given by the $A_{\infty}$-algebra with 
relations (\ref{algrels1}) and (\ref{algrels2}).

For completeness, let us write the generators of $D(\mathbb{P}^{{n+d \choose 
d}-1}, 
\cA_{0}\sharp \mathbb{Z}_d)$ as $\cA_{0}\sharp \mathbb{Z}_d$-modules. The 
B-brane $\mathcal{B}^{(0)}_{0}$ can be represented as the curved complex
\begin{equation}\label{mfB0}
	\xymatrix{
	{\cal O} \ar@<0.5ex>[r]^-{x \bar{\eta}} & {\cal O}(-\frac{1}{d})\otimes V 
\ar@<0.5ex>[l]^-{\frac{\partial W}{\partial x}\eta} \ar@<0.5ex>[r]^-{x 
\bar{\eta}} & 
{\cal O}(-\frac{2}{d})\otimes \wedge^2 V \ar@<0.5ex>[l]^-{\frac{\partial 
W}{\partial x}\eta} \ar@<0.5ex>[r]^-{ x\bar{\eta}} & 
\cdots \ar@<0.5ex>[l]^-{\frac{\partial W}{\partial x}\eta} \ar@<0.5ex>[r]^-{x 
\bar{\eta}} & 
{\cal O}(-\frac{n}{d})\otimes \wedge^n V \ar@<0.5ex>[l]^-{\frac{\partial 
W}{\partial x}\eta} \ar@<0.5ex>[r]^-{x \bar{\eta}}&
{\cal O}(-\frac{n+1}{d})\otimes \wedge^{n+1}V \ar@<0.5ex>[l]^-{\frac{\partial W}{\partial X}\eta},
	}
\end{equation}
where $\mathcal{O}(m)$ denotes orbibundles over $\mathbb{P}^{{n+d \choose 
d}-1}$. Define $\mathcal{B}^{(l)}_{0}$ to be the matrix factorization 
\eqref{mfB0} twisted by $\cO(l/d)$ for $l \in \mathbb{Z}$ and therefore the 
sheaf of $A_{\infty}$ modules corresponding to $\mathcal{B}^{(l)}_{0}$ is 
given by\footnote{Note that the $Hom$'s are taken over the orbifold category 
of the hybrid model defined by \eqref{targethyb}, \eqref{superhyb}.}
\begin{equation}\label{globalalghyb}
\cA^{(l)} := 
Hom\left(\bigoplus_{a=0}^{d-1}\mathcal{B}^{(a)}_{0},\mathcal{B}^{(l)}_{0}
\right)=\bigoplus_{a=0}^{d-1}\bigoplus_{k\geq 
0}\mathcal{O}\left(\frac{l-a}{d}-k\right)\otimes \wedge^{kd+a}V,
\end{equation}
where the sum over $k$ is such that $kd+a\leq n+1$
then, we can simplify (\ref{globalalghyb}) to
\begin{equation}
\cA^{(l)} = \bigoplus_{i=0}^{n+1} \cO \left( \frac{l-i}{d} \right) \otimes 
\wedge^i V.
\end{equation}
We can therefore write
\begin{equation}
D(\bP^{{n+d \choose d}-1},\cA_{0}\sharp \mathbb{Z}_d) = \langle 
\cA^{(1-C_{d,n})},\cdots,\cA^{(-1)},\cA^{(0)}\cong\mathcal{A}_{0} \rangle,
\end{equation}
where $C_{d,n} = d{n+d \choose d} - (n+1)$ is the expected number of factors 
obtained from the analysis of the Coulomb vacua in \cite{Chen:2020iyo}.



\subsection{HPD of Fano hypersurface in projective space}\label{sec:Fanohyper}

A degree $d\leq n$ Fano hypersurface in $\bP^n$, denoted $\mathbb{P}^{n}[d]$, 
can be described by a GLSM with $U(1)$ gauge group, $n+1$ chiral 
multiplets $x_i$ with gauge charge $1$, one chiral multiplet $p$ with gauge 
charge $-d$ and a superpotential
\begin{equation}
W_{\mathrm{Fano}} = p_0 F_{d}(x),
\end{equation}
where the polynomial $F_{d}(x)$, of degree $d$, is the defining equation 
of the hypersurface (which we assume to be smooth). We consider $x_{i}$ to be 
the 
coordinates on a complex vector space $V$ ($\dim 
V=n+1$), hence $\bP^n=\mathbb{P}(V)$. In this case, the equivalence 
(\ref{equivTX}) takes the form
\begin{equation}
D^bCoh(\bP^n[d]) = \langle 
MF(F_{d},\mathbb{Z}_{d}),\cO_{\mathbb{P}^{n}[d]},\cO_{\mathbb{P}^{n}[d]}(1),
\cdots,\cO_{\mathbb{P}^{n}[d]}(n-d) \rangle. 
\end{equation}
The associated Lefschetz decomposition is
\begin{equation}\label{LDpd}
D^b(\bP^n[d]) = \langle \cA_0,\cA_1,\cdots,\cA_{n-d} \rangle,
\end{equation}
where $\cA_0 := \langle MF(F_{d},\mathbb{Z}_{d}),\cO_{\mathbb{P}^{n}[d]} 
\rangle$, $\cA_i := \langle \cO_{\mathbb{P}^{n}[d]} \rangle$, $i>0$.
The small window category $\mathcal{W}^{\mathcal{L}}_{-,b}$ of this GLSM, 
defined in (\ref{constrwin}) consists of B-branes with charges $q$ 
satifying
\begin{equation}\label{chargescondt}
\left| q + \frac{\theta}{2 \pi} \right| < \frac{d}{2}.
\end{equation}
We choose the theta angle (i.e. the integer $b$) as $\theta=\pi d-\varepsilon$, 
with $0<\varepsilon\ll 1$, so $q$ takes values 
$-(d-1),-(d-2),\cdots,-2,-1,0$.

\begin{figure}[!h]
 \centering
\begin{tikzpicture}[inner sep=0in,outer sep=0in]
\node (n) {
 \centering  
    \begin{ytableau}
    \none[{\cal A}_0] & \none[{\cal A}_1]& \none[{\cal A}_2]& \none[{\cal A}_3]  & \none & \none[\cdots] &\none& \none[{\cal A}_{\scaleto{n-d-1\mathstrut}{4pt}}] & \none[{\cal A}_{\scaleto{n-d\mathstrut}{4pt}}]        & \none \\
    *(gray)           & *(gray) & *(gray) & *(gray)                  & \none & \none[\cdots]        &\none & *(gray) &  *(gray)    &  & \none  &  \none[\cdots]   &  \none &  &  \\
    *(gray)           &     & &              & \none & \none[\cdots] & \none         & &             &        & \none  &  \none[\cdots]   &  \none  &    &   \\
    *(gray)           &     & &              & \none & \none[\cdots] & \none         & &             &        & \none  &  \none[\cdots]   &  \none  &    &   \\
    *(gray)           &     & &              & \none & \none[\cdots] & \none         & &             &        & \none  &  \none[\cdots]   &  \none  &    & \\
    *(gray)           &      & &            & \none & \none[\cdots]  & \none        & &             &   & \none  &  \none[\cdots]   &  \none  &    & \\
    \none             &\none[{\cal B}_{\scaleto{n-1\mathstrut}{4pt}}]                 &\none[{\cal B}_{\scaleto{n-2\mathstrut}{4pt}}] &\none[{\cal B}_{\scaleto{n-3\mathstrut}{4pt}}]        &\none & \none[\cdots]   &  \none   &\none[{\cal B}_{d+1}] &\none[{\cal B}_{d}] &\none[{\cal B}_{d-1}] & \none  &  \none[\cdots]   &  \none & \none[{\cal B}_1] &\none[{\cal B}_0] \\
    \end{ytableau}
    };
\draw[thin,] (3.5,1.48) -- (-3,1.48);
\draw[thin,] (3.5,-1.48) -- (-3.3,-1.48);
\end{tikzpicture}
    \caption{Lefschetz decomposition of the hypersurface $\mathbb{P}^n[d]$ and the dual Lefschetz decomposition of the HPD.}\label{HPDCI}
\end{figure}
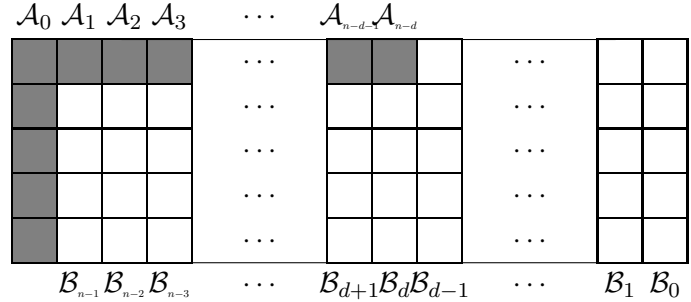

As shown in \cite{Chen:2020iyo}, the universal hyperplane section can be 
described by the geometric phase (with FI parameters lying in the first 
quadrant) of the GLSM $\mathcal{T}_{\mathcal{X}}$, this corresponds to a GLSM 
with gauge group $U(1)_{\mathcal{L}}\times U(1)$ and with matter content
\[
\begin{array}{ccccccccccc}
& x_0 & x_1 & \cdots & x_n & p_0 & p & \cdots & y_0 & \cdots & y_n \\
U(1)_{\mathcal{L}} & 1 & 1 & \cdots & 1 & -d & -1 & \cdots & 0 & \cdots & 0 \\
U(1)_{1} & 0 & 0 & \cdots & 0 & 0 &  -1   & \cdots & 1 & \cdots & 1
\end{array}
\]
and superpotential
\[
\widehat{W} = p_0 F_{d}(x) + p \sum_{i=0}^n x_i y_i.
\]
The HPD of $\bP^n[d]$ with Lefschetz decomposition given by \eqref{LDpd} can be 
described by the Higgs branch of the phase of 
$\mathcal{T}_{\mathcal{X}}$ corresponding to the FI parameters lying in the 
second quadrant: $(\zeta_{\mathcal{L}}\ll -1,\zeta_{1}\gg 1)$. This is 
a hybrid model with target space
\[
Y:=\mathrm{Tot}\left( \cO_{\check{\mathbb{P}}^{n}}^{\oplus(n+1)}\oplus 
\cO_{\check{\mathbb{P}}^{n}}(-1) \rightarrow \check{\bP}^n \right)/\mathbb{Z}_d 
\]
and superpotential
\begin{equation}\label{potentialCIHPD}
W = F_{d}(x) + p \sum_{i=0}^n x_i y_i,
\end{equation}
where $\check{\bP}^n:=\bP(V^{\vee})$, $x_i$'s are fiber coordinates of 
$\cO_{\check{\mathbb{P}}^{n}}^{\oplus(n+1)}$, $p$ is the fiber coordinate of 
$\cO_{\check{\mathbb{P}}^{n}}(-1)$ and the $y_i$'s are homogeneous 
coordinates on the base $\check{\bP}^n$. The $\mathbb{Z}_d$ orbifold acts with 
weight $1$ on the $x_{i}$'s and $-1$ on $p$. Denote the category of B-branes of 
this hybrid model as $D(Y,W)$.
Then $D(Y,W)$ has a (dual) Lefschetz decomposition that takes the form (as 
proposed in \cite{Chen:2020iyo}):
\begin{equation}
D(Y,W) = \langle \mathcal{B}_{n-1}(1-n), 
\mathcal{B}_{n-2}(2-n),\cdots, \mathcal{B}_2(-2), \mathcal{B}_1(-1), 
\mathcal{B}_0 \rangle,
\end{equation}
where we have the equivalence of categories 
$\mathcal{B}_{0}\cong\mathcal{A}_{0}$. Denote the functor 
implementing this equivalence by $\mathcal{F}$:
\[
\mathcal{F}: \cA_0 \rightarrow \mathcal{B}_0.
\]
Then $\mathcal{B}_i = \langle 
\mathcal{F}(MF(F_{d},\mathbb{Z}_{d})), \mathcal{F}(\mathcal{O}) \rangle$ for 
$ 0 \leq i \leq d$, $\mathcal{B}_j = \langle 
\mathcal{F}(MF(F_{d},\mathbb{Z}_{d})) \rangle$ for $ d+1 \leq j \leq n-1$.
The relationship between the Lefschetz decomposition and its dual decomposition 
is illustrated by figure \ref{HPDCI}.
Next we describe the functor $\mathcal{F}$ explicitly. Define the matrix 
factorization $Q'$ of $p \sum_{i=0}^n x_i y_i$ as
\[
\xymatrix{
\cO_{\check{\mathbb{P}}^{n}}(1) \ar@<0.5ex>[rr]^-p & & 
\cO_{\check{\mathbb{P}}^{n}} \ar@<0.5ex>[ll]^-{\sum_{i=0}^n x_i y_i}.
}
\]
Any matrix factorization $\mathcal{M} \in MF(F_{d},\mathbb{Z}_{d})$ can be 
lifted to a GLSM B-brane with $U(1)_{\mathcal{L}}$ charges in the small 
window \eqref{chargescondt}. On the other hand the category 
$\widehat{\mathcal{W}}^{\mathcal{L}}_{-,b}\cong D(Y,W)$ in \eqref{HPDC} can be 
chosen (by adjusting $b$) such that the $U(1)_{\mathcal{L}}$ charges $q'$ in 
the $\mathcal{T}_{\mathcal{X}}$ model satisfy $q'\in\{ 
-(d-1),-(d-2),\cdots,-1,0,1\}$. Then, the tensor product $\mathcal{M} \otimes 
Q'$ has $U(1)_{\mathcal{L}}$ charges belonging to 
$\widehat{\mathcal{W}}^{\mathcal{L}}_{-,b}$.
 The same is true for the B-brane\footnote{Here we should think of 
$\mathcal{O}_{\mathbb{P}^{n}[d]}$ as its lift to a matrix factorization for the 
GLSM $\mathcal{T}_{X}$, with theta angle $\theta=\pi d-\varepsilon$ and 
$U(1)_{\mathcal{L}}$ charges $q\in\{-d,0\}$. So, it does not belong 
to $\mathcal{W}^{\mathcal{L}}_{-,b}$, but it does belong to 
$\widehat{\mathcal{W}}^{\mathcal{L}}_{-,b}$ upon tensoring with $Q'$.} 
$\mathcal{O}_{\mathbb{P}^{n}[d]} \otimes Q'$. We conclude that the functor is 
given by
\[
\mathcal{F}(\mathcal{M}) = \mathcal{M} \otimes Q'.
\]

For every fixed point on the base, the superpotential along the fiber is given 
by \eqref{potentialCIHPD} with fixed $y_i$. The reference brane 
$\mathcal{B}^{(0)}_{0}$ can be written as
\[
Q_{0} = \sum_{i=0}^n (x_i \overline{\eta}_i) + p\overline{\eta}_{n+1} + 
\sum_{i=0}^n \left( \frac{1}{d}\frac{\partial F_{d}}{\partial x_i} + 
\frac{1}{2} p y_i \right) \eta_i + \frac{1}{2} \left( \sum_{i=0}^n x_i y_i 
\right) \eta_{n+1},
\]
where the trivial represenation of $\mathbb{Z}_{d}$ is chosen for the 
Clifford vacuum. Because the superpotential \eqref{potentialCIHPD} has a 
quadratic term $p \sum_{i=0}^n x_i y_i$ and a degree-$d$ term $F_{d}(x)$, the 
structure of the $\cA_{0}$ factor in the sheaf of $\Af$-algebras 
$\cA_{0}\sharp \mathbb{Z}_{d}$ is determined by the relations\footnote{At a 
generic point $y\in \mathbb{P}^{n}$ the superpotential $W|_{y}$ satisfies 
$dW|_{y}^{-1}(0)=\{ 0\}$, hence the assumed condition on the potential of 
LG orbifold is fulfilled.}
\begin{equation}\label{afmCI1}
m_2(\psi_i,\psi_j) + m_2(\psi_j,\psi_i) = \left.\frac{\partial^2 W}{\partial 
x_i \partial x_j}\right|_{x_i=0},
\end{equation}
\begin{equation}\label{afmCI2}
m_d(\psi_{i_1},\psi_{i_2},\cdots,\psi_{i_d}) = \left. \frac{\partial^d 
W}{\partial x_{i_1} \cdots \partial x_{i_d}} \right|_{x_i=0}
\end{equation}
at each point of the base, where we have identified $p$ with $x_{n+1}$. 

As in the case of Veronese embedding, the global sheaf structure of 
$\cA_{0}\sharp \mathbb{Z}_{d}$ can also be read off from the global behavior 
of $x_i$ and $p$, we present two examples for illustration.

\subsubsection*{Example: Quadrics}

In the case $d=2$, at each point of the base, the superpotential is quadratic in 
the fiber coordinates. Therefore, the sheaf of algebra is the Clifford algebra 
associated with the quadratic form given by $\frac{\partial^{2}W}{\partial 
x_{i}\partial x_{i}}$, $i=0,\ldots,n+1$, $x_{n+1}:=p$ at fixed $y_i$. We can 
take the reference brane $\mathcal{B}^{(0)}_{0}$ to be given by the matrix 
factorization
\[
Q_{0}=\sum_{i=0}^n (x_i \overline{\eta}_i) + p \overline{\eta}_{n+1} + 
\sum_{i=0}^n \left( \frac{1}{2}\frac{\partial F_{2}}{\partial x_i} + 
\frac{1}{2} p y_i \right) \eta_i + \frac{1}{2} \left( \sum_{i=0}^n y_i x_i 
\right) \eta_{n+1},
\]
hence, the curved complex associated to $\mathcal{B}^{(0)}_{0}$ is given by
\begin{equation}\label{B0qua}
	\xymatrix{
	{\cal O}_{+} \ar@<0.5ex>[r]^-{} & 
{\begin{array}{c} {\cal O}_{-}\otimes V \\ \oplus \\ \cO_{-}(-1) \end{array}} \ar@<0.5ex>[l]^-{} \ar@<0.5ex>[r]^-{} &  
\cdots \ar@<0.5ex>[l]^-{} \ar@<0.5ex>[r]^-{} & 
{\begin{array}{c}
\cO_{(-)^{n-1}}\otimes \wedge^{n+1} V \\ \oplus \\ \cO_{(-)^{n-1}}(-1)\otimes \wedge^n V \end{array}} \ar@<0.5ex>[l]^-{} \ar@<0.5ex>[r]^-{}&
{\cal O}_{(-)^n}(-1)\otimes \wedge^{n+1}V \ar@<0.5ex>[l]^-{},
	}
\end{equation}
where all the sheaves $\cO_{(-)^{i}}(a)$ in \eqref{B0qua} denote sheaves over 
$\mathbb{P}^{n}$ and the subindex $\pm$ indicates the $\mathbb{Z}_2$-weight of the sheaf. A similar computation as the 
one in section \ref{sec:HPDVeronese} let us conclude that
\begin{equation}
\cA_{0} \cong \left(\bigoplus_{i=0}^{n+1} \cO_{(-)^i} 
\otimes \wedge^i V \right) \oplus \left(\bigoplus_{i=0}^{n+1} 
\cO_{(-)^{i+1}}(-1) \otimes \wedge^{i} V \right)
\end{equation}
globally. So, we can write
\begin{equation}
D(Y,W)\cong D(\mathbb{P}^{n},\cA_{0}\sharp\mathbb{Z}_{2} )
\end{equation}
i.e., the hybrid model B-brane category $D(Y,W)$ is equivalent to the derived 
category of sheaves of $A_{\infty}$ $\cA_{0}\sharp\mathbb{Z}_{2}$-modules. 

\subsubsection*{Example: Cubic hypersurfaces}

The $\mathcal{B}^{(0)}_{0}$-brane is given by
\[
Q_{0} = \sum_{i=0}^n (x_i \overline{\eta}_i) + p \overline{\eta}_{n+1} + 
\sum_{i=0}^n \left( \frac{1}{3}\frac{\partial F_{3}}{\partial x_i} + 
\frac{1}{2} p y_i \right) \eta_i + \frac{1}{2} \left( \sum_{i=0}^n x_i y_i 
\right) \eta_{n+1},
\]
and its associated curved complex is
\begin{equation}\label{B0cub}
	\xymatrix{
	{\cal O}_{0} \ar@<0.5ex>[r]^-{} & 
{\begin{array}{c} {\cal O}_{1}\otimes V \\ \oplus \\ \cO_{-1}(-1) \end{array}} \ar@<0.5ex>[l]^-{} \ar@<0.5ex>[r]^-{} &  
\cdots \ar@<0.5ex>[l]^-{} \ar@<0.5ex>[r]^-{} & 
{\begin{array}{c}
\cO_{n+1}\otimes \wedge^{n+1} V \\ \oplus \\ \cO_{n-1}\otimes \wedge^n V \end{array}} \ar@<0.5ex>[l]^-{} \ar@<0.5ex>[r]^-{}&
{\cal O}_{n}\otimes \wedge^{n+1}V \ar@<0.5ex>[l]^-{},
	}
\end{equation}
where the subscripts of the line bundles are the $\mathbb{Z}_3$-weights. At each 
point of the base, the $\Af$-algebra is given by
\[
m_2(\psi_i,\psi_{n+1}) + m_2(\psi_{n+1},\psi_i) = y_i,
\]
\[
m_3(\psi_i,\psi_j,\psi_k) = \frac{\partial^3 F_{3}}{\partial x_i \partial x_j 
\partial x_k}.
\]
Globally, the sheaf of $\Af$-algebra $\cA_{D0}$ is
\[
\cA_{D0} = \left(\bigoplus_{i=0}^{n+1} \cO_i \otimes \wedge^i V \right) \oplus 
\left(\bigoplus_{i=-1}^{n-1} \cO_i(-1) \otimes \wedge^{i+1} V \right) \oplus 
\left( \cO_n \otimes \wedge^{n+1}V \right).
\]
Therefore, the HPD of $\bP^n[3]$ is the noncommutative space 
$(\check{\bP}^n,\cA_{0}\sharp \mathbb{Z}_{3})$.

\subsection{HPD of complete intersections}\label{sec:HPDCIII}

The method can also be applied to the HPD of Fano complete 
intersections of the form $\bP^n[d_1,d_2,\cdots,d_k]$, $\sum_{\alpha=1}^{k}d_{\alpha}<n+1$. 
From the GLSM construction, it is straightforward to see that the HPD can be described by the hybrid model on the space
\begin{equation}\label{compTot}
Y=\mathrm{Tot}\left( \mathcal{O}(-1,0)^{\oplus(n+1)} \oplus \mathcal{O}(1,-1) 
\rightarrow \mathrm{W}\mathbb{P}(d_1,\cdots,d_k) \times \check{\mathbb{P}}^n 
\right)
\end{equation}
with superpotential
\begin{equation}\label{compSp}
W = \sum _{\alpha=1}^k p_{\alpha} F_{d_{\alpha}}(x) +  p\sum_{i=0}^n x_i y_i,
\end{equation}
where $p_{\alpha}$ are homogeneous coordinates of the weighted projective space 
$\mathrm{W}\mathbb{P}(d_1,\cdots,d_k)$, $y_i$ are homogeneous coordinates of 
$\check{\mathbb{P}}^n$, $x_i$ and $p$ are coordinates along the fibers of 
$\mathcal{O}(-1,0)^{\oplus(n+1)}$ and $\mathcal{O}(1,-1)$ respectively.

As in the case of hypersurfaces, $\mathcal{A}_0$ is spanned by 
$\psi_{0},\ldots, \psi_{n+1}$ and the $\Af$-products of $\cA_0$ 
are determined by
\[
m_2(\psi_i,\psi_j) + m_2(\psi_j,\psi_i) = \left.\frac{\partial^2 W}{\partial 
x_i \partial x_j}\right|_{x_i=0},
\]
\[
m_{d_{\alpha}}(\psi_{i_1},\psi_{i_2},\cdots,\psi_{i_d}) = \left. 
\frac{\partial^d W}{\partial x_{i_1} \cdots \partial x_{i_d}} \right|_{x_i=0}
\]
at each point of the base $\mathrm{W}\mathbb{P}[d_1,\cdots,d_k] \times 
\check{\mathbb{P}}^n$, where we have identified $x_{n+1}:=p$. However, 
$\mathcal{A}_0$ is not the sheaf of algebra corresponding to the hybrid model 
under consideration, because we cannot ignore the orbifold singularity coming 
from the affine patches 
of $\mathrm{W}\mathbb{P}[d_1,\cdots,d_k]$ and, for generic $d_{\alpha}$, we 
cannot write the space $Y$ as a global orbifold. Exceptions are, for instance 
if $d_{\alpha}=d$ for all $\alpha$, then we have a noncommutative resolution of 
$\mathbb{P}^{k-1}\times\check{\mathbb{P}}^n$. Otherwise we have to deal with a 
sheaf of algebras over a singular space. Therefore we must resort to the general 
framework outlined in section \ref{sec:hybrid}. Here we just illustrate the idea 
by a case study, which can be easily genearalized.


\subsubsection*{Example: $\mathbb{P}^n[2,3]$}

We study the HPD of the complete intersection of a quadric hypersurface and a 
cubic hypersurface in $\mathbb{P}^n$, denoted by $X$. The GLSM for this 
complete intersection is a $U(1)$ gauge theory with the following matter content 
and gauge charges:
\[
\begin{array}{cccccc}
X_0 & X_1 & \cdots & X_n & P_1 & P_2 \\
1 & 1 & \cdots & 1 & -2 & -3
\end{array}
\]
and the superpotential is
\begin{equation}\label{WP23}
W_0 = P_1 G_2(X) + P_2 G_3(X),
\end{equation}
where $G_2$ and $G_3$ are quadric and cubic polynomials in $X_i$ respectively. We assume $n \geq 4$. Suppose that $\zeta$ is the FI parameter of this model. Then the geometric phase ($\zeta \gg 1$) is a nonlinear sigma model with target space the complete intersection defined by $G_2=G_3=0$ in $\mathbb{P}^n$. 

For $\zeta \ll -1$ and generic $G_2, G_3$, the Higgs branch of the GLSM above is a hybrid model on
\begin{equation}\label{totP23}
Y_0 \equiv \mathrm{Tot} \left( \mathcal{O}(-1)^{\oplus (n+1)} \rightarrow \mathrm{W}\mathbb{P}(2,3) \right)
\end{equation}
with the superpotential given by \eqref{WP23}, where $P_1, P_2$ are homogeneous 
coordinates on the weighted projective space $\mathrm{W}\mathbb{P}(2,3)$ and 
$X_i$'s are the fiber coordinates. This hybrid model can be viewed as two LG 
orbifold theories, each defined on a coordinate patch of 
$\mathrm{W}\mathbb{P}(2,3)$, glued together in a consistent way. Let's denote by 
$U_i$ the coordinate patch where $P_i \neq 0$. Then we have
\[
U_1 \cong \mathbb{C} / \mathbb{Z}_2, \quad U_2 \cong \mathbb{C} / \mathbb{Z}_3.
\]
Assume that $Z_1$ and $Z_2$ are local coordinates of $U_1$ and $U_2$ respectively. It is easy to see that when $P_1 \neq 0$, the hybrid model reduces to a LG model with $\mathbb{Z}_2$ orbifold on $\mathbb{C}^{\oplus(n+2)}$, with the $\mathbb{Z}_2$-action:
\[
e^{\pi i} \cdot (Z_1, X_i) = (e^{3 \pi i} Z_1, e^{-\pi i} X_i) = (-Z_1, -X_i) 
\]
and superpotential
\[
W_{01} = G_2(X) + Z_1 G_3(X).
\]
Similarly, when $P_2 \neq 0$, the hybrid model reduces to a LG model with $\mathbb{Z}_3$ orbifold on $\mathbb{C}^{\oplus(n+2)}$, with the $\mathbb{Z}_3$-action:
\[
e^{\frac{2 \pi i}{3}} \cdot (Z_2, X_i) = (e^{\frac{4 \pi i}{3}} Z_2, e^{-\frac{2 \pi i}{3}} X_i) = (e^{-\frac{2 \pi i}{3}} Z_2, e^{-\frac{2 \pi i}{3}} X_i) 
\]
and superpotential
\[
W_{02} = Z_2 G_2(X) + G_3(X).
\]
Matrix factorizations of the hybrid model are of the form $(\mathcal{E},Q)$, where $\mathcal{E}$ is a coherent sheave with $\mathbb{Z}_2$-grading on $Y_0$ defined in \eqref{totP23}, and $Q$ is an odd endomorphism of $\mathcal{E}$ that squres to $W_0 \cdot \mathrm{id}$. When restricted to a coordinate patch, the orbifold action on $\mathcal{E}$ can be read off from the sheaf structure. For example, sections of $\mathcal{O}(a)$ transform as $\exp(\pi i) \cdot \lambda = \exp(a \pi i) \lambda$ under $\mathbb{Z}_2$ in $U_1$ while they transform as $\exp(2 \pi i/3) \cdot \lambda = \exp(a 2 \pi i/3) \lambda$ under $\mathbb{Z}_3$ in $U_2$. Also, $Q(X,P_1,P_2)$ is replaced by $Q(X,1,Z_1)$ in $U_1$ and is replaced by $Q(X,Z_2,1)$ in $U_2$.

There is a semiorthogonal decomposition of the derived category of coherent sheaves of the complete intersection $X$:
\[
D(X) = \langle MF(Y_0,W_0), \mathcal{O}_X, \mathcal{O}_X(1),\cdots, \mathcal{O}_X(n-4) \rangle,
\]
therefore the Lefschetz decomposition associated with the embedding $X \hookrightarrow \mathbb{P}^n$ is\footnote{Note that $n \ge 5$ in the Fano case. In the Calabi-Yau case, $n = 4$, and $D(X) \cong MF(Y_0,W_0)$.}
\begin{equation}\label{Lef23}
D(X) = \langle \mathcal{A}_0, \mathcal{A}(1), \cdots, \mathcal{A}(n-5) \rangle,
\end{equation}
where $\mathcal{A}_0 = \langle MF(Y_0,W_0), \mathcal{O}_X \rangle$ if $n \geq 5$, $\mathcal{A}_0 = MF(Y_0,W_0)$ if $n =4$, and $\mathcal{A}_i = \langle \mathcal{O}_X \rangle$ for $i > 0$.

From our general construction, the GLSM for the universal hyperplane section $\mathcal{X}$ of the embedding $X \hookrightarrow \mathbb{P}^n$ is a $U(1) \times U(1)$ GLSM with the following matter content and gauge charges:
\[
\begin{array}{ccccccccccc}
& X_0 & X_1 & \cdots & X_n & P_1 & P_2 & P & Y_0 & \cdots & Y_n \\
U(1)_1 & 1 & 1 & \cdots & 1 & -2 & -3 & -1 & 0 & \cdots & 0 \\
U(1)_2 & 0 & 0 & \cdots & 0 & 0 & 0 & -1 & 1 & \cdots & 1
\end{array}
\]
together with the superpotential
\begin{equation}\label{WPP23}
W = P_1 G_2(X) + P_2 G_3(X) + P \sum_{i=0}^n X_i Y_i.
\end{equation}
The geometric phase ($\zeta_1 \gg 1, \zeta_2 \gg 1$) realizes the universal hyperplane section $\mathcal{X}$. The semiorthogonal decomposition of $D(\mathcal{X})$ corresponding to \eqref{Lef23} is
\[
D(\mathcal{X}) = \langle \mathcal{C}, \mathcal{A}_1(1) \boxtimes D(\check{\mathbb{P}}^n), \cdots \mathcal{A}_{n-5}(n-5) \boxtimes D(\check{\mathbb{P}}^n) \rangle,
\]
where $\mathcal{C}$ is the HPD category. 

From the general argument, $\mathcal{C}$ consists of the B-branes of the Higgs branch on the phase with $\zeta_1 \ll -1, \zeta_2 \gg 1$. This Higgs branch is described by a hybrid model on
\[
Y \equiv \mathrm{Tot}\left( \mathcal{O}(-1,0)^{\oplus(n+1)} \oplus \mathcal{O}(1,-1) 
\rightarrow \mathrm{W}\mathbb{P}(2,3) \times \check{\mathbb{P}}^n 
\right)
\]
with superpotential \eqref{WPP23}, where $P_1, P_2$ are homogeneous coordinates of the weighted projective space 
$\mathrm{W}\mathbb{P}(2,3)$, $Y_i$ are homogeneous coordinates of 
$\check{\mathbb{P}}^n$, $X_i$ and $P$ are coordinates along the fibers of 
$\mathcal{O}(-1,0)^{\oplus(n+1)}$ and $\mathcal{O}(1,-1)$ respectively. Again, we can think of this hybrid model as a family of LG orbifolds defined on coordinate patches of $\mathrm{W}\mathbb{P}(2,3) \times \check{\mathbb{P}}^n$ glued together consistently.
The reference brane $\mathcal{B}_0$ has the tachyon profile
\[
Q_{0} = \sum_{i=0}^n X_i \bar{\eta}_i + P \bar{\eta}_{n+1} + \sum_{i=0}^n \left( \frac{1}{2} P_1 \frac{\partial G_2}{\partial X_i} + \frac{1}{3} P_2 \frac{\partial G_3}{\partial X_i} \right) \eta_i + \left( \sum_{i=0}^n X_i Y_i \right) \eta_{n+1},
\]
where the vacuum state of the Clifford algebra is identified with $\mathcal{O}$ 
on $\mathrm{W}\mathbb{P}(2,3) \times \check{\mathbb{P}}^n$. Notice that $X_i$'s 
are sections of $\mathcal{O}(-1,0)$ and $P$ is a section of $\mathcal{O}(1,-1)$, 
we see that the sheaf of algebras $\mathcal{A}=End_{MF(Y,W)}(\mathcal{B}_0)$ 
can be written globally as
\[
\mathcal{A} = \bigoplus_{m=0}^{n+1} \left( \mathcal{O}(-m,0) \oplus \mathcal{O}(1-m,-1) \right) \otimes \wedge^m V,
\]
where $V = \mathbb{C}^{n+1}$. The $A_\infty$ structure of $\mathcal{A}$ can be analyzed in the same way as the example shown in section \ref{sec:hybrid}.

\section{The functor $D(B,Mod-\mathcal{A}) \rightarrow 
D(\mathcal{X})$}\label{sec:functor}

In this section we give a description of the functor from the category 
$D(B,Mod-\mathcal{A}):=D(Y,W)$ used to describe the B-brane category of the 
hybrid models on $Y=\mathrm{Tot}(\mathcal{V}\rightarrow B)$ that arises as HPD 
of Fano embeddings. Given a projective embedding, the HPD category $\mathcal{C}$ 
is a subcategory of 
$D(\mathcal{X})$ by definition, where $\mathcal{X}$ is the universal hyperplane 
section. Therefore there exists a fully faithful functor from the proposed HPD 
category $D(B,Mod-\mathcal{A})$ to $D(\mathcal{X})$, realizing the 
equivalence $D(B,Mod-\mathcal{A}) \cong \mathcal{C}$, where 
$D(B,Mod-\mathcal{A})$ is the derived category of $A_\infty$ 
$\mathcal{A}$-modules. The GLSM construction gives a simple description of this 
functor.

First, we have a functor $\mathcal{F}: D(B, Mod-\mathcal{A}) \rightarrow 
MF(\mathrm{Tot}(\mathcal{V}\rightarrow B),W)$. This functor can be defined by 
generalizing the functor from $D(Mod-\mathcal{A}_{D0})$ to the category of 
matrix factorizations of LG models described in section \ref{sec:Af_homo} (cf. 
eq. \eqref{QM}). Suppose we have an object $\mathcal{N}$ in $D(B, 
Mod-\mathcal{A})$, i.e. $\mathcal{N}$ is a sheaf of $A_\infty$ module of 
$\mathcal{A}$ with $\mathcal{A}_\infty$ actions given by $m^{\mathcal{N}}_i$, $i 
\geq 1$. The matrix factorization $\mathcal{F}(\mathcal{N})$ can be defined as 
follows. The Chan-Paton sheaf of $\mathcal{F}(\mathcal{N})$ is
\[
\mathcal{M} = \mathcal{N} \otimes \mathrm{Sym} \mathcal{V}^\vee.
\]
Let $x_i, i=1,\cdots,\mathrm{rank}(\mathcal{V})$ be a set of linearly independent sections\footnote{If $B=\cup_i U_i$ is an open cover, one can choose a basis $\{ e^{(i)}_1, \cdots, e^{(i)}_{\mathrm{rank}\mathcal{V}}\}$ of $\mathcal{V}|_{U_i}$ for every open patch $U_i$, then $x_j$ can be defined in $U_i$ as the linear function such that $x^{(i)}_j(e^{(i)}_k) = \delta_{jk}$.} of $\mathcal{V}^\vee$, then $x_{i_1} x_{i_2} \cdots x_{i_k}$ is a section of $\mathrm{Sym} \mathcal{V}^\vee$ for any set of indices $i_1,\cdots,i_k$. Consequently, the Tachyon profile $Q_{\mathcal{M}}$ of $\mathcal{F}(\mathcal{N})$ can be defined by
\[
Q_{\mathcal{M}}(\phi) = \sum_{k} \sum_{i_1,\cdots,i_k} m^{\mathcal{N}}_{k+1}(\phi,\psi_{i_1},\cdots,\psi_{i_k})x_{i_1} x_{i_2} \cdots x_{i_k},
\]
where $\phi$ is a section of $\mathcal{N}$. This functor is an equivalence with the inverse functor given by $\mathcal{B} \rightarrow Hom_{MF(\mathrm{Tot}(\mathcal{V}\rightarrow B),W)}(\mathcal{B}_0,\mathcal{B})$, where $\mathcal{B}_0$ is the reference brane that is point-like along each fiber of $\mathcal{V}$. In particular, $\mathcal{A} = End_{MF(\mathrm{Tot}(\mathcal{V}\rightarrow B),W)}(\mathcal{B}_0)$.


Now we can use the GLSM $\mathcal{T}_\mathcal{X}$ defined in eq.\eqref{exGLSM} to construct a fully faithful functor $\mathcal{G}$: \[
\mathcal{G}: MF(\mathrm{Tot}(\mathcal{V}\rightarrow B),W) \rightarrow D(\mathcal{X}).
\]
Recall that the hybrid model on $\mathrm{Tot}(\mathcal{V}\rightarrow B)$ with 
superpotential $W$ is the Higgs branch of the GLSM $\mathcal{T}_\mathcal{X}$ in 
the phase with $\zeta_\mathcal{L} \ll -1, \zeta' \gg 1$ (LG phase), while the 
Higgs branch of the phase with $\zeta_\mathcal{L} \gg 1, \zeta' \gg 1$ 
(geometric phase) is the nonlinear sigma model with target space $\mathcal{X}$. 
Consequently, the functor $\mathcal{G}$ can be implemented by the brane 
transport across the classical phase boundary at $\zeta_\mathcal{L} = 0, 
\zeta' \gg 1$. More precisely, for each matrix factorization in the 
category $MF(\mathrm{Tot}(\mathcal{V}\rightarrow B),W)$, we first lift it to a 
matrix factorization of the GLSM $\mathcal{T}_\mathcal{X}$. Then, after possibly 
combining this matrix factorization with several branes that are empty in the LG 
phase, we can transport it to the geometric phase and project it to an object in 
$D(\mathcal{X})$. Here the grade restriction rule states that the combined brane 
must be in the small window category of the local model associated with the 
phase boundary. Thus we see the brane transport realizes the functor 
$\mathcal{G}$. There are examples in \cite{Chen:2020iyo} demonstrating how this 
procedure works.

In conclusion, the fully faithful functor from $D(B, Mod-\mathcal{A})$ to 
$D(\mathcal{X})$ is given by $\mathcal{G} \circ \mathcal{F}$, which is an 
equivalence between $D(B,Mod-\mathcal{A})$ and the HPD category $\mathcal{C}$.

\section*{Acknowledgements}

We would like to thank Zhuo Chen for collaboration at an early stage
of this work. We also would like to thank Nils Carqueville, Will Donovan,
David Favero, Daniel Pomerleano, Johanna Knapp, Tsung-Ju Lee, Yun Shi,
Eric Sharpe, Emanuel Scheidegger and Philsang Yoo for useful discussions and 
comments. JG acknowledges support from the China Postdoctoral Science Foundation 
No. 2020T130353. MR thanks Fudan U. for hospitality. MR acknowledges support 
from the National Key Research and Development Program of China, grant No. 
2020YFA0713000, and the Research Fund for International Young Scientists, NSFC 
grant No. 11950410500.

\appendix

\section{Checkings on the $(Mod-\mathcal{A}_{D0})\rightarrow MF(W)$ 
functor}\label{app:checks}

Here we provide several checks for the proposed functor \eqref{QM}. 

{\bf Check 1. $d=2$ Case.} When $d=2$, we have $\cA_{D0} = Cl(n,\mathbb{C})$, 
the 
Clifford algebra defined by the quadratic form $\frac{\partial^2 W}{\partial 
x_i 
\partial x_j}$. The correspondence between the matrix factorization for 
quadratic superpotentials and Clifford modules is given in 
\cite{Kapustin:2002bi,Bertin:2008osh}, which matches \eqref{QM}.

{\bf Check 2. $\mathbf{N}=\mathcal{A}_{D0}$.} It is straightforward to check 
\eqref{QM} for the case the module is $\mathcal{A}_{D0}$ itself. Then \eqref{QM}
corresponds to $Q_{D0}$. In fact, as shown in \cite{ballard2014derived}, the 
fermionic generators 
$\psi_i, i=1,\cdots,n$, satisfy
\begin{itemize}
\item If $d=2$, $\cA_{D0}$ is the Clifford algebra:
\begin{equation}\label{2m2}
m_2(\psi_i,\psi_j)+m_2(\psi_j,\psi_i) = \frac{\partial^2 W}{\partial x_i 
\partial x_j}.
\end{equation}
\item If $d>2$, $\cA_{D0}$ is an $\Af$-algebra, where $m_2$ satisfies
\begin{equation}\label{dm2}
m_2(\psi_i,\psi_j)+m_2(\psi_j,\psi_i)=0,
\end{equation} 
$m_k=0$ for $k=1,\cdots,d-1$ and
\begin{equation}\label{dmd}
m_d(\psi_{i_1},\psi_{i_2},\cdots,\psi_{i_d}) = \frac{1}{d!}\frac{\partial^d 
W}{\partial 
x_{i_1} \cdots \partial x_{i_d}}.
\end{equation}
\end{itemize}
Thus when $d>2$, if we identify $m_2(\cdot, v_i)$ with 
$\{\overline{\eta}_i,\cdot\}$ according to \eqref{dm2}, then \eqref{dmd} tells 
us that $m_d(\cdot, \psi_{i_1},\psi_{i_2},\cdots,\psi_{i_{d-1}})$ should be 
identified 
with 
\[
\frac{1}{d!} \sum_{i=1}^n \frac{\partial^d W}{\partial x_i \partial x_{i_1} 
\cdots \partial x_{i_{d-1}}} \{ \eta_i, \cdot \}.
\]
Because $W$ is homogeneous with degree $d$, we see that $Q_M$ defined by 
\eqref{QM} is exactly $Q_{D0}$ defined by \eqref{D0} in this case.

{\bf Check 3. $Q_M^2 = W \cdot \mathrm{id}$.}. Here we will show that the 
object $Q_M^2$ is indeed a matrix factorization of $W$. We will make the 
assumption that $m^\mathbf{N}_s = 
0$ for $s>d$ (which can be shown below to be true for the case $n=1$), it can 
be shown that $Q_M^2 = W \cdot 
\mathrm{id}$. 
For example, when $d=3$
\[
Q_M(\phi) = \sum_{ij} m^\mathbf{N}_3(\phi,\psi_i,\psi_j)x_ix_j + \sum_i 
m_2^\mathbf{N}(\phi,\psi_i)x_i.
\]
Therefore
\begin{equation}\label{QM3ext}
\begin{split}
Q_M^2(\phi) &= \sum_{ijkl} 
m_3^\mathbf{N}(m_3^\mathbf{N}(\phi,\psi_i,\psi_j),\psi_k,\psi_l)x_ix_jx_kx_l + 
\sum_{ijk} m_3^\mathbf{N}(m_2^\mathbf{N}(\phi,\psi_i),\psi_j,\psi_k)x_ix_jx_k \\
&- \sum_{ijk} m_2^\mathbf{N}(m^\mathbf{N}_3(\phi,\psi_i,\psi_j),\psi_k)x_ix_jx_k 
+ \sum_{ij} m_2^\mathbf{N} 
(m_2^\mathbf{N}(\phi,\psi_i),\psi_j)x_ix_j.
\end{split}
\end{equation}
From \eqref{ainftyrels}, we get
\[
m^\mathbf{N}_2(m^\mathbf{N}_2(\phi,\psi_i),\psi_j) = 
m^\mathbf{N}_2(\phi,m_2(\psi_i,\psi_j)),
\]
then the last term of \eqref{QM3ext} vanishes due to \eqref{dm2}. The first 
term 
of \eqref{QM3ext} also vanishes because of \eqref{ainftyrels}, \eqref{dmd} and 
$m_k(\cdots,1,\cdots) = 0$ for $k>2$. Also from \eqref{ainftyrels} and 
\eqref{dmd}, the second and third terms of \eqref{QM3ext} yield
\[
\begin{split}
&\sum_{ijk} \left( m_3^\mathbf{N}((m_2^\mathbf{N}((\phi,\psi_i),\psi_j,\psi_k) 
- 
m_2^\mathbf{N}((m^\mathbf{N}(_3(\phi,\psi_i,\psi_j),v_k) \right)x_ix_jx_k\\
& = \sum_{ijk} m_2^\mathbf{N}((\phi,m_3(\psi_i,\psi_j,\psi_k))x_ix_jx_k = 
\frac{1}{3!} \phi 
\sum_{ijk} \frac{\partial^3 W}{\partial x_i \partial x_j \partial x_k} 
x_ix_jx_k 
= W \cdot \phi,
\end{split}
\]
which shows $Q_M^2 = W \cdot \mathrm{id}$.

{\bf Check 4. $n=1$ case.} Finally, we show that the functor reproduces the 
matrix factorizations for the 
case $n=1$, i.e. $W = x^d$. In this case, the $D0$-brane 
is 
given by the matrix factorization
\begin{equation}
Q_{D0} = x \overline{\eta} + x^{d-1} \eta.
\end{equation}
The fermionic generator of $\cA_{D0} = Hom(\mathcal{B}_{D0},\mathcal{B}_{D0})$ 
is $\psi = \overline{\eta} - x^{d-2} \eta$. Let $\mathcal{B}_l$ be the matrix 
factorization with
\begin{equation}\label{MFmm}
Q_M = x^l \overline{\pi} + x^{d-l} \pi
\end{equation}
where $1<l<d$ and $\{\pi,\pi\} = \{ \overline{\pi}, \overline{\pi} \} = 0$. 
Next, we will show that \eqref{QM} recovers $Q_{M}$. Start by considering the 
bosonic state $\phi_0 \in Hom_0(\mathcal{B}_{D0},\mathcal{B}_l)$ and the 
fermionic state $\phi_1 \in Hom_1(\mathcal{B}_{D0},\mathcal{B}_l)$. If the 
vacuum state of $M$ is denoted as $|\Omega \rangle$, then
\[
\phi_0 |0\rangle = |\Omega \rangle, \quad \quad \phi_0 \overline{\eta}|0\rangle 
= x^{l-1} \overline{\pi} |\Omega \rangle,
\]
and
\[
\phi_1 |0\rangle = \overline{\eta} |\Omega \rangle, \quad \quad \phi_1 
\overline{\eta}|0\rangle = -x^{d-l-1} |\Omega \rangle.
\]
In matrix form,
\[
\psi = \left( \begin{array}{cc} 0 & -x^{d-2} \\ 1 & 0 \end{array} \right),\quad
\phi_0 = \left( \begin{array}{cc} 1 & 0 \\ 0 & x^{l-1} \end{array} \right),\quad
\phi_1 = \left( \begin{array}{cc} 0 & -x^{d-l-1} \\ 1 & 0 \end{array} \right).
\]
Using the algorithm reviewed in section \ref{sec:LGAinf}, one can compute 
($\iota \psi = v$)\footnote{Another computation for this result can be found 
in 
the example at the end of section \ref{sec:Af_homo}.}
\[
m_k(\psi^{\otimes k}) = 0,\quad f_k(\psi^{\otimes k}) = (-1)^k x^{d-k-1} 
\eta,\quad 
1<k<d,
\] 
and
\[
m_d(\psi^{\otimes d}) = 1,\quad f_d(\psi^{\otimes d}) = 0.
\]
By composing the homomorphisms, one gets
\[
\phi_0 \circ \psi = x^{l-1} \phi_1,\quad \phi_1 \circ \psi = -x^{d-l-1} \phi_0.
\]
Therefore,
\[
\phi_0 \circ \psi = \mathrm{d} \tilde{\phi}^{(1)}_0 = \tilde{\phi}^{(1)}_0 
Q_{D0} - 
Q_M \tilde{\phi}^{(1)}_0,\quad \phi_1 \circ \psi = -\mathrm{d} 
\tilde{\phi}^{(d-l-1)}_1 = -\tilde{\phi}^{(d-l-1)}_1 Q_{D0} - Q_M 
\tilde{\phi}^{(d-l-1)}_1,
\]
where
\[
\tilde{\phi}^{(1)}_0 = \left( \begin{array}{cc} 0 & 0 \\ 0 & x^{l-2} 
\end{array} 
\right), \quad
\tilde{\phi}^{(d-l-1)}_1 = \left( \begin{array}{cc} 0 & x^{d-l-2} \\ 0 & 0 
\end{array} \right),
\]
from which one can deduce that\footnote{We use $\phi_i$ to denote the 
cohomology 
class and the representative, the meaning should be clear from the context.}
\[
m^\mathbf{N}_2(\phi_0,v) = 0, \quad f^\mathbf{N}_2(\phi_0,v) = 
-\tilde{\phi}^{(1)}_0,
\]
and
\[
m^\mathbf{N}_2(\phi_1,\psi) = 0, \quad f^\mathbf{N}_2(\phi_1,\psi) = 
\tilde{\phi}^{(d-l-1)}_1.
\]
It can be shown by induction that
\[
m^\mathbf{N}_{k+1}(\phi_0,\psi^{\otimes k}) = 0, \quad 
f^\mathbf{N}_{k+1}(\phi_0,\psi^{\otimes k}) = 
-\tilde{\phi}^{(k)}_0,\quad 1<k<l,
\]
where
\[
\tilde{\phi}^{(k)}_0 = \left( \begin{array}{cc} 0 & 0 \\ 0 & x^{l-k-1} 
\end{array} \right).
\]
Similarly,
\[
m^\mathbf{N}_{k+1}(\phi_1,\psi^{\otimes k}) = 0, \quad 
f^\mathbf{N}_{k+1}(\phi_1,\psi^{\otimes k}) = 
\tilde{\phi}^{(d-l-k)}_1,\quad 1<k<d-l,
\]
where
\[
\tilde{\phi}^{(k)}_1 = \left( \begin{array}{cc} 0 & x^{k-1} \\ 0 & 0 
\end{array} 
\right).
\]
Now, one can compute
\[
\iota m^\mathbf{N}_{l+1}(\phi_0,\psi^{\otimes l}) = 
-f^\mathbf{N}_l(\phi_0,\psi^{\otimes (l-1)}) \circ 
\psi - \phi_0 \circ f_l(\psi^{\otimes l}) = \phi_1,
\]
similarly $\iota m^N_{d-l+1}(\phi_1,\psi^{\otimes (d-l)}) = \phi_0$,
and all the higher order multiplications vanish. Therefore, in the basis 
$\{\phi_0,\phi_1\}$, \eqref{QM} yields
\[
Q_M = \left( \begin{array}{cc} 0 & 0 \\ 1 & 0 \end{array} \right) x^l + \left( 
\begin{array}{cc} 0 & 1 \\ 0 & 0 \end{array} \right) x^{d-l} = \left( 
\begin{array}{cc} 0 & x^{d-l} \\ x^l & 0 \end{array} \right),
\]
which is exactly the matrix factorization \eqref{MFmm} we started with.

\section{$\Af$-algebras defined by ribbon trees}\label{app:ribbon}

The structure of the $\Af$-algebra $\cA=\mathrm{End}(\mathcal{B}_{D0})$ 
corresponding to a Landau-Ginzburg model with 
homogeneous superpotential was derived in \cite{ballard2014derived} using the 
method of summing over ribbon trees. In this appendix, we review the idea of 
\cite{ballard2014derived} and generalize it to LG models with inhomogeneous 
superpotentials.

Let $\iota$ be an embedding of $H(\cA):=H_{m^{\mathcal{A}}_{1}}(\cA)$ into 
$\cA$. If we define the 
projection $\pi: \cA \rightarrow H(\cA)$ such that $\pi \circ \iota = 1$ and 
there is a map $h: \cA \rightarrow \cA$ of degree $-1$ such that $1 - \iota 
\circ \pi = m^{\mathcal{A}}_{1}\circ h+ h \circ m^{\mathcal{A}}_{1}$ and $h^2 = 
\pi h = h \iota = 0$, then the 
$\Af$ products $m_{k}:H(\cA)^{\otimes k}\rightarrow H(\cA)$, $k\geq 2$ can be 
defined by summing over the contributions from 
ribbon trees \cite{Kontsevich:2000yf}:
\begin{equation}\label{sumtree}
m_k = \sum_T m_{k,T}.
\end{equation}
For a LG model with degree-$d$ superpotential, the ribbon trees contributing to 
the sum have one root and $d$ leaves such that the valency of any vertex is 2 or 
3  \cite{ballard2014derived}. \eqref{sumtree} is a solution to the defining 
relations \eqref{ainftyrels}.

In our convention, $\iota(\psi_i) = v_i$ defined by \eqref{generator}, 
consequently $h$ can be defined to be $h = \sum_i \eta_i 
\frac{\partial}{\partial x_i}$ where $\eta_i$ acts by multiplication in the Clifford algebra.

By definition, a ribbon tree is a tree $T$ with a collection of vertices, external edges and internal edges such that: (a) Each external edge is incident to a single vertex. (b) Each internal edge is incident to exactly two vertices. (c) One of the external edge is the root, the other external edges are the leaves. Every ribbon tree $T$ with one root and $k$ leaves determines a term $m_{k,T}$ in \eqref{sumtree}.

Given a tree $T$, to compute $m_{k,T}(\psi_{i_1},\psi_{i_2},\cdots,\psi_{i_k})$ 
we put $\psi_{i_1},\psi_{i_2},\cdots,\psi_{i_k}$ on the leaves from left to 
right and then act on them a series of maps as follows:
\begin{itemize}
\item Each leaf gives a map $\iota$;
\item Each bivalent vertex gives a map $f$;
\item Each internal edge gives a map $h$;
\item Each trivalent vertex corresponds to the multiplication in $\cA$;
\item The root gives the map $\pi$
\end{itemize} 
while reading from the top to the bottom. Here $f$ is defined by
\[
\left\{ \sum_i \frac{\partial W}{\partial x_i} \eta_i, \cdot \right] - \sum_i 
\frac{\partial W}{\partial x_i} \eta_i ,
\]
where $\{ ]$ denotes commutator/anticommutator depending on whether the second argument is of even/odd degree and the second term is the usual multiplication of the Clifford algebra.

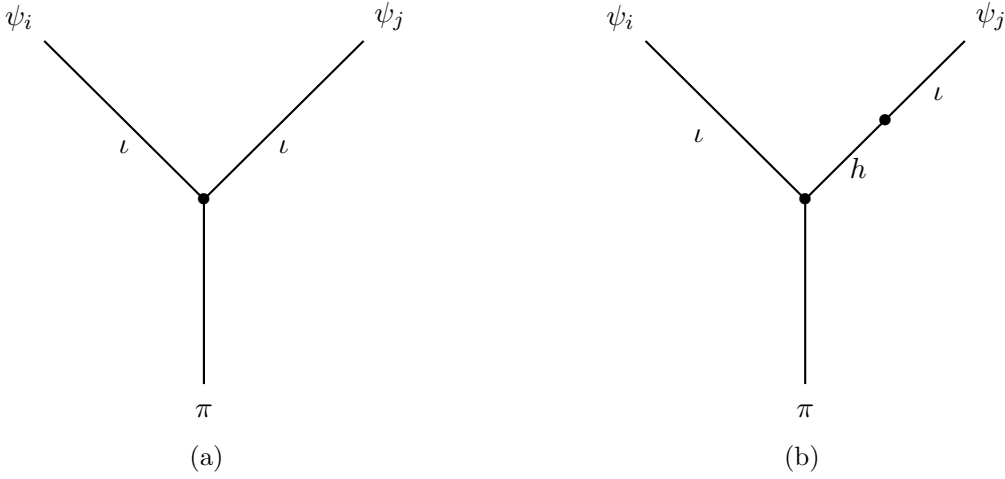
\begin{figure}
\centering
 \begin{subfigure}[b]{0.5\textwidth}
 \begin{tikzpicture}[scale=0.7]
  	\draw[thick] (0,0) -- (3,3) node[anchor=south west]{$\psi_j$};
	\draw[thick] (0,0) -- (-3,3) node[anchor=south east]{$\psi_i$};
	\draw [thick] (0,0) -- (0,-3.5);
    \node at (0,-4) {$\pi$};
    \node at (0,0) {$\bullet$};
    \node at (-1.5,1) {$\iota$};
    \node at (1.5,1) {$\iota$};
 \end{tikzpicture}
 \caption{\qquad\qquad\qquad\qquad}\label{fig:tree2_a}
\end{subfigure}
\begin{subfigure}[b]{0.4\textwidth}
 \begin{tikzpicture}[scale=0.7]
  	\draw[thick] (0,0) -- (3,3) node[anchor=south west]{$\psi_j$};
	\draw[thick] (0,0) -- (-3,3) node[anchor=south east]{$\psi_i$};
	\draw[thick] (0,0) -- (0,-3.5);
    \node at (0,-4) {$\pi$};
    \node at (0,0) {$\bullet$};
    \node at (1.5,1.5) {$\bullet$};
    \node at (-2,1.2) {$\iota$};
    \node at (1,0.6) {$h$};
    \node at (2.5,2) {$\iota$};
 \end{tikzpicture}
 \caption{\qquad\qquad}\label{fig:tree2_b}
\end{subfigure}
\caption{(a) Ribbon tree contributing to $m_2$. (b) Another contribution to $m_2$ when $d=2$.}
\end{figure}

\begin{figure}
\centering
 \begin{tikzpicture}[scale=0.8]
  	\draw[thick] (0,0) -- (1,1);
    \draw[thick] (1,1) -- (-4,6) node[anchor=south east]{$\psi_{i_2}$};
	\draw[thick] (0,0) -- (-6,6) node[anchor=south east]{$\psi_{i_1}$};
	\draw[thick] (0,0) -- (0,-3);
    \draw[thick, dotted] (1,1) -- (3,3);
    \draw[thick] (3,3) -- (6,6) node[anchor=south west]{$\psi_{i_k}$};
    \draw[thick] (3,3) -- (0,6) node[anchor=south]{$\psi_{i_{k-2}}$};
    \draw[thick] (4,4) -- (2,6) node[anchor=south]{$\psi_{i_{k-1}}$};
    \node at (1,1) {$\bullet$};
    \node at (5,5) {$\bullet$};
    \node at (4,4) {$\bullet$};
    \node at (3,3) {$\bullet$};
    \node at (0,-3.5) {$\pi$};
    \node at (0,0) {$\bullet$};
    \node at (-2.3,6.3) {$\cdots\cdots$};
 \end{tikzpicture}
 \caption{Ribbon tree contributing to $m_k$.}\label{fig:treek}
\end{figure}
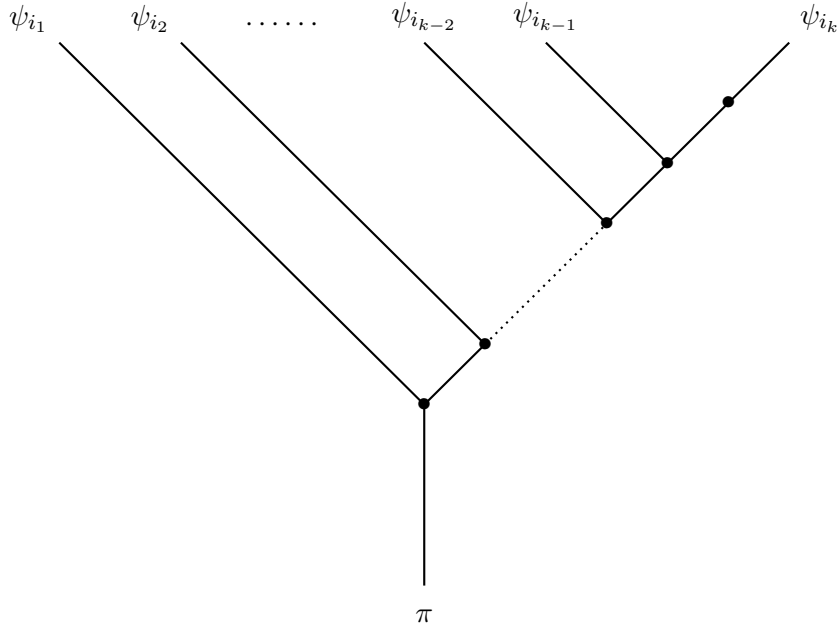

It is shown in \cite{ballard2014derived} that there is always a tree given by 
Figure.\ref{fig:tree2_a} making a nontrivial contribution to $m_2$. For $d>2$, 
this is the only contribution and it makes $m_2$ to satisfy 
$m_2(\psi_i,\psi_j)+m_2(\psi_j,\psi_i)=0$. When $d=2$, there is another 
nontrivial contribution from the tree given by Figure.\ref{fig:tree2_b}. The 
effect of Figure.\ref{fig:tree2_b} is to modify $m_2$ such that 
$m_2(\psi_i,\psi_j)+m_2(\psi_j,\psi_i) = \frac{\partial^2 W}{\partial x_i 
\partial x_j}$.
In general, other than the tree in Figure.\ref{fig:tree2_a}, the only ribbon 
tree that can make a nonzero contribution is the one in Figure.\ref{fig:treek}. 
If the input of the tree is $\psi_{i_1},\psi_{i_2},\cdots,\psi_{i_k}$, then 
before hitting the root, the image of the set of maps encoded in the tree is 
$\frac{1}{k!}\frac{\partial^k W}{\partial x_{i_1}\partial x_{i_2}\cdots\partial 
x_{i_k}}$ plus some $Q$-exact terms. When $k \neq d$, this image is $Q$-exact 
and annihilated by the projection $\pi$, so the output of the tree is zero. When 
$k=d$, the output is $\frac{1}{k!}\frac{\partial^k W}{\partial x_{i_1}\partial 
x_{i_2}\cdots\partial x_{i_k}}$ because it is not $Q$-exact.    
In summary, we have
\begin{itemize}
\item If $d=2$, $\cA$ is a Clifford algebra given by:
\[
m_2(\psi_i,\psi_j)+m_2(\psi_j,\psi_i) = \frac{\partial^2 W}{\partial x_i 
\partial x_j}.
\]
\item If $d>2$, $m_2$ is the wedge product, $m_k=0$ for $k \neq 2$ and $k \neq d$.
\[
m_d(\psi_{i_1},\psi_{i_2},\cdots,\psi_{i_d}) = \frac{\partial^d W}{\partial 
x_{i_1} \cdots \partial x_{i_d}}.
\]
\end{itemize}
Now assume we have a inhomogeneous superpotential of the form
\[
W = \sum_{l=2}^d W^{(l)},
\]
where $\deg W^{(l)} = l$. Because now every nonvanishing derivative 
$\frac{\partial^l W^{(l)}}{\partial x_{i_1} \partial x_{i_2} \cdots \partial 
x_{i_l}}$ is not $Q$-exact, we can use the same argument to deduce that
\begin{equation}\label{eqsinhomog1}
m_2 (\psi_i,\psi_j) + m_2(\psi_j,\psi_i) = \frac{\partial^2 W^{(2)}}{\partial 
x_i \partial x_j}
\end{equation}
and
\begin{equation}\label{eqsinhomog2}
m_l (\psi_{i_1},\psi_{i_2},\cdots,\psi_{i_l}) = \frac{1}{l!} \frac{\partial^l 
W^{(l)}}{\partial x_{i_1} \partial x_{i_2} \cdots \partial x_{i_l}}
\end{equation}
for $3 \leq l \leq d$.

\section{Orbibundle}\label{app:orbibundle}

Let $X$ be a smooth manifold admitting a $G$-action, where $G$ is a group.
An orbibundle on the quotient stack $[X/G]$ is a fiber bundle $E 
\stackrel{\pi}{\rightarrow} X/G $ with each fiber an orbifold. Explicitly, let 
$V$ be a vector space admitting a representation of $G$:
\[
\rho:\quad G \rightarrow GL(V).
\]
The fibre of $E$ is $V/\rho(G)$. If $\{U_\alpha:~\alpha \in I\}$ is an open 
cover of $X/G$ and
\[
\phi_\alpha: U_\alpha \times V/\rho(G) \rightarrow \pi^{-1}(U_\alpha)
\]
are the corresponding local trivializations. Then the transition functions 
$g_{\alpha\beta} = \phi^{-1}_\alpha \circ \phi_\beta$ take values in 
$GL(V)/\rho(G)$. A local section of $E$ is given by a $\rho(G)$-invariant 
function $s_\alpha: U_\alpha \rightarrow V$ so the relation
\[
s_\alpha = g_{\alpha\beta}\cdot s_\beta
\]
is well defined on $U_\alpha \cap U_\beta$. Given a representation of $G$ as 
above, the orbibundles on $[X/G]$ are classified by $H^1(X,GL(V)/\rho(G))$. When 
the representation is trivial, the orbibundle is just an ordinary vector bundle. 
When $\dim_{\mathbb{C}}V=1$, we call it a line bundle.

A morphism between two orbibundles $E_1\stackrel{\pi_1}{\rightarrow} X/G$ and 
$E_2\stackrel{\pi_2}{\rightarrow} X/G$ is a bundle map $f: E_1 \rightarrow E_2$, 
i.e. $\pi_2 \circ f = \pi_1$. Given local trivializations of $E_1$ and $E_2$ in 
an open set $U$:
\[
\phi_1: U \times V_1/\rho_1(G) \rightarrow \pi_1^{-1}(U),
\]
\[
\phi_2: U \times V_2/\rho_2(G) \rightarrow \pi_2^{-1}(U),
\]
and for each $x \in U$,
$f_U(x) := \phi_2^{-1} \circ f \circ \phi_1|_x$ is a linear map from $V_1$ to 
$V_2$ satisfying
\[
f_U(x) \circ \rho_1(g) = \rho_2(g) \circ f_U(x)
\]
for all $g \in G$.

\section*{Conflict of interest statement}

On behalf of all authors, the corresponding author states that there is no conflict of interest.










\bibliographystyle{fullsort}
\bibliography{ncref.bib}
\end{document}